\documentclass[aps,prb,preprint,groupedaddress,superscriptaddress]{revtex4-1}
\usepackage{graphicx,longtable}
\usepackage{rotating,color}
\usepackage{textcomp}
\usepackage{amsmath}
\usepackage[utf8]{inputenc}

\usepackage{array}
\newcolumntype{L}[1]{>{\raggedleft\arraybackslash}m{#1}}
\newcolumntype{C}[1]{>{\centering\arraybackslash}m{#1}}
\newcolumntype{R}[1]{>{\raggedright\arraybackslash}m{#1}}

\newcommand{\ntc}{black}

\begin{document}

\title{\textcolor{\ntc}{Accurate description of charged excitations in molecular solids from embedded many-body perturbation theory} }

\author{Jing Li}
\author{Gabriele D'Avino}
\email{gabriele.davino@neel.cnrs.fr}
\affiliation{Institut N\'{e}el, CNRS and Grenoble Alpes University, F-38042 Grenoble, France}

\author{Ivan Duchemin}
\affiliation{INAC, SP2M/L$\_$Sim, CEA/UJF Cedex 09, 38054 Grenoble, France}

\author{David Beljonne}
\affiliation{Laboratory for Chemistry of Novel Materials, University of Mons, BE-7000 Mons, Belgium}

\author{Xavier Blase}
\affiliation{Institut N\'{e}el, CNRS and Grenoble Alpes University, F-38042 Grenoble, France}
\email{xavier.blase@neel.cnrs.fr}

\date{\today}

\begin{abstract}
We present a novel hybrid quantum/classical (QM/MM) approach to the calculation of charged excitations in molecular solids based on the many-body Green's function $GW$ formalism.
Molecules described at the $GW$ level are embedded into the crystalline environment modeled with an accurate classical polarizable scheme.
This allows the calculation of electron addition and removal energies in the bulk and at crystal surfaces where charged excitations are probed in photoelectron experiments.
By considering the paradigmatic case of pentacene and perfluoropentacene crystals, we discuss the different contributions from intermolecular interactions to electronic energy levels, distinguishing between polarization, which is accounted for combining quantum and classical polarizabilities, and crystal field effects, that can impact energy levels by up to $\pm0.6$ eV.
After introducing band dispersion, we achieve quantitative agreement (within 0.2 eV) on the ionization potential and electron affinity measured at pentacene and perfluoropentacene crystal surfaces characterized by standing molecules. 
\end{abstract}


\maketitle


\section{Introduction}
\label{s:intro} \label{s:}

The ability to accurately predict the energies of charged excitation from first principles is of primary importance for the computational study of organic conjugated materials that find applications in electronic and opto-electronic devices,
since phenomena such as charge injection from a metal electrode, electron-hole separation in solar cells or their radiative recombination in light-emitting diodes do all crucially depend on the energetics of electronic energy levels.\cite{Ish99,Cah03,Hei11,Dav16_rev}

The achievement of quantitative accuracy on quantities such as the ionization potential (IP) or the electron affinity (EA) is definitely a challenging task for theory.
Well-grounded approximations and efficient implementations are both required to make calculations on systems counting a large number of atoms accurate and feasible at the same time.
An additional hurdle comes from the subtle effect of intermolecular interactions in the solid state.
In fact, while IP and EA are well-defined properties of a given (isolated) molecule, the same quantities can present variations that can exceed 0.6 eV between different solid samples of the same compound.\cite{Duh08,Sal08,Che08}
Such a large variability originates from intermolecular interactions of electrostatic nature and reflects in charged excitations energetics that depend on morphology and, in crystals, on the facet though which electron are injected or extracted.\cite{Duh08}
IP and EA are therefore not intrinsic properties of a given compound but instead depend on the molecular organization, e.g. amorphous vs. crystal or standing vs. lying molecules.\cite{Hei11,Dav16_rev} 

Many-body perturbation theory (MBPT) techniques, such as the Green's function $GW$ formalism,\cite{Hed65,Oni02} stand as the state-of-the-art for the first principles description of charged (quasiparticle) excitations in condensed matter.
Originally developed within the solid state physics community, the $GW$ formalism has been extensively applied to inorganic solids in the last decades,\cite{Hyb86,God88} leading to a substantial improvement over density functional theory (DFT) in the description of the electronic band gap.
The $GW$ formalism is gaining increasing attention also in the context of organic systems, with several applications to extended solids reported recently using periodic boundary condition implementations,\cite{Tia03,Sha12,Ran16,Yan14,Wan16} 
\textcolor{\ntc}{
including studies of the band structure of the prototypical molecular semiconductor pentacene (PEN)\cite{Tia03,Sha12,Ran16} and subsequent investigations of optical properties within the Bethe-Salpeter formalism \cite{Tia03,Sha12,Ran16,Dra09,Sha13,Cud15,Wan16,Len16}. }

A severe limitation of periodic bulk calculations is that they cannot attain the absolute value of charged excitations due to the missing internal reference for the energy of a free-electron.
The use of slab geometries allows the definition of a consistent vacuum level and hence the calculation of absolute values for IP and EA.
However, such calculations come at a high computational cost and are difficult to converge with respect to slab and vacuum thickness, making this route impractical especially for organic solids with many atoms in the unit cell.
Kang {\it et al.} \cite{Kan16} very recently proposed a strategy to obtain IP and EA combining bulk $GW$ calculations with DFT slab calculations, the latter permitting to refer $GW$ quasiparticle energies to the appropriate surface-specific vacuum level.\cite{Kan16}
Such an approach is, however, loosely consistent since it does not account for the reduced screening at crystal surfaces, hence results in a systematic underestimation of the gap.\cite{Kan16}

The development of $GW$ implementations on Gaussian atomic orbital bases greatly facilitated $GW$ calculations of finite, aperiodic, systems.\cite{Bla11a,Bau12,Bru12,Set13,Kap15}
The $GW$ formalism achieved a very accurate description of quasiparticle energies and gap in isolated molecules, as demonstrated by extensive benchmarks against gas-phase experiments \cite{Tia08,Pal09,Ma09,Bla11a,Foe11,Mar12,Laf15,Anh13}
and high-level quantum chemistry calculations.\cite{Set15,Ran17,Fab11a,Klo15,Kap16,Kni16}
Thanks to efficient algorithms and parallel implementations, $GW$ calculations enabled accurate calculations on systems exceeding hundred atoms.\cite{Duc12,Nie15,Li17,Bau12,Neu14,Gov15}
Yet, charged excitations in extended systems are largely governed by long-range electrostatic interactions that are not amenable to a full QM treatment.

In this paper, we present a novel hybrid quantum/classical (QM/MM) approach  to quasiparticle excitations combining a state-of-the-art implementation of the $GW$ formalism \cite{Bla11a} for the QM subsystem with a discrete polarizable model of atomistic resolution.
Such an approach goes beyond pioneering implementations that describe the MM dielectric medium as a regular grid of polarizable centers\cite{Bau14} or with the  polarizable continuum model.\cite{Duc16}  
In the present work the MM subsystem is described by the charge response (CR) model by Tsiper and Soos\cite{Tsi01} that provides a careful description of the static dielectric response of molecular solids.\cite{Tsi01b,Dav14,Dav16c}
The CR and related microelectrostatic models \cite{Bou79,Bou81} greatly contributed to the comprehension of the role of intermolecular electrostatic and polarization interactions on photoelectron spectroscopy measurements~\cite{Tsi02,Top11} and on the energetics of charge carriers in organic solar cells.\cite{Dav13,Poe15_nmat,Sch16}

Our hybrid formalism has been very recently applied to the pristine\cite{Li16} and doped\cite{Li17} PEN crystal.
A first validation of our hybrid scheme has been demonstrated by comparing our results for the electronic band gap of PEN to experimental data\cite{Sal08,Yos15} and values provided by plane-wave $GW$ calculations.\cite{Sha12}
In the present work the embedded $GW$ framework is extended to account for electrostatic crystal field effects, hence providing access not only to the band gap, but also the absolute values of IP and EA at specific crystal surfaces.
Our hybrid methodology is applied to model photoemission spectra of PEN and perfluoropentacene (PFP), two widely studied molecular semiconductors for which accurate photoemission data are available for solids of well-defined surface structure.\cite{Sal08,Yos15}
The quantitative (within 0.2 eV) agreement on IP and EA for both compounds in the gas
and solid phase demonstrates the accuracy and the internal consistency of our approach.

The paper is organized as follows.
Section \ref{s:theory} describes our hybrid formalism in full detail.
Results for PEN and PFP are presented in Section \ref{s:results}, where we highlight the importance of crystal field effects on the charged excitations of individual embedded molecules in the bulk and at crystal surface.
Our results are discussed and compared to experiments in Section \ref{s:disc}, where 
we dissect the different contributions from intermolecular interactions to IP and EA in PEN and PFP solids, i.e. polarization, crystal field and band dispersion, the latter being accounted for with an {\it ab initio} parametrized tight binding model.  
The main conclusions are finally drawn in Section \ref{s:conclu}.

\section{QM/MM methodology}
\label{s:theory}

As in other hybrid QM/MM approaches, our embedded MBPT calculations are defined by the level of theory employed for the QM and MM subsystems and by the formalization of the interaction between the two parts.
We provide below the details of the theory along with approximations and expedients employed to make our hybrid framework feasible and computationally efficient.

\subsection{MM embedding of the ground-state DFT calculation}
\label{ss:DFT}

The anisotropic charge densities of neutral organic molecules are source of intense and inhomogeneous electric fields within the crystals that can affect IP or EA by several tenths of an eV, as summarized in recent review papers.~\cite{Hei11,Dav16_rev} 
We account for crystal field effects at the DFT level (providing the starting point for the subsequent $GW$ treatment) by computing the Kohn-Sham (KS) eigenstates $\{\phi_n\}$ and eigenvalues $\{\varepsilon_n\}$ of the QM subsystem in the field of the surrounding neutral molecules in the crystal, described at classical MM level.
The MM embedding implies a modification of the electrostatic potential experienced by the QM system, i.e. where the charged excitation is created. 
Such an  electrostatic effect is well described in a ground-state DFT calculation and should not be confused with the dynamic reaction of the system to the ionization, which is accounted for within the $GW$ formalism (see Section \ref{ss:GWMM}).

Such a strategy, namely starting MBPT calculations with  DFT eigenstates obtained in the electric field of the classical environment, was also recently applied for the study of the optical properties of molecular systems in condensed phases.\cite{Mos09,Bag16,Var17} 
As shown below, the effect on the absolute position of the occupied/virtual electronic energy levels is significantly larger than that on the optical excitations relying on energy difference. 

An accurate classical description of the MM subsystem can be obtained with discrete polarizable models of atomistic resolution describing the static (zero-frequency) dielectric response of systems of interacting molecules.
\textcolor{\ntc}{Our calculations can in principle account for different contributions (i.e. ionic, vibrational, electronic) to the dielectric susceptibility, yet, in the present case of organic crystals of neutral molecules, only the leading electronic response is considered.\cite{Dav14,Dav16c} }
Specifically, in the present work we resort to the charge response (CR) model by Tsiper and Soos,~\cite{Tsi01} describing the anisotropic linear molecular response to electric fields in terms of induced atomic charges and induced dipoles.
The CR model is entirely parametrized with quantum-chemical calculations.
It has been shown that CR models, including the very similar charge response kernel theory by Morita and Kato,~\cite{Mor97} provide a quantitative description of the static permittivity tensor of several molecular crystals.~\cite{Tsi01b,Tsu09,Dav14,Dav16c}
A careful description of the electrostatic potential of isolated neutral molecules is of particular importance for an accurate assessment of crystal field effects within the MM model.
In our CR scheme we rely on point atomic charges obtained from the fitting of the electrostatic potential generated by the DFT electron density.~\cite{esp}

From a practical point of view, an iterative scheme consisting of cross-coupled DFT and CR calculations is set-up to obtain  KS orbitals in the self-consistent field of permanent and induced multipoles in the MM region. 
We start from a gas-phase DFT calculation on the QM subsystem and compute the electric potential and fields generated by the DFT electron density at the atomic sites of the MM region using efficient and accurate Coulomb-fitting resolution-of-the-identity (RI-V) techniques.
Fields and potentials generated by the QM electron density are then used to compute the induced charges and dipoles within the MM region, accounting for mutual interactions between MM molecules.
The DFT calculation for the QM subsystem is then repeated in the field of permanent and induced multipoles in the MM region, and the whole procedure is iterated until achieving self-consistency.
For the crystalline materials considered in this work, the energies of occupied and virtual molecular orbitals converge (within 1 meV) in 3 iterations.

Electrostatic interactions are notably long-ranged and special care is required when truncated sums are employed to approximate results for infinite systems.\cite{Dav16_rev,Poe16}
Moreover, the potential generated by ordered arrays of quadrupolar molecules does not  only depend on its size but also on its shape, leading to different values in crystalline bulk (3D geometry) and in 2D slabs, the latter depending on the crystallographic facet exposed to the vacuum.
Results presented in this paper are obtained with a MM embedding of DFT calculations ensuring an electrostatic potential on the QM molecule converged within 50 meV (average over atomic positions).
For bulk calculations on PEN and PFP this corresponds to spherical clusters of 4 nm radius, while for surfaces the criterion is reached for cylinders of 15 nm radius and a height of two molecular layers.

\subsection{The $GW$ formalism as the QM method}
\label{ss:GW}

We describe briefly the $GW$ formalism on which hinges the chosen QM framework within the QM/MM approach developed in this study, mostly emphasizing the main features related to the embedding strategy.
More details about MBPT can be found in review articles devoted to the $GW$ approach.\cite{Mar59,Hed65,Str82,Hyb86,God88,Ary98,Oni02}

Our starting point is the  time-ordered one-body Green's function $G$ describing 
the propagation in time of an added (removed) electron to  (from) the $N$-electron system in its ground-state. 
More precisely,  $G$ reads:
\begin{eqnarray*}
i{\hbar}G({\bf r},t;{\bf r}',t') &=& \theta(t-t') 
      \left\langle \psi_{GS}(N) \left| {\hat \psi}({\bf r}',t') {\hat \psi}^{\dagger}({\bf r},t) \right| \psi_{GS}(N) \right\rangle     \\
                             &-& \theta(t'-t) 
      \left\langle \psi_{GS}(N) \left| {\hat \psi}^{\dagger}({\bf r},t) {\hat \psi}({\bf r}',t')  \right| \psi_{GS}(N) \right\rangle      
\end{eqnarray*}
where $\psi_{GS}(N)$ is the $N$-electron ground-state wave function and $\left\{ {\hat \psi}({\bf r},t) , {\hat \psi}^{\dagger}({\bf r},t) \right\}$ are the destruction/creation field-operators in the Heisenberg representation.
Alternatively, it can be shown that $G$ adopts in a frequency representation the form:
\begin{eqnarray}
G({\bf r},{\bf r}'; \omega) &=& \sum_a  { g_a({\bf r}) g_a^*({\bf r}')  \over \omega - E_a + i0^+}  
        +  \sum_s  { g_s({\bf r}) g_s^*({\bf r}')  \over \omega - E_s - i0^+} 
        \label{eqn1}
\end{eqnarray}
where $E_a = E_a(N+1)-E_0(N) \;$ is an addition energy, with the index $a$ labeling the eigenstates of the ($N+1$)-electron system, while $E_s = E_0(N) - E_s(N-1) \;$ span the removal energies. 
The $g_{a/s}$ are called Lehman amplitudes and the infinitesimally small ($0^+$) positive parameter indicates that the Green's function can only be analytically continued in the first and third quadrants of the $\omega$-frequency complex plane.\cite{Far99} 
This is a crucial feature of the present formalism that aims at providing the true addition and removal energies, including the interaction energy of the added charge with the  $N$-electron system. 

To proceed further and obtain the $G$ operator in practice, one should solve the following equation-of-motion:

\begin{eqnarray}
  \left[ \omega - h_0({\bf r}) \right]  G({\bf r},{\bf r}_0; \omega) - \int d{\bf r}' \; \Sigma({\bf r},{\bf r}';\omega) G({\bf r}',{\bf r}_0; \omega) =  \delta({\bf r}-{\bf r}_0)
  \label{eqn2}
\end{eqnarray}
where the one-body Hamiltonian $h_0$ contains the kinetic energy operator and the ionic and Hartree potential.
The self-energy operator $\Sigma({\bf r},{\bf r}';\omega)$ represents  all exchange and correlation effects.  
Note that it is non-local and energy dependent, in contrast e.g. with adiabatic and semi-local DFT exchange-correlation functionals. 
While such formal developments are exact, an accurate approximation for the self-energy $\Sigma$ is required.  
Within the $GW$ formalism, which can be considered as the lowest-order  approximation to  $\Sigma$ in terms of the screened Coulomb potential $W$,\cite{Mar59,Hed65} the quantities to be calculated read:
\begin{eqnarray}
 {\Sigma^{GW}}({\bf r},{\bf r}';E) &=& {i \over 2\pi} \int d{\omega} e^{i{\omega}{0^+}}
            G({\bf r},{\bf r}';E+\omega) W({\bf r},{\bf r}';\omega)     \label{eqn3}  \\
  G({\bf r},{\bf r}';E) &=& \sum_n { \phi_n({\bf r}) \phi_n^*({\bf r}') \over E - \varepsilon_n + i{0^+} \times sgn(\varepsilon_n - E_F) }  \label{eqn4}    \\
  W({\bf r},{\bf r}';\omega) &=&  v({\bf r},{\bf r}') + \int d{\bf r}_1 d{\bf r}_2 \;  v({\bf r},{\bf r}_1)
               {\chi}^0({\bf r}_1,{\bf r}_2; \omega) W({\bf r}_2,{\bf r}';\omega),      \label{eqn5}    \\
 \chi^0({\bf r},{\bf r}'; \omega) &=& \sum_{n,m} (f_n - f_m) { \phi_n^*({\bf r}) \phi_m({\bf r}) \phi_m^*({\bf r}') \phi_n({\bf r}') 
               \over \varepsilon_i - \varepsilon_j - \omega -i{0^+}  \times sgn(  \varepsilon_n - \varepsilon_m  ) }         \label{eqn6}
\end{eqnarray}
where $v({\bf r},{\bf r}')=({\bf r}-{\bf r}')^{-1}$  is the bare Coulomb potential, 
$\chi^0$ the independent-electron susceptibility and $W$ the screened Coulomb potential. 
The $\lbrace f_n \rbrace$ are occupation numbers and 
$\lbrace  \varepsilon_n \rbrace$ are the energies of the KS eigenstates (orbitals) $\lbrace  \phi_n \rbrace$ that will be corrected within the present $GW$ formalism. 

While the Green's function $G$ can be calculated with the above set of equations,  in practice, and since $\Sigma$ contains all effects related to exchange and correlation,
the most common approach consists in  replacing in a perturbative fashion the DFT exchange-correlation potential $V^{DFT}_{XC}$ contribution to the KS eigenstates by its self-energy analog, namely:
\begin{eqnarray}
  E_n^{GW} &=& \varepsilon_n + 
\langle \phi_n |  \Sigma^{GW}(E_n^{GW}) - V^{DFT}_{XC} | \phi_n \rangle.
\label{eqn8}
\end{eqnarray}

Starting from the KS eigenstates obtained with a given exchange and correlation functional, $\Sigma^{GW}$ is constructed following Eqs.~\ref{eqn3}-\ref{eqn6}.
The $GW$ quasiparticle excitations $E_n^{GW}$ are then obtained 
by correcting the KS energy levels according to Eq.~\ref{eqn8}.
Such a scheme is labeled $G_0W_0$, where the ``0'' subscript indicates that $G$ and the screened-Coulomb potential $W$ are built from the zero-ordered (uncorrected) KS eigenstates.
The $G_0W_0$ scheme provides improved quasiparticle excitations with respect to DFT, 
leading, however, to results that depend on the starting exchange and correlation functional.\cite{Kor12,Bru13,Klo15,Kni16,Ran17}

A more accurate, although computationally more expensive, approach consists in achieving a partial self-consistency on the eigenvalues only. 
In the so-called ev$GW$ approach, the many-body corrected energies are in fact
re-injected in Eqs.~\ref{eqn3}-\ref{eqn6}  ($E_n^{GW} \rightarrow \varepsilon_n $) building the self-energy operator corrected to the next order,  and the whole procedure is iterated until convergence of $\{E_n^{GW}\}$.
The dependence on the starting functional is significantly reduced in the ev$GW$ scheme, leading to quasiparticle excitations in quantitative agreement with experimental values or higher level CCSD(T) calculations.\cite{Fab11a,Ran17,Kap16}

\subsection{MM dielectric contribution to the screened Coulomb potential $W$}
\label{ss:GWMM}
We now show how the contribution of the MM subsystem dielectric response can be merged with the $GW$ formalism to properly contribute to the energy required for adding or removing an electron to the QM subsystem.
On general grounds, the analysis of Eq.~\ref{eqn6} shows that if two subsystems,
hereafter labeled  ``1'' and ``2'',  have non-overlapping orbitals, the independent-electron polarizability cannot couple the two systems, namely $\chi^0({\bf r}_1,{\bf r}_2)$ 
is zero for any pair of positions ${\bf r}_1$ and ${\bf r}_2$ in systems 1 and 2, respectively. 
As a consequence, the screened Coulomb potential restricted to the QM subsystem (1), $W_{11}$, reads:
\begin{eqnarray}
W_{11} &=& v_{11} + v_{11} \chi^0_{11}  W_{11} + v_{12} \chi^0_{22} W_{21}       \label{eqn9} \\
W_{21} &=& v_{21} + v_{21} \chi^0_{11}  W_{11} + v_{22} \chi^0_{22} W_{21}       \label{eqn10}
\end{eqnarray}
using a block notation where index 1 (2) corresponds to points located in area 1 (2). After some algebra, one obtains the following set of equations: 

\begin{eqnarray}
W_{11} &=& {\tilde v}_{11} + {\tilde v}_{11} \chi^0_{11} W_{11}    \label{eqn11} \\
{\tilde v}_{11} &=& v_{11} + v_{12} \chi^{*}_{22}  v_{21}  = v_{11} + v^{reac}            \label{eqn12} \\
\chi^{*}_{22} &=& \chi^0_{22} + \chi^0_{22} v_{22} \chi^{*}_{22}              \label{eqn13}
\end{eqnarray}
%
%
where ${\tilde v}_{11}$ is the Coulomb potential in the QM cavity screened by the MM subsystem only (namely without the response of the QM section itself that is incorporated in the $\chi^0_{11}$ susceptibility in Eq.~\ref{eqn11}) and $\chi^{*}_{22}$ is the interacting polarizability of system 2 alone, i.e. \textit{in the absence of system 1}.
Upon introducing real-space coordinates, the  Coulomb potential within the QM region is renormalized by adding
\begin{eqnarray} \label{v_reac} 
v^{reac}({\bf r}_1,{\bf r}^{'}_1;\omega) = \int d{\bf r}_2 d{\bf r}^{'}_2  \; v({\bf r}_1,{\bf r}_2) \chi^{*}_{22} ({\bf r}_2,{\bf r}^{'}_2;\omega)   v({\bf r}^{'}_2,{\bf r}^{'}_1),
\end{eqnarray}
representing the reaction field generated  in ${\bf r}^{'}_1$ by the MM subsystem  in response to a charge added in ${\bf r}_1$, with both ${\bf r}_1$ and ${\bf r}^{'}_1$ pointing in the QM subsystem 1.

The reaction field $v^{reac}$ in Eq.~\ref{v_reac}  is therefore the key quantity through which classical polarizabilities of molecules belonging to the MM region enter the embedded $GW$ calculation restricted to the QM subsystem.
Such a calculation can then be performed as a standard $GW$ calculation in the gas phase, but with the bare Coulomb potential substituted by the renormalized 
(MM-screened)   potential  ${\tilde v}_{11}$.



The construction of the $GW$ self-energy actually requires the knowledge of the dynamically screened $W(\omega)$ Coulomb potential in Eq.~\ref{eqn5}, accounting for the fact that the system  dielectric response is frequency-dependent in the optical range.
Indeed, in a full $GW$ calculation the dispersion of the QM subsystem susceptibility is accounted for in Eq.~\ref{eqn6}, while a possible frequency dependence of the MM subsystem polarizabilities would result in a frequency-dependent reaction field.

Although the dependence of the optical dielectric properties on photon frequency is experimentally well documented in organic solids,\cite{Dre08} the classical polarizable models we rely on focus on reproducing correctly the correct optical dielectric response in the low-frequency, $\omega \rightarrow 0$, limit. 
While generalizing MM polarizable models to dynamical response may be considered, we describe here an alternative strategy that  consists in the merging of static MM polarizable models with the static limit of the $GW$ formalism. 
Such an approach has been recently applied to couple the $GW$ formalism to continuum \cite{Duc16} and discrete polarizable models.\cite{Li16}

The static formulation of the $GW$ formalism, the so-called static Coulomb-hole plus screened exchange (COHSEX) approximation, was discussed in the seminal paper by Hedin \cite{Hed65} and was shown to be very efficient within the framework of simplified self-consistent $GW$ calculations. \cite{Bru06}
COHSEX calculations are known to be less accurate than the full $GW$ ones in determining e.g. the band gap of semiconductors. 
However, within the purpose of the present QM/MM scheme, the COHSEX approximation will only be adopted to obtain the contribution of the MM environment to quasiparticle excitations.
Using a symbolic notation, we decompose the self-energy operator for  the embedded QM system as follows:
\begin{eqnarray}
\label{sigma_gwmm}
\Sigma^{GW/\mathrm{MM}}  &=&  \Sigma^{GW} + \left[  \Sigma^{GW/\mathrm{MM}}   - \Sigma^{GW}  \right] \nonumber  \\ &\simeq & 
\Sigma^{GW} + \left[  \Sigma^{COHSEX/\mathrm{MM}}   - \Sigma^{COHSEX}\right].
\end{eqnarray}
This formula approximates the self-energy operator of the embedded system by the sum of its analogue for the isolated QM system,  plus a correction calculated at the COHSEX level, namely as the difference between the self-energy obtained including or not the MM reaction field contribution.
The reason for such a formulation is that the use of the COHSEX approximation in the form of a difference allows to reduce the error 
introduced by replacing the frequency-dependent optical dielectric constant by its low-frequency limit. 
Even though the screening potential $W$ in the quantum-mechanical region is modified by the MM response, one here assumes that the dynamical screening contribution entering  in the difference ($ \Sigma^{GW/\mathrm{MM}}  - \Sigma^{\mathrm{COHSEX}/\mathrm{MM}}$) largely 
cancels with the one in ($\Sigma^{GW}  -  \Sigma^{\mathrm{COHSEX}})$.  

The approximated $GW$/MM self-energy in Eq.~\ref{sigma_gwmm} can be finally used to 
compute the quasiparticle energies \textcolor{\ntc}{
\begin{eqnarray}
\label{dch}
E_n^{GW/\mathrm{MM}} &=& E_n^{GW} + \Delta_n^{\mathrm{COHSEX}} \nonumber \\
& = & E_n^{GW} + \langle \phi_n  | \Sigma^\mathrm{COHSEX/MM} - \Sigma^\mathrm{COHSEX} | \phi_n \rangle
\end{eqnarray}
}
where $E_n^{GW}$ are quasiparticle excitation energies of the full $GW$ calculation 
(possibly accounting for electrostatic effects through the starting DFT calculations as described in Section~\ref{ss:DFT}) and $\Delta_n^{\mathrm{COHSEX}}$ is the state-specific polarization energy accounting for the screening provided by induced dipoles in the MM region.

It is worth remarking that the full-$GW$ quasiparticle energies $E_n^{GW}$ and the COHSEX polarization contribution $\Delta_n^{\mathrm{COHSEX}}$ present different convergence behavior with respect to the direct and auxiliary basis employed in the calculations. 
While large basis sets are required to converge $E_n^{GW}$ in the gas phase, the calculation of the COHSEX polarization energy is less demanding.
For instance, the difference in $\Delta_n^{\mathrm{COHSEX}}$
between calculations using cc-pVTZ or 6-311G* (cc-pVTZ-RI or the universal Weigend Coulomb Fitting) as principal (auxiliary) basis is lower than 10 meV in the case of PEN.\cite{Li16}
Both the long-range nature of the reaction-field, and the fact that polarization energy is calculated as an energy difference, may explain this observation.

\subsection{Reaction field matrix on a Gaussian basis}
\label{ss:gauss}

In our Gaussian atomic orbital implementation we do not calculate $v^{reac}$ on a $({\bf r},{\bf r}^{'})$ real-space grid but look for the following matrix elements:
\begin{equation}
v^{reac}(\beta,\beta') = \int d{\bf r}\, d{\bf r}^{'} \; \beta({\bf r}) v^{reac}({\bf r},{\bf r}^{'}) {\beta}'({\bf r}^{'}) 
\end{equation}
namely the two-center two-electron Coulomb integrals between auxiliary Gaussian orbitals 
$\lbrace\beta\rbrace$ located at the atomic sites in the QM region. 
This auxiliary basis stems from the Coulomb-fitting resolution-of-the-identity (RI-V) formulation of the $GW$ implementation we adopt.

\textcolor{\ntc}{
In practice, before performing the $GW$ calculation, we compute the self-consistent rearrangement of MM charges and dipoles induced by the potential generated by the charge density associated to each orbital $\beta$ of the auxiliary basis, namely
\begin{equation}
V_{\beta}(\mathbf{r}_{MM}) = \int d\mathbf{r}\; \frac{\beta(\mathbf{r})}{| \mathbf{r}_{MM} - \mathbf{r} |},  
\end{equation}
where $\mathbf{r}_{MM}$ are the positions of MM atoms. 
The self-consistent calculation of induced charges and dipoles within the MM system under the $V_{\beta}(\mathbf{r}_{MM})$ external potential is performed with the CR model.\cite{Tsi01,Dav14} 
These induced MM charges and dipoles generate in return the reaction potential
$ V_{\beta}^{reac}({\bf r})$ acting on the QM subsystem. 
The energy of the \emph{probe} auxiliary orbital $\beta'$ in the field of the MM system polarized by the \emph{source} charge density $\beta$ is finally computed as: 
\begin{equation}
  v^{reac}(\beta,\beta') = \int d\mathbf{r}\; V_{\beta}^{reac}({\bf r}) \beta’({\bf r}).
\end{equation}
Since the $\{\beta\}$ orbitals serve as a basis set to represent charge densities, the reaction field $v^{reac}(\beta,\beta')$ allows thus to describe the contribution of the MM environment to the field generated by any charge variation in the QM subsystem.
Each self-consistent calculation of induced polarization in the MM subsystem scales as $\mathcal{O}(N_{MM}^2)$, with $N_{MM}$ the number of MM atoms.\cite{Li16} 
Since this has to be done for each of the auxiliary basis orbitals, of which the number scales as the number of QM atoms $N_{QM}$, the evaluation of the reaction field matrix scales as $\mathcal{O}(N_{MM}^2 N_{QM})$.
}

We emphasize that only electrical multipoles induced in the MM region by the addition/removal of charges in the QM subsystem contribute to the reaction field matrix $v^{reac}(\beta,{\beta}')$. 
Fixed charges in the MM part, as well as the multipoles induced by MM molecules by the DFT ground state electron density of QM molecule(s), do not contribute to $v^{reac}(\beta,{\beta}')$. 

\begin{figure}[ht]
\centering
\includegraphics[width=10cm]{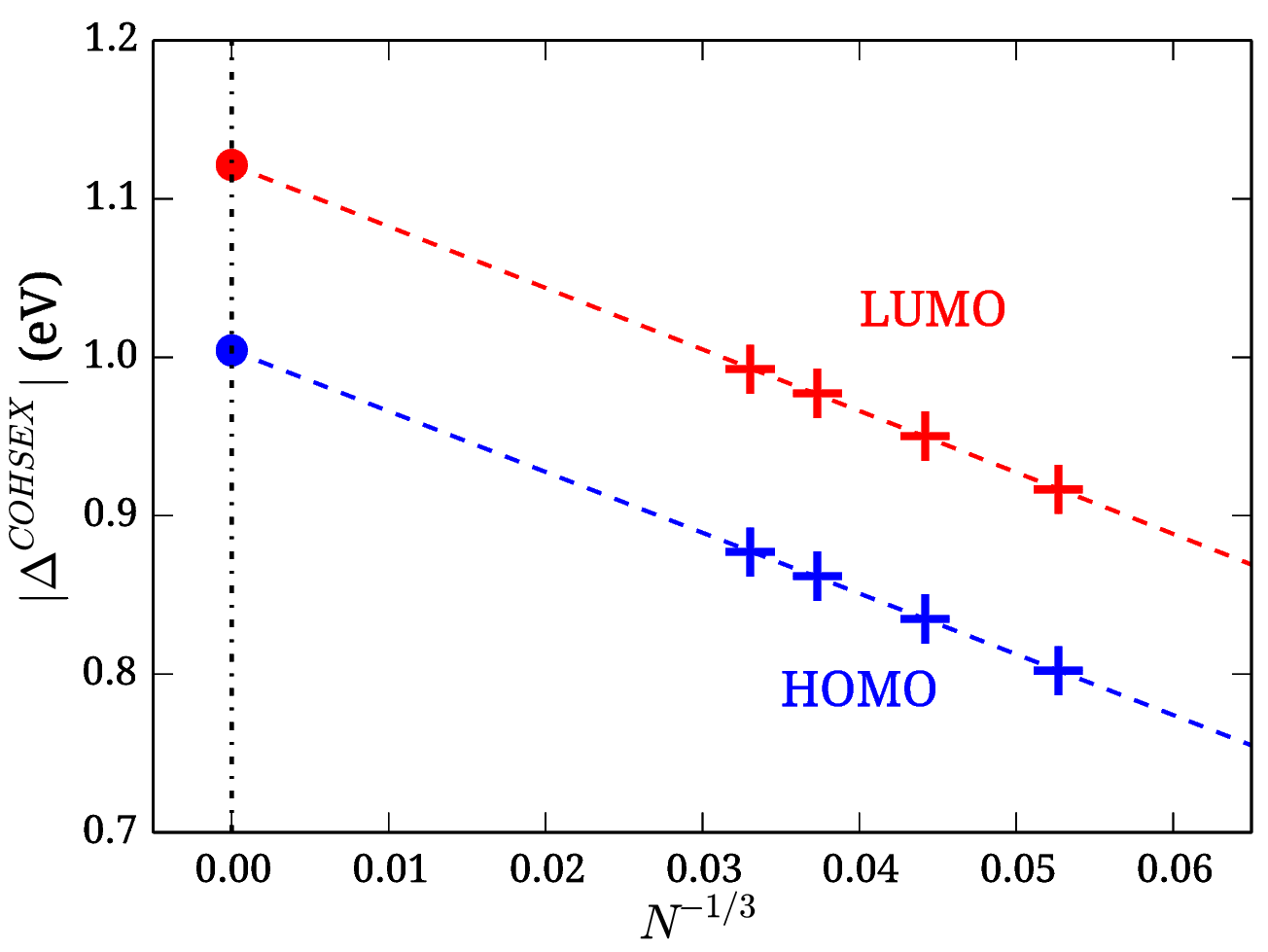}
\caption{  (a) Polarization energy $\Delta^{\mathrm{COHSEX}}$ for PEN HOMO and LUMO levels as a function of the inverse radius of the embedding MM spherical cluster  ($N$ is the number of MM atoms).
  Pluses correspond to  $\Delta^{\mathrm{COHSEX}}$ calculated with reaction field obtained for specific values of the MM cluster radius, namely $R=$25, 30, 35, 40 \AA.
Dotted lines are linear extrapolation to the infinite radius limit.
Full circles are the $\Delta^{\mathrm{COHSEX}}$ values computed with the extrapolated reaction field  matrix.
The two extrapolation techniques give the same results.
}
\label{f:extrap}
\end{figure}

The calculation of the symmetric reaction field matrix is practically performed on a MM subsystem of finite size, although appropriate extrapolation techniques can be used to obtain $\Delta_n^{\mathrm{COHSEX}}$ polarization energies for an infinite system.
A first approach consists in the explicit calculation of the matrix $v^{reac}(\beta,{\beta}')$, and then $\Delta_n^{\mathrm{COHSEX}}$, for spherical MM embedding clusters of increasing radii, and 
finally extrapolate the polarization energy to the infinite bulk crystal limit. 
This is illustrated  in Figure~\ref{f:extrap} for a hole (electron) in the PEN HOMO (LUMO).  
As expected, the polarization energy scales linearly with the inverse radius of the MM spherical cluster (crosses) allowing a direct extrapolation.
However, this comes at the  cost of performing several COHSEX calculations.

Another strategy is the extrapolation of the reaction field matrix elements $v^{reac}(\beta,{\beta}')$ to the infinite MM cluster limit before performing a single
COHSEX calculation that will directly target the infinite crystal. 
\textcolor{\ntc}{
In the case of auxiliary $s$ orbitals (angular momentum $l=0$), representing electrical monopoles in the RI-V formalism, the reaction field matrix scales as $R^{-1}$ in three dimensions.
For $\beta$ ($\beta'$) orbitals of arbitrary angular momentum, the reaction field matrix elements scale as $R^{-(1+l+l')}$, allowing  straightforward extrapolation.}
\footnote{\textcolor{\ntc}{
We recall that auxiliary functions $\{\beta\}$ describe charge densities corresponding to an electrical monopoles ($l=0$), dipoles ($l=1$), quadrupoles ($l=2$), etc.
The $R^{-(1+l+l')}$ dependence of reaction field matrix elements follows from classical electrostatics under the hypothesis that $R$ largely exceeds the size of the QM region} }
The decay of $v^{reac}(\beta,{\beta}')$ elements is faster and faster for high $l$, leading in some cases to values that are practically already converged in clusters of relatively small size.
Additional details on the efficient calculation of the reaction field matrix are given in Appendix \ref{a:vr}.


Once the extrapolated reaction field is obtained, the polarization energy can be directly obtained in the infinite cluster limit by performing a single COHSEX calculation. 
The result obtained with this approach for the HOMO and LUMO polarization energies of PEN (filled dots in Figure~\ref{f:extrap}) are practically identical to those extrapolated in the first brute force scheme. 

In the following we will describe both calculations of charged excitations in a bulk material and at its surface. 
The reaction field matrix for surface calculations is extrapolated in the limit of an infinite semi-sphere, strictly describing the polarization response of a semi-infinite crystal to the charging of a molecules at the surface.
Such an approach can be considered a good approximation also for molecular films on insulating substrates of comparable dielectric constant, such as SiO$_\mathrm{x}$.

\clearpage
\subsection{Technical details}
\label{ss:details}
Our calculations have been performed with the {\sc{Fiesta}} package that implements the $GW$ \cite{Bla11a,Fab11a,Fab11b} and Bethe-Salpeter \cite{Bla11b,Jac15a,Jac15b} formalisms for a Gaussian atomic orbital basis.
The code relies on Coulomb-fitting resolution-of-the-identity (RI-V) techniques and a contour deformation approach to perform the frequency integration in Eq.~\ref{eqn3}.
The KS eigenstates needed to start the ev$GW$ calculations are obtained with the {\sc{NWChem}} computational package adopting the PBE0 functional and the cc-pVQZ basis.~\cite{Dun89}
Within the RI-V technique, we adopt the universal Weigend Coulomb Fitting auxiliary basis. \cite{Wei06}

The MM region is described  with the CR model\cite{Tsi01} as implemented in the {\sc{Mescal}} code.\cite{Dav14}
The molecular polarizability tensors are computed at the DFT level, while  atom-atom polarizabilities governing intramolecular charge flows are evaluated with semi-empirical Hartree-Fock calculations (ZINDO parametrization).\cite{zindo}
The DFT calculations for the parametrization of the CR model were performed with the \textsc{Gaussian} suite,~\cite{g09d1} using the B3LYP functional and the 6-311++G** basis set.

In our calculations we considered the crystal structure of common polymorphs of the two compounds, both presenting two molecules in the unit cells arranged in a herringbone fashion. 
For PEN we considered the triclinic structure by Siegriest {\it et al.} \cite{Sie01} ($a=6.265$\AA, $b=7.786$ \AA, $c=14.511$ \AA, $\alpha$=76.65$^{\circ}$, $\beta$=87.50$^{\circ}$, $\gamma$=84.61$^{\circ}$; CCDC no. 145333), for PFP we adopted the monoclinic cell by Sakamoto {\it et al.} \cite{Sak04} ($a=15.510$ \AA, $b=4.490$ \AA, $c=11.449$ \AA, $\beta$=91.57$^{\circ}$;  CCDC no. 234729).
\textcolor{\ntc}{
Surface structures were obtained by cutting bulk crystals along given crystal planes and employed in calculations without performing any structural relaxation.}

\clearpage
\section{Charged excitations in pentacene and perfluoropentacene}
\label{s:results}

Our novel Green's function QM/MM formalism is here applied to calculate charged excitations in PEN and PFP crystals, considering the ionization of molecules in the bulk and at the crystal surface. 
The photoelectron spectra of PEN and PFP have been extensively studied experimentally \cite{Sal08,Yos15,Ker13} and theoretically,\cite{Top11,Hei11,Yos15}
as they represent an ideal case study for dissecting the different contributions to charge transport levels arising from intermolecular interactions in the solid state.\cite{Dav16_rev}
Photoelectron spectra of PEN and PFP are therefore chosen here to demonstrate the accuracy and internal consistency of our hybrid formalism. 

The two molecules have indeed similar chemical and electronic structure 
(the elementary H\"uckel model for $\pi$-electrons cannot distinguish between the two), leading to comparable frontier orbitals \cite{Che05,Del09} and  nearly identical polarizability tensors.\cite{Top11}
Both compounds crystallize in a layered structure characterized by planes of
nearly standing (slightly tilted) molecules arranged in a herringbone fashion.
Such a layers are parallel to the (001) and (100) planes in PEN and PFP, respectively, which are low energy crystal facets usually found also in molecular films grown on insulating substrates such as SiO$_\mathrm{x}$.\cite{Sal08}

A crucial difference between the two molecules resides in their ground-state electrical quadrupole moment, which has principal components of comparable magnitude but opposite sign, owing to the different polarity of C-H vs. C-F bonds.\cite{Top11,Ryn13}
The availability of accurate experimental data from ultraviolet photoemission spectroscopy (UPS) and low-energy inverse photoemission spectroscopy (LEIPS) for crystalline films of standing molecules\cite{Sal08,Yos15} make these systems ideal for benchmarking purpose. 
UPS and LEIPS spectra have been also reported for films of laying molecules on metal or graphite substrates.\cite{Sal08,Yos15}
We will not address such measurements here as their calculation would also require the modeling of the interaction with the conducting substrate (i.e. image charge effects), which goes beyond the scope of the present work. 

Table \ref{t:bulk} compares HOMO and LUMO energies for PEN and PFP obtained at DFT and $GW$ level for an isolated (gas-phase) molecule and for a molecule embedded in the bulk crystal.
In order to disentangle the different contributions to the energy levels in the solid state, we report results for different DFT starting points, namely  performed either in the gas phase or in the presence of a MM embedding with its proper crystal field.  
In the following we will adopt the ``g'' or ``e'' subscripts to label calculations that are performed for gas-phase or MM-embedded molecules, with e.g. $GW_\mathrm{e}|$DFT$_\mathrm{g}$ corresponding to a $GW$ calculation on an embedded molecule (i.e. accounting for the $\Delta^{COHSEX}$ term in Eq.~\ref{dch}) initiated with a gas-phase DFT calculation, namely including MM polarization effects upon excitation but without ground-state crystal field.
A handy reference to the effects accounted for in different calculations is provided in Table~\ref{t:scheme}.

\begin{table}
\begin{tabular}{p{22mm} | C{27mm} C{20mm} C{20mm}}
\hline \hline 
level of & many-body  & crystal & polarization  \\
theory      & correlations   & field   &  \\
\hline 
DFT$_\mathrm{g}$ & no & no & no\\
DFT$_\mathrm{e}$ & no & yes & no \\
$GW_\mathrm{g}|$DFT$_\mathrm{g}$ & yes & no & no \\
$GW_\mathrm{e}|$DFT$_\mathrm{g}$ & yes & no & yes \\
$GW_\mathrm{e}|$DFT$_\mathrm{e}$ & yes & yes & yes \\
\hline \hline 
\end{tabular}
\caption{Summary table of the effects accounted for at the different levels of theory, namely many-body electronic correlations within the QM subsystem, crystal field effects and polarization of the MM dielectric environment.}
\label{t:scheme}
\end{table}

\begin{table}
  \begin{tabular}{l C{13mm} C{13mm} C{13mm} | C{13mm} C{13mm} C{13mm} |  C{13mm} }
\hline \hline 
  Starting from:  & \multicolumn{3}{c|}{DFT$_\mathrm{g}|$} & \multicolumn{3}{c|}{DFT$_\mathrm{e}|$} & \\
  & KS & $GW_\mathrm{g}$ & $GW_\mathrm{e}$ &  KS & $GW_\mathrm{g}$ & $GW_\mathrm{e}$   & $\Delta^{\mathrm{cf}}$ \\
\hline    
\multicolumn{4}{l|}{\textbf{pentacene}} & & & & \\
HOMO & -5.08 & -6.48 & -5.48 & -4.92 & -6.32 & -5.32 &  0.16  \\
LUMO & -2.63 & -1.45 & -2.58 & -2.49 & -1.30 & -2.42 &  0.16 \\
gap  &  2.45 &  5.03	&  2.90	&  2.43 &  5.02 &  2.90 & 0.00 \\
\hline 
\multicolumn{4}{l|}{\textbf{perfluoropentacene}} & & & &  \\
HOMO & -5.97 & -7.39 & -6.48 & -6.28 & -7.70 & -6.79 &  -0.31  \\
LUMO & -3.79 & -2.65 & -3.64 & -4.09 & -2.94 & -3.93 &  -0.29 \\
gap  &  2.18 &  4.74 &  2.84 &  2.19 &  4.76 &  2.86 &   0.02 \\
\hline \hline
\end{tabular}
  \caption{Evolution of HOMO and LUMO levels and gap of PEN and PFP from the gas phase to a molecule embedded into the bulk crystal structure. Energies are in eV.
Kohn-Sham, gas-phase and embedded $GW$ results are reported for calculations initiated with DFT orbitals obtained with and without MM embedding (see text for the definition of the notation), highlighting the importance of crystal field effects on the transport levels. 
The crystal field shift $\Delta^{\mathrm{cf}}$ is the difference between the $GW_\mathrm{e}|$DFT$_\mathrm{e}$ and the $GW_\mathrm{e}|$DFT$_\mathrm{g}$ results.
}
\label{t:bulk}
\end{table}

We start our analysis by considering results obtained starting from a gas-phase DFT calculation (left columns in Table~\ref{t:bulk}).
The well-known effect of non-local many-body electronic correlations is the large increase of the HOMO-LUMO gap with respect to the KS value obtained with a functional presenting a small amount of exact exchange (25\% in the PBE0 case).\cite{Fab14} 
As reported in Table \ref{t:bulk}, the $GW_\mathrm{g}$ gap is approximately 2.5 eV larger than the PBE0 one for both PEN and PFP, irrespectively on the presence of MM embedding in the ground-state DFT calculation.
The frontier orbitals of PFP are found to be about 1 eV deeper in PFP than in PEN, as a result of the electron-depleting effect exerted by fluorine atoms on the $\pi$-electron system.

The inclusion of the MM dielectric embedding in the $GW$ calculation closes the gap by approximately 2 eV for both molecules, as we have shown in a very recent paper where we first applied the $GW_\mathrm{e}|$DFT$_\mathrm{g}$ methodology to bulk PEN.\cite{Li16}
\footnote{The difference in the $GW_\mathrm{e}|$DFT$_\mathrm{g}$ gap with respect to the value we recently reported (3.05 eV in Ref.~\onlinecite{Li16})
is due to the different atomic coordinates used for the QM molecule.
In this work atomic positions are used as provided in the cif file,\cite{Sie01} while in Ref.~\onlinecite{Li16} hydrogen atoms were optimized in a preliminary gas-phase DFT calculation}
Such gap reduction originates from the dielectric screening provided by the MM  environment, i.e. the microscopic dipoles induced in the polarizable environment by the charge, hole or electron, created in the QM subsystem.
\textcolor{\ntc}{
The magnitude of such a \emph{polarization} contribution is consistent with (differences within 15\%)  earlier results from classical polarizable models of atomistic resolution,\cite{Tsi03,Ryn13,Dav14}
and from $GW$ calculations using the polarizable continuum model for embedding.\cite{Duc16} }
\footnote{ \textcolor{\ntc}{
A polarization energy of 0.96 (1.31) eV has been obtained for PEN HOMO (LUMO) with our $GW$ approach
combined with the standard polarizable continuum model,\cite{Duc16} using an isotropic dielectric constant of 3.5, representative of bulk PEN.\cite{Dav14}
This compares well with the present embedded-$GW$ results employing the CR model\cite{Tsi01} for the MM subsystem: $\Delta^{\mathrm{COHSEX}}=$1.00 and 1.13 eV for PEN HOMO and LUMO, respectively.
We emphasize, however, that the polarizable continuum model cannot account for the electrostatic crystal field effects that can shift the energy levels by as much as one eV with respect to the vacuum level
}
}
\textcolor{\ntc}{
We notice that a quantitative account 
of the gap renormalization in the solid state has been recently achieved also at the DFT level using optimally tuned screened range-separated hybrid functionals.\cite{Abr13}}

We now turn our attention to the results of calculations that start from a ground state DFT calculations performed for molecules embedded in the MM environment, hence experiencing the microscopic \emph{crystal field} exerted by the surrounding molecules in the bulk solid.
First, we remark that the HOMO-LUMO gap, either obtained at the KS, $GW_\mathrm{g}$  or $GW_\mathrm{e}$ level, is to a very good approximation insensitive to crystal field effects,
in contrast to the energies of individual orbitals that are significantly affected (see Table \ref{t:bulk}).

Such a result can be rationalized by considering, to first approximation, the superposition of the quadrupolar fields of MM molecules as a uniform potential acting on the QM region, which implies a rigid shift of all occupied and virtual molecular orbitals with respect to their gas-phase values, with negligible effects on the gap and on neutral optical excitations.
This is actually the leading effect in the crystalline materials considered in this work, where the crystal field shifts all energy levels by approximately the same amount $\Delta^{\mathrm{cf}}\sim 0.15$ and $\sim-0.30$ eV in bulk PEN and PFP, respectively.
The crystal field shifts $\Delta^{\mathrm{cf}}$ in Table~\ref{t:bulk} are quantified by the difference between the DFT$_\mathrm{e}|GW_\mathrm{e}$ and the DFT$_\mathrm{g}|GW_\mathrm{e}$ results.
Such level shifts are already present in the starting DFT electronic structure and are then reflected in the subsequent $GW$ calculations, as can be inferred from the data in Table \ref{t:bulk}.
The opposite signs of $\Delta^{\mathrm{cf}}$,  defined as the difference between DFT$_\mathrm{e}|GW_\mathrm{e}$ and  DFT$_\mathrm{g}|GW_\mathrm{e}$ energy levels, in PEN and PFP is imputable to the opposite signs of the principal components of the molecular quadrupole moments.\cite{Hei11,Top11}

\begin{figure}[hb]
\centering
\includegraphics[width=10cm]{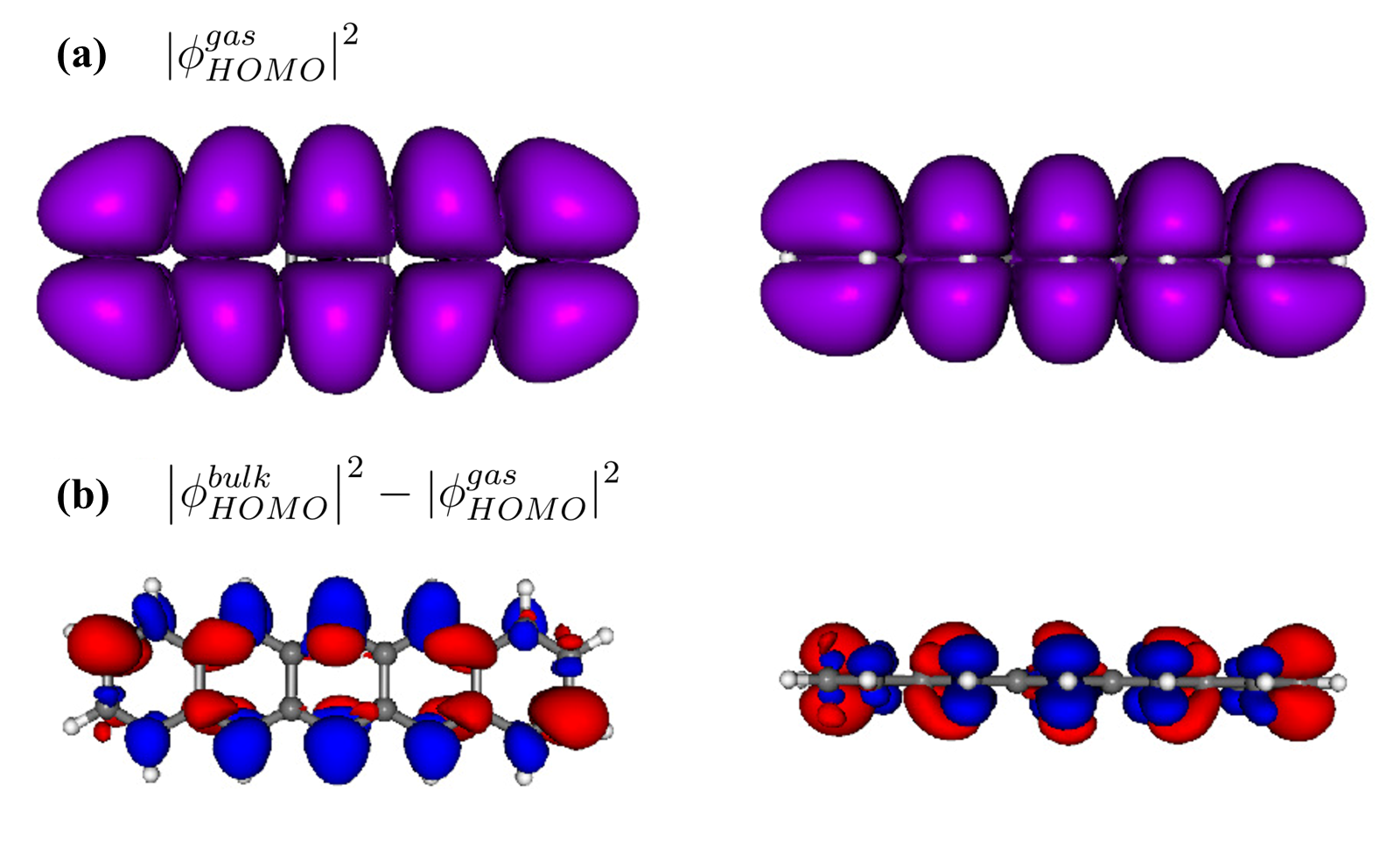}
\caption{(Color online)
(a) Isocontour plot of the squared HOMO amplitude of the pentacene molecule in the gas phase. 
(b) Isocontour of the difference between the squared amplitude of the pentacene HOMO in the bulk crystal and in the gas phase. Blue and red colors are chosen for positive and negative isovalues. 
All plots are obtained for an isovalue of 10$^{-5}$. 
}
\label{f:HOMOpen}
\end{figure}

The inhomogeneity of the crystal field  at the atomic scale can also alter the shape and spatial extension of the orbitals, further affecting their energies.
Such an effect is illustrated in Figure \ref{f:HOMOpen}, showing the amplitude of the pentacene HOMO in the gas phase ($|\phi^{gas}_{HOMO}|^2$) and its difference with the HOMO in the pentacene crystal. 
In this case, the effect of neighboring molecules is to stretch the HOMO from the $\pi$-conjugated region towards C-H bonds.
The relaxation of molecular orbital in the crystal field is found to affect orbital energies by a few tens of meV in crystalline PEN and PFP, although a larger influence is expected in disordered environments or in the case of dipolar molecules.
The orbital relaxation in the crystal field does also affect the intermolecular charge transfer couplings and band dispersion, as discussed in Appendix \ref{a:tb}.

\textcolor{\ntc}{
Concerning bulk PEN, our result for the gap in Table \ref{t:bulk} (2.9 eV) compares well 
with the \emph{band center-to-center gap} values from $GW$ calculations for periodic bulk systems.\cite{Tia03,Sha12,Kan16,Ran16}  
$GW$ gap values reported in the literature present small variations stemming from different polymorphs, 
starting DFT functional and the level of self-consistency.
Recent estimates for the PEN center-to-center gap range between the 2.8 eV reported by 
Sharifzadeh \textit{et al.} for the solution polymorph \cite{Cam62} at the $G_0W_0$@HSE level,
and the 2.9 eV for the thin-film structure \cite{Sch07} by Rangel \textit{et al.} ($G_0W_0$@PBE)\cite{Ran16} and by Kang \textit{et al.} ($GW_0$@PBE).\cite{Kan16}}

Both the polarization and the crystal field effects depend on the shape of the sample, leading to different charged excitations in the bulk and at the crystal surfaces where actually these quantities are experimentally measured.
In order to approach photoelectron spectroscopy experiments, we hence explicitly computed quasiparticle excitations for a molecule at the specific crystal surface probed in the experiment we aim at modeling, i.e. the (001) and (100) face for PEN and PFP, respectively.
Our results for ionization energies at crystal surfaces are reported in Table~\ref{t:surf}.

The magnitude of the gap essentially depends on polarization effects.
As we discussed in a very recent paper, the gap is $\sim0.2$ eV larger for a molecule at the surface than in the bulk, as a result of the less effective screening at the interface to vacuum.\cite{Li17}
The 10\% decrease of polarization from bulk to surface seems to be characteristic for films of standing elongated molecules as PEN and PFP.
Very similar gap values are found at the  $GW_\mathrm{e}|$DFT$_\mathrm{g}$ and $GW_\mathrm{e}|$DFT$_\mathrm{e}$ level, confirming that the gap is almost insensitive to the crystal field also at the crystal surface.

HOMO and LUMO levels are instead rigidly shifted by the crystal field. 
$\Delta^{\mathrm{cf}}$ is found to be larger at the surfaces we considered (Table \ref{t:surf}) than in the bulk (Table \ref{t:bulk}), as already reported on the basis of classical electrostatic modeling.\cite{Top11,Dav16_rev}
We recall that the dependence of $\Delta^{\mathrm{cf}}$ on the macroscopic shape of the sample and on the crystal facet in 2D slabs both originate from the conditional convergence of charge-quadrupole interactions, which leads to surface-dependent charged excitations.\cite{Hei11,Dav16_rev}
We remark that $\Delta^{\mathrm{cf}}$ has opposite signs for PEN and PFP both at the surface and in the bulk.

\begin{table}
\begin{tabular}{l  C{22mm} C{22mm} C{22mm}  }
\hline \hline 
& $GW_\mathrm{e}|$DFT$_\mathrm{g}$ &$GW_\mathrm{e}|$DFT$_\mathrm{e}$ & $\Delta^{\mathrm{cf}}$ \\
\hline
\multicolumn{4}{l}{\textbf{pentacene (001)}} \\
HOMO & -5.57  & -5.19  &  0.38  \\
LUMO & -2.47  & -2.10  &  0.37   \\
gap  &  3.10  &  3.09  & -0.01   \\
\hline
\multicolumn{4}{l}{\textbf{perfluoropentacene (100)}} \\
HOMO & -6.57  &  -7.19 & -0.62  \\
LUMO & -3.55  &  -4.14 & -0.59  \\
gap  &  3.02  &   3.05 &  0.03  \\
\hline \hline
\end{tabular}
\caption{HOMO and LUMO levels and gap of a molecule at the crystal surface, (001) for PEN and (100) for PFP. Energies are in eV.
Results from embedded $GW$ calculations with one molecule as QM subsystem.
$GW_\mathrm{e}|$DFT$_\mathrm{e}$ ($GW_\mathrm{e}|$DFT$_\mathrm{g}$) labels results obtained considering (neglecting) the MM embedding in the starting DFT calculations. 
$\Delta^{\mathrm{cf}}$ is the difference between 
$GW_\mathrm{e}|$DFT$_\mathrm{e}$ and $GW_\mathrm{e}|$DFT$_\mathrm{g}$ results, quantifying the magnitude of the crystal field contribution.
}
\label{t:surf}
\end{table}

\clearpage
\section{Discussion}
\label{s:disc}

\begin{figure}[ht]
\centering
\includegraphics[trim={0 0 0 10},clip,width=12cm]{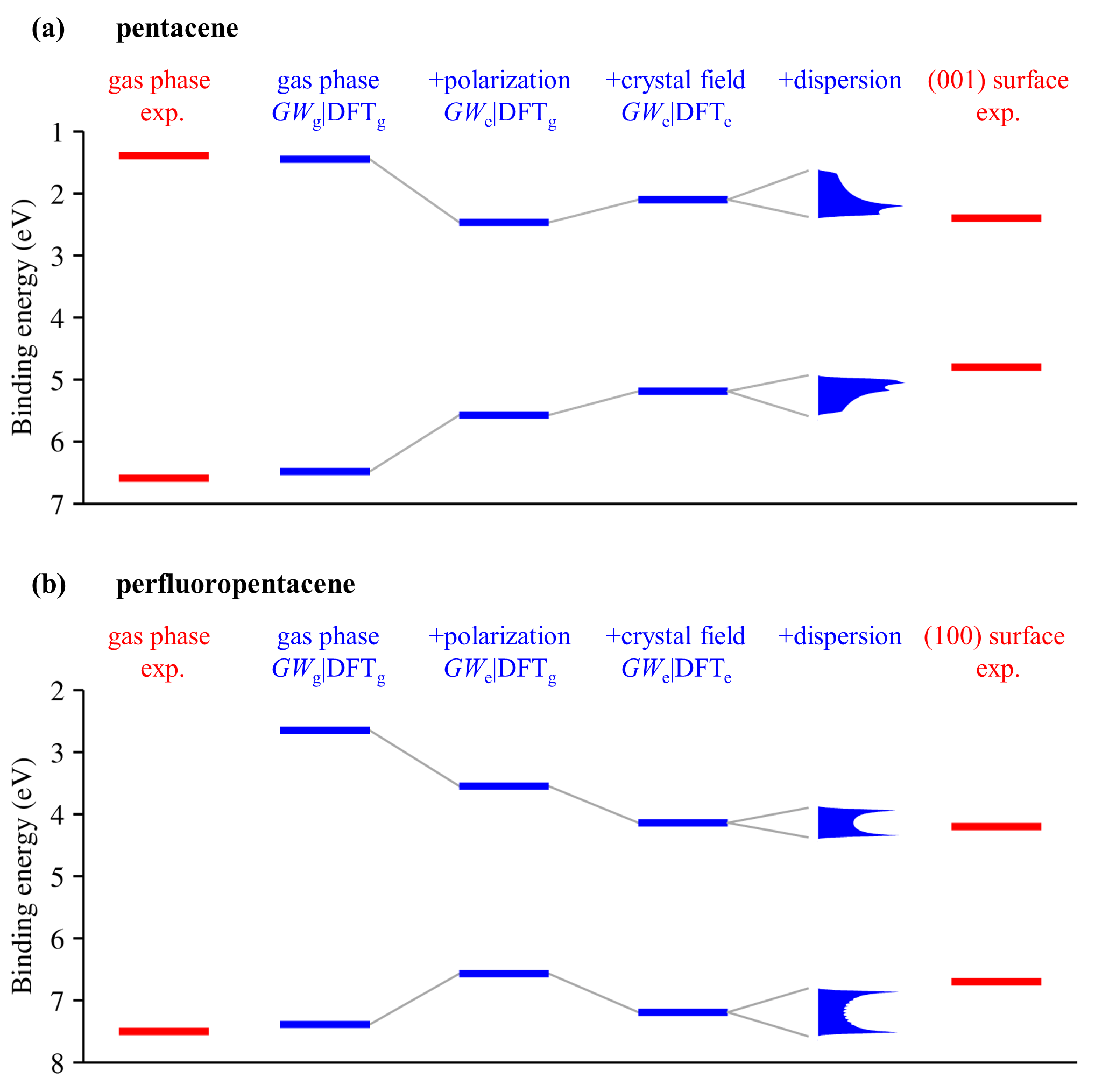}
\caption{Evolution of IP and EA from the gas phase to crystal surfaces for (a) PEN and (b) PFP.
Calculation results (in blue) are presented by progressively introducing the different contributions from intermolecular interactions (polarization, crystal field, band dispersion) to the final excitation energies at the crystal surface.
The agreement with available experimental data (in red) is very good both in the gas and in the solid state. This highlights the consistency and accuracy of the proposed computational scheme.
}
\label{f:last}
\end{figure}

The evolution of the HOMO and LUMO levels in PEN and PFP from the isolated gas-phase molecule to crystal surfaces is summarized in Figure \ref{f:last}.
We recall that our results for ionization at surfaces do apply also to molecular films on insulating substrate (e.g. SiO$_\mathrm{x}$) with same molecular orientations.

The calculated gas-phase levels ($GW_\mathrm{g}|$DFT$_\mathrm{g}$ with cc-pVQZ basis) are in excellent agreement with experimental data for both PEN (IP=6.59 eV,\cite{Cor02} EA=1.35 eV \cite{Cro93}) and PFP (IP=7.50 eV \cite{Del09}).
Our gas-phase ev$GW$ EA for PFP of 2.65 eV, closely compares with earlier DFT ($\Delta$SCF approach) values.\cite{Del09,Top11}

The effects of the environment are then progressively added, starting from polarization that closes the gap by approximately 2 eV in both compounds.
This effect is accounted for at the $GW_\mathrm{e}|$DFT$_\mathrm{g}$ level through the
polarization contribution $\Delta^{COHSEX}$ (see Eq.~\ref{dch}).
The latter has the same physical meaning of the charge-induced dipole interaction 
in microelectrostatic models,\cite{Bou79} although the two calculations follow completely different schemes.
Classical microelectrostatic calculations compute the interaction between a localized charge and the dipole induced in the medium, 
while in embedded $GW$ calculations one computes the MM dielectric contribution to the screened Coulomb potential $W$, allowing to obtain the polarization energy to all occupied/virtual energy levels. 

Crystal field shifts HOMO and LUMO levels by approximately the same amount, in opposite directions for PEN and PFP, owing to the opposite sign of the quadrupole components.
This contribution is the equivalent of the charge-quadrupole interaction in classical
microelectrostatics, lifting the so-called electron-hole symmetry in polarization energies.\cite{Bou81}
Such an effect is accounted for in our hybrid calculations by obtaining the KS orbitals in the self-consistent field of permanent and induced multipoles in the MM environment.

So far we have considered quasiparticle excitations of PEN and PFP films in the limit of charges fully localized on a single molecule, corresponding to the QM subsystem in our hybrid formalism.
A fair comparison with experiments requires, however, to account for band dispersion, a phenomenon that has been experimentally observed for a few crystalline molecular solids, including PEN.\cite{Kak07,Ciu12}
The interplay between charge delocalization and polarization effects has been discussed in a very recent study \cite{Li16}, where, upon comparing systems including a different number of molecules within the QM region, we have shown that the intermolecular charge transfer couplings can be safely treated as a perturbation to transport levels obtained for charges localized on  individual molecules in a relaxed dielectric environment.
Band dispersion is hence introduced by means of a tight binding model for HOMO and LUMO bands, that is fully parametrized \textit{ab initio}.  
Namely, site energies are obtained from embedded $GW$ calculations in Table \ref{t:surf}, while charge transfer couplings are calculated at the DFT level, fully accounting for crystal field effects -- see appendix \ref{a:tb} for details.
Our tight binding band structures are in excellent agreement with literature DFT analogues (see Figure~\ref{f:tb}),
\textcolor{\ntc}{
yet it has been reported that nonlocal correlations lead to $GW$ quasiparticle bands that are up to 20\% wider than DFT ones.\cite{Tia03,Sha12,Yan14} 
The latter effect is expect to lead to a small reduction of the gap by less than 50 meV. }

The edges of the densities of states obtained from tight binding calculations, shown as blue areas in Figure \ref{f:last},
define our theoretical estimates of the IP and EA of PEN and PFP crystals, compared to experimental values obtained at the same surfaces in Table \ref{t:exp}. 
The agreement between our calculations and accurate experiments \cite{Sal08,Yos15} is within 0.2 eV for both IP and EA, a value comparable to the experimental uncertainty, but also to the spread of IP values obtained from different UPS experiments on films of standing PEN (see Ref. \onlinecite{Yos15} for a compilation of experimental data).
Structural and energetic disorder, specific interactions with a given substrate and polaronic relaxation are additional factors impacting photoelectron measurements in organic materials, which can all source errors in the comparison between theory and experiment.

\begin{table}
\begin{tabular}{l  C{20mm} C{20mm}|  C{20mm} C{20mm}  }
\hline \hline 
& \multicolumn{2}{c|}{\textbf{pentacene}} & \multicolumn{2}{c}{\textbf{perfluoropentacene}} \\
 & calc. & exp.  & calc. & exp. \\
\hline
IP   & 4.93 & 4.8 & 6.81 & 6.7  \\
EA   & 2.41 & 2.4 & 4.39 & 4.2  \\
gap  & 2.52 & 2.4 & 2.42 & 2.5  \\
\hline \hline
\end{tabular}
\caption{Comparison of calculated and experimental IP and EA for PEN and PFP. Energies are in eV. The calculated IP (EA) is defined as the edge of the  valence (conduction) obtained from our embedded $GW$-parametrized tight binding model (see text and appendix~\ref{a:tb}). 
  Experimental data from \cite{Sal08,Yos15} with typical uncertainty of 0.1 eV on IP and EA and 0.2 eV on the gap.
}
\label{t:exp}
\end{table}

\clearpage
\section{Conclusions}
\label{s:conclu}

We have presented a novel computational approach to the calculation of charged excitations in organic solids based on a hybrid quantum/classical scheme merging the Green's function $GW$ formalism with classical polarizable models.
The high accuracy of the description of the electronic structure of the ionized molecule and of its interaction with the environment allowed us to obtain electron addition or removal energies that are in quantitative agreement with photoelectron spectroscopy experiments for the two molecular semiconductors pentacene and perfluoropentacene. 

Four ingredients are found to be important for a quantitative first-principles calculation of IP and EA in the solid state:
(i) nonlocal many-body interactions at the molecular level, which are key to gas-phase IP and EA, (ii) polarization, i.e. the screening of excitations provided by the solid-state environment, (iii) the crystal field exerted by the charge densities of neutral molecules in the solid and (iv) band dispersion.

The hybrid formalism presented herein can be applied to compute from first principles surface-specific charged excitations in molecular crystals  with  quantitative accuracy.
Moreover, such a general tool can also be applied to more complex molecular solids, including disordered or heterogeneous systems for the accurate evaluation of charge transport levels or as a starting point for the calculation of optical excitations within the framework of the Bethe-Salpeter equation.

\appendix

\section{Additional information on the calculation of the reaction field matrix}
\label{a:vr}

\begin{figure}[ht]
\centering
\includegraphics[width=12cm]{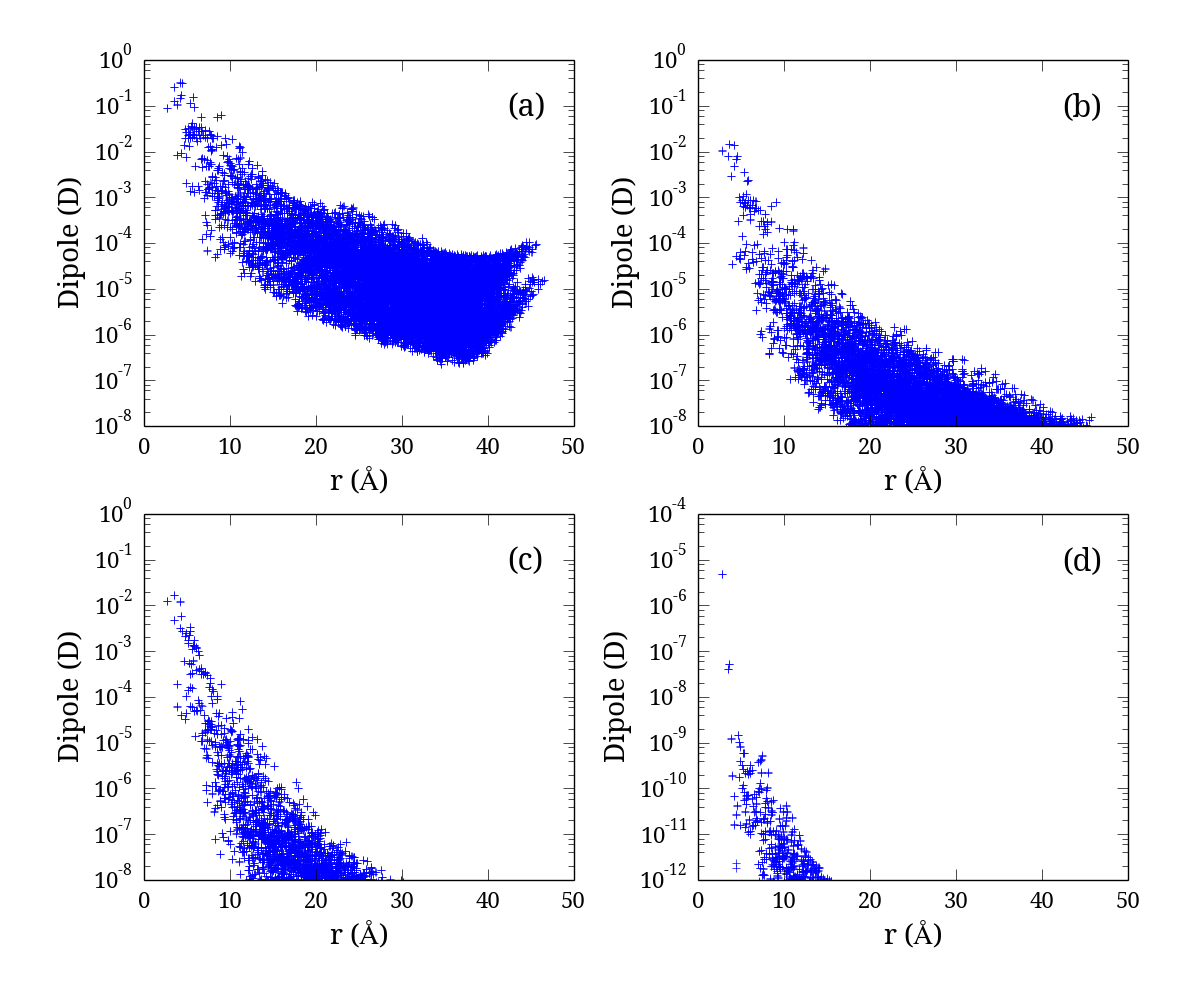}
\caption{Dipoles (in Debye) induced on MM molecules by auxiliary orbitals of type (a) $s$, (b) $p$ and (c) $d$ within QM subsystem.
Dipole moments are plotted against the distance $r$ between the source charge and the centroids of MM molecules.
Panel (d) shows the dipoles induced by a charge distribution that is the superposition of two $s$ orbitals centered on the same atom with zero net charge, showing a very quick distance decay.
}
\label{f:rf}
\end{figure}

Figure \ref{f:rf} provides an overview of the dipoles induced on MM molecules by charges within the QM region.
Panel (a) shows the distance dependence of the dipoles induced by an auxiliary $s$ orbitals as a function of the distance between the source charge and the polarized molecules.
Panels (b) and (c) show the dipoles induced by auxiliary $p$ and $d$ orbitals, characterized by a faster and faster distance decay.

An important ingredient in the calulation of $v^{reac}$ in the common case of  multiple-zeta basis is that we only calculate the reaction field associated with one $s$-orbital (say $s_j$) per atom, calculating for the other $s_i$ orbitals on the same atom the response to the  ($s_i -  c_i s_{ji}$) difference, with the $c_{ji}$ coefficient chosen so that the net charge of the formed linear combination vanishes.
The field exerted by such a combination decays extremely fast [see dipoles induced in MM molecules in Figure~\ref{f:rf}(d)]  and vanishes at distances where the differential charge density has completely decayed to zero, according to the Gauss theorem. 

Exploiting the linear response property of the MM subsystem, the reation field matrix elements corresponding to a given orbitals $s_i$ on a given atom  can be written as
\begin{equation}
v^{reac}(s_i,{\beta}')= \frac{1}{c_{ji}} \left[v^{reac}(s_j,{\beta}') - v^{reac}(s_j-c_{ji}s_i,{\beta}') \right].
\end{equation}
In practice, only the reaction field associated with one $s$ orbital per atom 
need to be calculated in the limit of large MM clusters, allowing to dramatically speed up the calculation of the extrapolated reaction field matrix.

A further reduction of the computational burden could be obtained generalizing the strategy outlined above to $s$ orbitals on different atoms, hence computing fewer 
$s$ orbitals (e.g. one per molecule) and obtaining the other $v^{reac}(s_i,{\beta}')$ elements through the reaction to the dipole field created by appropriate linear combination of charges.

\section{Tight binding band structure from first principles inputs}
\label{a:tb}

The electronic band structure of PEN and PFP is here described with a tight binding model fully parametrized from first-principles. 
The model accounts for HOMO and LUMO orbitals of the two molecules in the unit cell and considers a two dimensional lattice corresponding to the herringbone plane, i.e. (001) for PEN and (100) for PFP.
Dispersion along the plane normal is neglected, owing to the very small intermolecular hopping terms. 

The model is parametrized with orbital site energies from $GW_\mathrm{e}|$DFT$_\mathrm{e}$ calculations at crystal surfaces (Table~\ref{t:surf}) and intermolecular transfer integrals obtained at the DFT level (Table~\ref{t:cti}) fully accounting for crystal field effect.
Specifically, HOMO-HOMO and LUMO-LUMO couplings are calculated at the PBE0/6-31G* level with the dimer projection approach,\cite{Val06} employing molecular orbitals and dimer Hamiltonian obtained in the self-consistent field of embedding MM atoms (DFT$_\mathrm{e}$ level). 
Table~\ref{t:cti} reports both values from MM-embedded calculations and those obtained from the standard dimer projection technique where DFT calculations are performed in vacuum (in parentheses).
The two set of values differ by few meV for PEN and up to 10 meV in PFP.
Consistent signs of the orbital couplings are obtained by probing the phase of the orbitals with  fictitious $s$ orbitals.\cite{Li17}

\begin{figure}[ht]
\centering
\includegraphics[width=8cm]{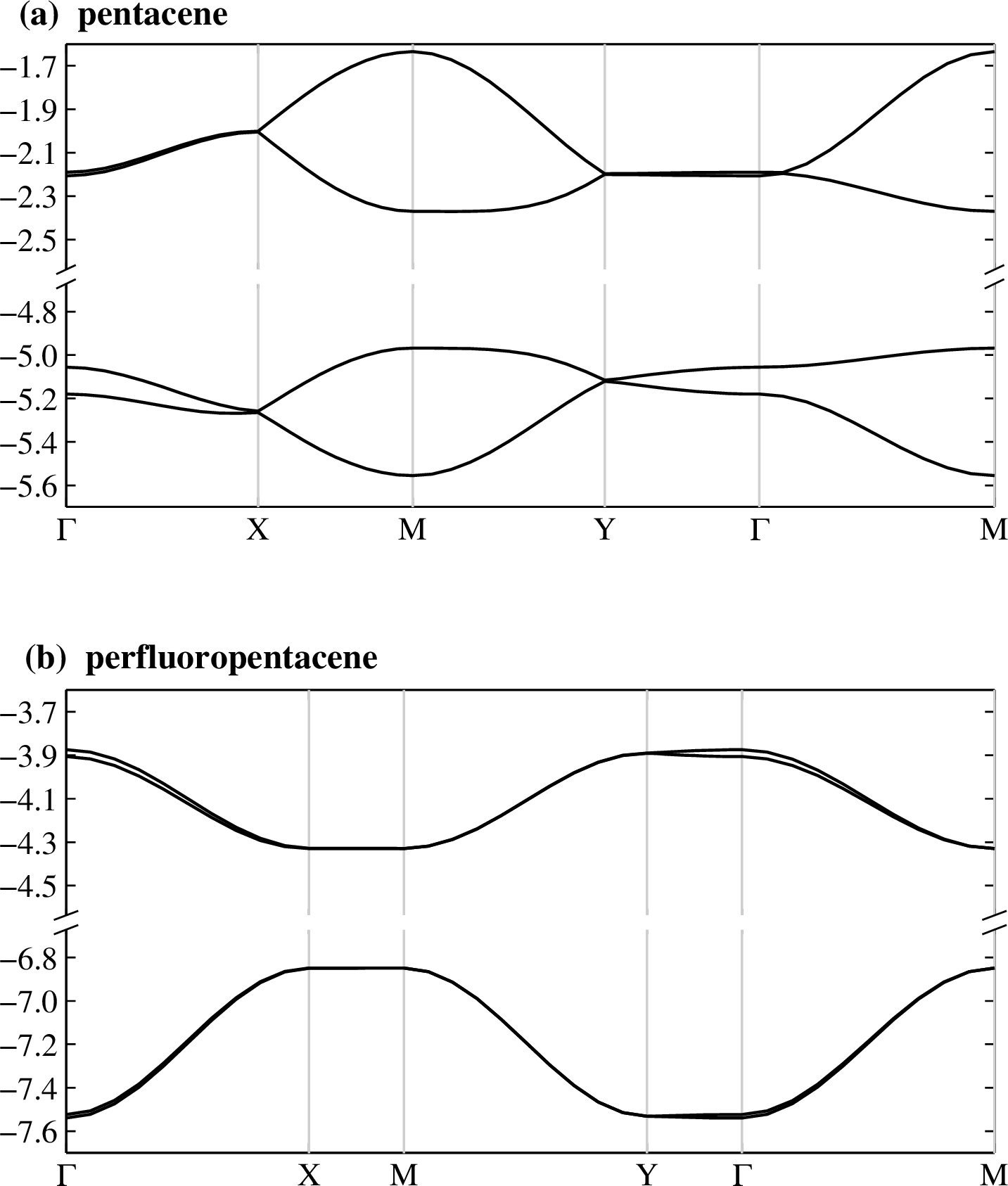}
\caption{
  Tight binding band structure for HOMO and LUMO bands of (a) PEN and (b) PFP as obtained with orbital energies from $GW_\mathrm{e}|$DFT$_\mathrm{e}$ calculations at crystal surfaces (Table~\ref{t:surf}) and DFT transfer integrals in Table~\ref{t:cti}.
  The high-symmetry points in the Brillouin zone of PEN correspond to:
  $\Gamma=$(0,0,0), $X=$(\textonehalf,0,0), $M=$(\textonehalf,\textonehalf,0),
  $Y=$(0,\textonehalf,0).
For PFP:   $\Gamma=$(0,0,0), $X=$(0,\textonehalf,0), $M=$(0,\textonehalf,\textonehalf), $Y=$(0,0,\textonehalf).
Band structure obtained with a $41{\times}41$ sampling of the Brillouin zone.
}
\label{f:tb}
\end{figure}

Figure~\ref{f:tb} shows our tight binding band structures for PEN and PFP.  
The tight binding band structure for PEN is in full quantitative agreement with \emph{ab initio} results obtained from periodic calculations at a comparable level of theory (B3LYP/6-31G).\cite{Li12}
The dispersion of PFP bands  compares well with earlier plane-wave results,\cite{Del09,Yos15} although our bandwidths (0.69 and 0.46 eV for HOMO and LUMO bands, respectively) are considerably larger.
Such a discrepancy can be attributed to the generalized gradient approximation (GGA) functionals employed in the plane wave calculations.\cite{Del09,Yos15}
Charge transfer integrals calculated with the GGA PBE functional are in fact found to be about 20\% smaller than PBE0 ones in Table~\ref{t:cti} and, when employed in the tight binding calculations, they lead to bandwidths in very good agreement  with plane-wave results.\cite{Del09,Yos15}

It is interesting to note that the dispersion in PFP is almost one-dimensional, as testified by the flat bands along the $X-M$ and $Y-\Gamma$ paths in Figure~\ref{f:tb}(b).
This results from the very small transfer integrals for both HOMO and LUMO involving pairs with herringbone ``T''-like arrangement (last line in Table~\ref{t:cti}).

\begin{table}
\begin{tabular}{l c rr}
  \hline \hline 
 & dimer  & $t_{HOMO}$   & $t_{LUMO}$  \\
\hline
\textbf{PEN} & (0,0,0) - (1,0,0) & 35 (34) & -48 (-47) \\
& (0,0,0) - (\textonehalf,\textonehalf,0) & -58 (-53) & -94 (-86) \\
& (0,0,0) - (-\textonehalf,\textonehalf,0) & 89 (85) & 90 (86) \\
& (-\textonehalf,-\textonehalf,0) - (\textonehalf,\textonehalf,0) & 37 (39) & -50 (-54) \\
\hline
\textbf{PFP} & (0,0,0) - (0,0,1) & -171 (-161) & 110 (106) \\
& (0,0,0) - (0,\textonehalf,\textonehalf) & 2 (2) & -4 (-3) \\
\hline \hline
\end{tabular}
\caption{Charge transfer integrals ($t$ in meV) between dimers of neighboring molecules in the PEN and PFP crystal. 
Transfer integrals are computed as intermolecular orbital couplings
($t_k=\langle \phi_k |H|\phi_k \rangle $, $k=$HOMO, LUMO)  at the PBE0/6-31G* level.
The effect of the crystalline environment on $t$ is accounted for by computing molecules and dimers in the self-consistent field of embedding MM atoms (DFT$_\mathrm{e}$ level). 
$t$ values obtained with the standard dimer projection approach (DFT$_\mathrm{g}$ level, i.e. without MM embedding) are given in parentheses.
PFP features a monoclinic cell with two equivalent molecules, leading to only two independent transfer integrals. 
}
\label{t:cti}
\end{table}

\clearpage 
\begin{acknowledgements}
G.D. is indebted to Z.~G. Soos for enlightening discussions on CR and photoelectron spectroscopy of organic systems.
This project has received funding from the European Union Horizon 2020 research and innovation programme under grant agreement No 646176 (EXTMOS) and No 676629 (EoCoE).
D.B. is FNRS research director. 
This research used resources from the GENCI French national supercomputing resources. 
\end{acknowledgements}


\bibliography{gwmm}

\begin{thebibliography}{102}%
\makeatletter
\providecommand \@ifxundefined [1]{%
 \@ifx{#1\undefined}
}%
\providecommand \@ifnum [1]{%
 \ifnum #1\expandafter \@firstoftwo
 \else \expandafter \@secondoftwo
 \fi
}%
\providecommand \@ifx [1]{%
 \ifx #1\expandafter \@firstoftwo
 \else \expandafter \@secondoftwo
 \fi
}%
\providecommand \natexlab [1]{#1}%
\providecommand \enquote  [1]{``#1''}%
\providecommand \bibnamefont  [1]{#1}%
\providecommand \bibfnamefont [1]{#1}%
\providecommand \citenamefont [1]{#1}%
\providecommand \href@noop [0]{\@secondoftwo}%
\providecommand \href [0]{\begingroup \@sanitize@url \@href}%
\providecommand \@href[1]{\@@startlink{#1}\@@href}%
\providecommand \@@href[1]{\endgroup#1\@@endlink}%
\providecommand \@sanitize@url [0]{\catcode `\\12\catcode `\$12\catcode
  `\&12\catcode `\#12\catcode `\^12\catcode `\_12\catcode `\%12\relax}%
\providecommand \@@startlink[1]{}%
\providecommand \@@endlink[0]{}%
\providecommand \url  [0]{\begingroup\@sanitize@url \@url }%
\providecommand \@url [1]{\endgroup\@href {#1}{\urlprefix }}%
\providecommand \urlprefix  [0]{URL }%
\providecommand \Eprint [0]{\href }%
\providecommand \doibase [0]{http://dx.doi.org/}%
\providecommand \selectlanguage [0]{\@gobble}%
\providecommand \bibinfo  [0]{\@secondoftwo}%
\providecommand \bibfield  [0]{\@secondoftwo}%
\providecommand \translation [1]{[#1]}%
\providecommand \BibitemOpen [0]{}%
\providecommand \bibitemStop [0]{}%
\providecommand \bibitemNoStop [0]{.\EOS\space}%
\providecommand \EOS [0]{\spacefactor3000\relax}%
\providecommand \BibitemShut  [1]{\csname bibitem#1\endcsname}%
\let\auto@bib@innerbib\@empty
\bibitem [{\citenamefont {Ishii}\ \emph {et~al.}(1999)\citenamefont {Ishii},
  \citenamefont {Sugiyama}, \citenamefont {Ito},\ and\ \citenamefont
  {Seki}}]{Ish99}%
  \BibitemOpen
  \bibfield  {author} {\bibinfo {author} {\bibfnamefont {H.}~\bibnamefont
  {Ishii}}, \bibinfo {author} {\bibfnamefont {K.}~\bibnamefont {Sugiyama}},
  \bibinfo {author} {\bibfnamefont {E.}~\bibnamefont {Ito}}, \ and\ \bibinfo
  {author} {\bibfnamefont {K.}~\bibnamefont {Seki}},\ }\href {\doibase
  10.1002/(SICI)1521-4095(199906)11:8<605::AID-ADMA605>3.0.CO;2-Q} {\bibfield
  {journal} {\bibinfo  {journal} {Adv. Mater.}\ }\textbf {\bibinfo {volume}
  {11}},\ \bibinfo {pages} {605} (\bibinfo {year} {1999})}\BibitemShut
  {NoStop}%
\bibitem [{\citenamefont {Cahen}\ and\ \citenamefont {Kahn}(2003)}]{Cah03}%
  \BibitemOpen
  \bibfield  {author} {\bibinfo {author} {\bibfnamefont {D.}~\bibnamefont
  {Cahen}}\ and\ \bibinfo {author} {\bibfnamefont {A.}~\bibnamefont {Kahn}},\
  }\href {\doibase 10.1002/adma.200390065} {\bibfield  {journal} {\bibinfo
  {journal} {Adv. Mater.}\ }\textbf {\bibinfo {volume} {15}},\ \bibinfo {pages}
  {271} (\bibinfo {year} {2003})}\BibitemShut {NoStop}%
\bibitem [{\citenamefont {Heimel}\ \emph {et~al.}(2011)\citenamefont {Heimel},
  \citenamefont {Salzmann}, \citenamefont {Duhm},\ and\ \citenamefont
  {Koch}}]{Hei11}%
  \BibitemOpen
  \bibfield  {author} {\bibinfo {author} {\bibfnamefont {G.}~\bibnamefont
  {Heimel}}, \bibinfo {author} {\bibfnamefont {I.}~\bibnamefont {Salzmann}},
  \bibinfo {author} {\bibfnamefont {S.}~\bibnamefont {Duhm}}, \ and\ \bibinfo
  {author} {\bibfnamefont {N.}~\bibnamefont {Koch}},\ }\href {\doibase
  10.1021/cm1021257} {\bibfield  {journal} {\bibinfo  {journal} {Chem. Mater.}\
  }\textbf {\bibinfo {volume} {23}},\ \bibinfo {pages} {359} (\bibinfo {year}
  {2011})}\BibitemShut {NoStop}%
\bibitem [{\citenamefont {D'Avino}\ \emph
  {et~al.}(2016{\natexlab{a}})\citenamefont {D'Avino}, \citenamefont
  {Muccioli}, \citenamefont {Castet}, \citenamefont {Poelking}, \citenamefont
  {Andrienko}, \citenamefont {Soos}, \citenamefont {Cornil},\ and\
  \citenamefont {Beljonne}}]{Dav16_rev}%
  \BibitemOpen
  \bibfield  {author} {\bibinfo {author} {\bibfnamefont {G.}~\bibnamefont
  {D'Avino}}, \bibinfo {author} {\bibfnamefont {L.}~\bibnamefont {Muccioli}},
  \bibinfo {author} {\bibfnamefont {F.}~\bibnamefont {Castet}}, \bibinfo
  {author} {\bibfnamefont {C.}~\bibnamefont {Poelking}}, \bibinfo {author}
  {\bibfnamefont {D.}~\bibnamefont {Andrienko}}, \bibinfo {author}
  {\bibfnamefont {Z.~G.}\ \bibnamefont {Soos}}, \bibinfo {author}
  {\bibfnamefont {J.}~\bibnamefont {Cornil}}, \ and\ \bibinfo {author}
  {\bibfnamefont {D.}~\bibnamefont {Beljonne}},\ }\href {\doibase
  10.1088/0953-8984/28/43/433002} {\bibfield  {journal} {\bibinfo  {journal}
  {J. Phys.: Condens. Matter}\ }\textbf {\bibinfo {volume} {28}},\ \bibinfo
  {pages} {433002} (\bibinfo {year} {2016}{\natexlab{a}})}\BibitemShut
  {NoStop}%
\bibitem [{\citenamefont {Duhm}\ \emph {et~al.}(2008)\citenamefont {Duhm},
  \citenamefont {Heimel}, \citenamefont {Salzmann}, \citenamefont {Glowatzki},
  \citenamefont {Johnson}, \citenamefont {Vollmer}, \citenamefont {Rabe},\ and\
  \citenamefont {Koch}}]{Duh08}%
  \BibitemOpen
  \bibfield  {author} {\bibinfo {author} {\bibfnamefont {S.}~\bibnamefont
  {Duhm}}, \bibinfo {author} {\bibfnamefont {G.}~\bibnamefont {Heimel}},
  \bibinfo {author} {\bibfnamefont {I.}~\bibnamefont {Salzmann}}, \bibinfo
  {author} {\bibfnamefont {H.}~\bibnamefont {Glowatzki}}, \bibinfo {author}
  {\bibfnamefont {R.~L.}\ \bibnamefont {Johnson}}, \bibinfo {author}
  {\bibfnamefont {A.}~\bibnamefont {Vollmer}}, \bibinfo {author} {\bibfnamefont
  {J.~P.}\ \bibnamefont {Rabe}}, \ and\ \bibinfo {author} {\bibfnamefont
  {N.}~\bibnamefont {Koch}},\ }\href@noop {} {\bibfield  {journal} {\bibinfo
  {journal} {Nat. Mater.}\ }\textbf {\bibinfo {volume} {7}},\ \bibinfo {pages}
  {326} (\bibinfo {year} {2008})}\BibitemShut {NoStop}%
\bibitem [{\citenamefont {Salzmann}\ \emph {et~al.}(2008)\citenamefont
  {Salzmann}, \citenamefont {Duhm}, \citenamefont {Heimel}, \citenamefont
  {Oehzelt}, \citenamefont {Kniprath}, \citenamefont {Johnson}, \citenamefont
  {Rabe},\ and\ \citenamefont {Koch}}]{Sal08}%
  \BibitemOpen
  \bibfield  {author} {\bibinfo {author} {\bibfnamefont {I.}~\bibnamefont
  {Salzmann}}, \bibinfo {author} {\bibfnamefont {S.}~\bibnamefont {Duhm}},
  \bibinfo {author} {\bibfnamefont {G.}~\bibnamefont {Heimel}}, \bibinfo
  {author} {\bibfnamefont {M.}~\bibnamefont {Oehzelt}}, \bibinfo {author}
  {\bibfnamefont {R.}~\bibnamefont {Kniprath}}, \bibinfo {author}
  {\bibfnamefont {R.~L.}\ \bibnamefont {Johnson}}, \bibinfo {author}
  {\bibfnamefont {J.~P.}\ \bibnamefont {Rabe}}, \ and\ \bibinfo {author}
  {\bibfnamefont {N.}~\bibnamefont {Koch}},\ }\href {\doibase
  10.1021/ja804793a} {\bibfield  {journal} {\bibinfo  {journal} {J. Am. Chem.
  Soc.}\ }\textbf {\bibinfo {volume} {130}},\ \bibinfo {pages} {12870}
  (\bibinfo {year} {2008})}\BibitemShut {NoStop}%
\bibitem [{\citenamefont {Chen}\ \emph {et~al.}(2008)\citenamefont {Chen},
  \citenamefont {Huang}, \citenamefont {Chen}, \citenamefont {Huang},
  \citenamefont {Gao},\ and\ \citenamefont {Wee}}]{Che08}%
  \BibitemOpen
  \bibfield  {author} {\bibinfo {author} {\bibfnamefont {W.}~\bibnamefont
  {Chen}}, \bibinfo {author} {\bibfnamefont {H.}~\bibnamefont {Huang}},
  \bibinfo {author} {\bibfnamefont {S.}~\bibnamefont {Chen}}, \bibinfo {author}
  {\bibfnamefont {Y.~L.}\ \bibnamefont {Huang}}, \bibinfo {author}
  {\bibfnamefont {X.~Y.}\ \bibnamefont {Gao}}, \ and\ \bibinfo {author}
  {\bibfnamefont {A.~T.~S.}\ \bibnamefont {Wee}},\ }\href {\doibase
  10.1021/cm8016352} {\bibfield  {journal} {\bibinfo  {journal} {Chem. Mater.}\
  }\textbf {\bibinfo {volume} {20}},\ \bibinfo {pages} {7017} (\bibinfo {year}
  {2008})}\BibitemShut {NoStop}%
\bibitem [{\citenamefont {Hedin}(1965)}]{Hed65}%
  \BibitemOpen
  \bibfield  {author} {\bibinfo {author} {\bibfnamefont {L.}~\bibnamefont
  {Hedin}},\ }\href@noop {} {\bibfield  {journal} {\bibinfo  {journal} {Phys.
  Rev. A}\ }\textbf {\bibinfo {volume} {139}},\ \bibinfo {pages} {796}
  (\bibinfo {year} {1965})}\BibitemShut {NoStop}%
\bibitem [{\citenamefont {Onida}\ \emph {et~al.}(2002)\citenamefont {Onida},
  \citenamefont {Reining},\ and\ \citenamefont {Rubio}}]{Oni02}%
  \BibitemOpen
  \bibfield  {author} {\bibinfo {author} {\bibfnamefont {G.}~\bibnamefont
  {Onida}}, \bibinfo {author} {\bibfnamefont {L.}~\bibnamefont {Reining}}, \
  and\ \bibinfo {author} {\bibfnamefont {A.}~\bibnamefont {Rubio}},\
  }\href@noop {} {\bibfield  {journal} {\bibinfo  {journal} {Rev. Mod. Phys.}\
  }\textbf {\bibinfo {volume} {74}},\ \bibinfo {pages} {601} (\bibinfo {year}
  {2002})}\BibitemShut {NoStop}%
\bibitem [{\citenamefont {Hybertsen}\ and\ \citenamefont
  {Louie}(1986)}]{Hyb86}%
  \BibitemOpen
  \bibfield  {author} {\bibinfo {author} {\bibfnamefont {M.~S.}\ \bibnamefont
  {Hybertsen}}\ and\ \bibinfo {author} {\bibfnamefont {S.~G.}\ \bibnamefont
  {Louie}},\ }\href {\doibase 10.1103/PhysRevB.34.5390} {\bibfield  {journal}
  {\bibinfo  {journal} {Phys. Rev. B}\ }\textbf {\bibinfo {volume} {34}},\
  \bibinfo {pages} {5390} (\bibinfo {year} {1986})}\BibitemShut {NoStop}%
\bibitem [{\citenamefont {Godby}\ \emph {et~al.}(1988)\citenamefont {Godby},
  \citenamefont {Schl\"uter},\ and\ \citenamefont {Sham}}]{God88}%
  \BibitemOpen
  \bibfield  {author} {\bibinfo {author} {\bibfnamefont {R.~W.}\ \bibnamefont
  {Godby}}, \bibinfo {author} {\bibfnamefont {M.}~\bibnamefont {Schl\"uter}}, \
  and\ \bibinfo {author} {\bibfnamefont {L.~J.}\ \bibnamefont {Sham}},\
  }\href@noop {} {\bibfield  {journal} {\bibinfo  {journal} {Phys. Rev. B}\
  }\textbf {\bibinfo {volume} {37}},\ \bibinfo {pages} {10159} (\bibinfo {year}
  {1988})}\BibitemShut {NoStop}%
\bibitem [{\citenamefont {Tiago}\ \emph {et~al.}(2003)\citenamefont {Tiago},
  \citenamefont {Northrup},\ and\ \citenamefont {Louie}}]{Tia03}%
  \BibitemOpen
  \bibfield  {author} {\bibinfo {author} {\bibfnamefont {M.~L.}\ \bibnamefont
  {Tiago}}, \bibinfo {author} {\bibfnamefont {J.~E.}\ \bibnamefont {Northrup}},
  \ and\ \bibinfo {author} {\bibfnamefont {S.~G.}\ \bibnamefont {Louie}},\
  }\href {\doibase 10.1103/PhysRevB.67.115212} {\bibfield  {journal} {\bibinfo
  {journal} {Phys. Rev. B}\ }\textbf {\bibinfo {volume} {67}},\ \bibinfo
  {pages} {115212} (\bibinfo {year} {2003})}\BibitemShut {NoStop}%
\bibitem [{\citenamefont {Sharifzadeh}\ \emph {et~al.}(2012)\citenamefont
  {Sharifzadeh}, \citenamefont {Biller}, \citenamefont {Kronik},\ and\
  \citenamefont {Neaton}}]{Sha12}%
  \BibitemOpen
  \bibfield  {author} {\bibinfo {author} {\bibfnamefont {S.}~\bibnamefont
  {Sharifzadeh}}, \bibinfo {author} {\bibfnamefont {A.}~\bibnamefont {Biller}},
  \bibinfo {author} {\bibfnamefont {L.}~\bibnamefont {Kronik}}, \ and\ \bibinfo
  {author} {\bibfnamefont {J.~B.}\ \bibnamefont {Neaton}},\ }\href {\doibase
  10.1103/PhysRevB.85.125307} {\bibfield  {journal} {\bibinfo  {journal} {Phys.
  Rev. B}\ }\textbf {\bibinfo {volume} {85}},\ \bibinfo {pages} {125307}
  (\bibinfo {year} {2012})}\BibitemShut {NoStop}%
\bibitem [{\citenamefont {Rangel}\ \emph {et~al.}(2016)\citenamefont {Rangel},
  \citenamefont {Berland}, \citenamefont {Sharifzadeh}, \citenamefont
  {Brown-Altvater}, \citenamefont {Lee}, \citenamefont {Hyldgaard},
  \citenamefont {Kronik},\ and\ \citenamefont {Neaton}}]{Ran16}%
  \BibitemOpen
  \bibfield  {author} {\bibinfo {author} {\bibfnamefont {T.}~\bibnamefont
  {Rangel}}, \bibinfo {author} {\bibfnamefont {K.}~\bibnamefont {Berland}},
  \bibinfo {author} {\bibfnamefont {S.}~\bibnamefont {Sharifzadeh}}, \bibinfo
  {author} {\bibfnamefont {F.}~\bibnamefont {Brown-Altvater}}, \bibinfo
  {author} {\bibfnamefont {K.}~\bibnamefont {Lee}}, \bibinfo {author}
  {\bibfnamefont {P.}~\bibnamefont {Hyldgaard}}, \bibinfo {author}
  {\bibfnamefont {L.}~\bibnamefont {Kronik}}, \ and\ \bibinfo {author}
  {\bibfnamefont {J.~B.}\ \bibnamefont {Neaton}},\ }\href {\doibase
  10.1103/PhysRevB.93.115206} {\bibfield  {journal} {\bibinfo  {journal} {Phys.
  Rev. B}\ }\textbf {\bibinfo {volume} {93}},\ \bibinfo {pages} {115206}
  (\bibinfo {year} {2016})}\BibitemShut {NoStop}%
\bibitem [{\citenamefont {Yanagisawa}\ \emph {et~al.}(2014)\citenamefont
  {Yanagisawa}, \citenamefont {Morikawa},\ and\ \citenamefont
  {Schindlmayr}}]{Yan14}%
  \BibitemOpen
  \bibfield  {author} {\bibinfo {author} {\bibfnamefont {S.}~\bibnamefont
  {Yanagisawa}}, \bibinfo {author} {\bibfnamefont {Y.}~\bibnamefont
  {Morikawa}}, \ and\ \bibinfo {author} {\bibfnamefont {A.}~\bibnamefont
  {Schindlmayr}},\ }\href {\doibase 10.7567/JJAP.53.05FY02} {\bibfield
  {journal} {\bibinfo  {journal} {Jpn. J. Appl. Phys.}\ }\textbf {\bibinfo
  {volume} {53}},\ \bibinfo {pages} {05FY02} (\bibinfo {year}
  {2014})}\BibitemShut {NoStop}%
\bibitem [{\citenamefont {Wang}\ \emph {et~al.}(2016)\citenamefont {Wang},
  \citenamefont {Garcia}, \citenamefont {Monaco}, \citenamefont
  {Schatschneider},\ and\ \citenamefont {Marom}}]{Wan16}%
  \BibitemOpen
  \bibfield  {author} {\bibinfo {author} {\bibfnamefont {X.}~\bibnamefont
  {Wang}}, \bibinfo {author} {\bibfnamefont {T.}~\bibnamefont {Garcia}},
  \bibinfo {author} {\bibfnamefont {S.}~\bibnamefont {Monaco}}, \bibinfo
  {author} {\bibfnamefont {B.}~\bibnamefont {Schatschneider}}, \ and\ \bibinfo
  {author} {\bibfnamefont {N.}~\bibnamefont {Marom}},\ }\href {\doibase
  10.1039/C6CE00873A} {\bibfield  {journal} {\bibinfo  {journal}
  {CrystEngComm}\ }\textbf {\bibinfo {volume} {18}},\ \bibinfo {pages} {7353}
  (\bibinfo {year} {2016})}\BibitemShut {NoStop}%
\bibitem [{\citenamefont {Ambrosch-Draxl}\ \emph {et~al.}(2009)\citenamefont
  {Ambrosch-Draxl}, \citenamefont {Nabok}, \citenamefont {Puschnig},\ and\
  \citenamefont {Meisenbichler}}]{Dra09}%
  \BibitemOpen
  \bibfield  {author} {\bibinfo {author} {\bibfnamefont {C.}~\bibnamefont
  {Ambrosch-Draxl}}, \bibinfo {author} {\bibfnamefont {D.}~\bibnamefont
  {Nabok}}, \bibinfo {author} {\bibfnamefont {P.}~\bibnamefont {Puschnig}}, \
  and\ \bibinfo {author} {\bibfnamefont {C.}~\bibnamefont {Meisenbichler}},\
  }\href {\doibase 10.1088/1367-2630/11/12/125010} {\bibfield  {journal}
  {\bibinfo  {journal} {New J. Phys.}\ }\textbf {\bibinfo {volume} {11}},\
  \bibinfo {pages} {125010} (\bibinfo {year} {2009})}\BibitemShut {NoStop}%
\bibitem [{\citenamefont {Sharifzadeh}\ \emph {et~al.}(2013)\citenamefont
  {Sharifzadeh}, \citenamefont {Darancet}, \citenamefont {Kronik},\ and\
  \citenamefont {Neaton}}]{Sha13}%
  \BibitemOpen
  \bibfield  {author} {\bibinfo {author} {\bibfnamefont {S.}~\bibnamefont
  {Sharifzadeh}}, \bibinfo {author} {\bibfnamefont {P.}~\bibnamefont
  {Darancet}}, \bibinfo {author} {\bibfnamefont {L.}~\bibnamefont {Kronik}}, \
  and\ \bibinfo {author} {\bibfnamefont {J.~B.}\ \bibnamefont {Neaton}},\
  }\href {\doibase 10.1021/jz401069f} {\bibfield  {journal} {\bibinfo
  {journal} {J. Phys. Chem. Lett.}\ }\textbf {\bibinfo {volume} {4}},\ \bibinfo
  {pages} {2197} (\bibinfo {year} {2013})}\BibitemShut {NoStop}%
\bibitem [{\citenamefont {Cudazzo}\ \emph {et~al.}(2015)\citenamefont
  {Cudazzo}, \citenamefont {Sottile}, \citenamefont {Rubio},\ and\
  \citenamefont {Gatti}}]{Cud15}%
  \BibitemOpen
  \bibfield  {author} {\bibinfo {author} {\bibfnamefont {P.}~\bibnamefont
  {Cudazzo}}, \bibinfo {author} {\bibfnamefont {F.}~\bibnamefont {Sottile}},
  \bibinfo {author} {\bibfnamefont {A.}~\bibnamefont {Rubio}}, \ and\ \bibinfo
  {author} {\bibfnamefont {M.}~\bibnamefont {Gatti}},\ }\href {\doibase
  10.1088/0953-8984/27/11/113204} {\bibfield  {journal} {\bibinfo  {journal}
  {J. Phys.: Cond. Matt.}\ }\textbf {\bibinfo {volume} {27}},\ \bibinfo {pages}
  {113204} (\bibinfo {year} {2015})}\BibitemShut {NoStop}%
\bibitem [{\citenamefont {Leng}\ \emph {et~al.}(2016)\citenamefont {Leng},
  \citenamefont {Feng}, \citenamefont {Chen}, \citenamefont {Liu},\ and\
  \citenamefont {Ma}}]{Len16}%
  \BibitemOpen
  \bibfield  {author} {\bibinfo {author} {\bibfnamefont {X.}~\bibnamefont
  {Leng}}, \bibinfo {author} {\bibfnamefont {J.}~\bibnamefont {Feng}}, \bibinfo
  {author} {\bibfnamefont {T.}~\bibnamefont {Chen}}, \bibinfo {author}
  {\bibfnamefont {C.}~\bibnamefont {Liu}}, \ and\ \bibinfo {author}
  {\bibfnamefont {Y.}~\bibnamefont {Ma}},\ }\href {\doibase 10.1039/C6CP05902C}
  {\bibfield  {journal} {\bibinfo  {journal} {Phys. Chem. Chem. Phys.}\
  }\textbf {\bibinfo {volume} {18}},\ \bibinfo {pages} {30777} (\bibinfo {year}
  {2016})}\BibitemShut {NoStop}%
\bibitem [{\citenamefont {Kang}\ \emph {et~al.}(2016)\citenamefont {Kang},
  \citenamefont {Jeon}, \citenamefont {Cho},\ and\ \citenamefont
  {Han}}]{Kan16}%
  \BibitemOpen
  \bibfield  {author} {\bibinfo {author} {\bibfnamefont {Y.}~\bibnamefont
  {Kang}}, \bibinfo {author} {\bibfnamefont {S.~H.}\ \bibnamefont {Jeon}},
  \bibinfo {author} {\bibfnamefont {Y.}~\bibnamefont {Cho}}, \ and\ \bibinfo
  {author} {\bibfnamefont {S.}~\bibnamefont {Han}},\ }\href {\doibase
  10.1103/PhysRevB.93.035131} {\bibfield  {journal} {\bibinfo  {journal} {Phys.
  Rev. B}\ }\textbf {\bibinfo {volume} {93}},\ \bibinfo {pages} {035131}
  (\bibinfo {year} {2016})}\BibitemShut {NoStop}%
\bibitem [{\citenamefont {Blase}\ \emph {et~al.}(2011)\citenamefont {Blase},
  \citenamefont {Attaccalite},\ and\ \citenamefont {Olevano}}]{Bla11a}%
  \BibitemOpen
  \bibfield  {author} {\bibinfo {author} {\bibfnamefont {X.}~\bibnamefont
  {Blase}}, \bibinfo {author} {\bibfnamefont {C.}~\bibnamefont {Attaccalite}},
  \ and\ \bibinfo {author} {\bibfnamefont {V.}~\bibnamefont {Olevano}},\ }\href
  {\doibase 10.1103/PhysRevB.83.115103} {\bibfield  {journal} {\bibinfo
  {journal} {Phys. Rev. B}\ }\textbf {\bibinfo {volume} {83}},\ \bibinfo
  {pages} {115103} (\bibinfo {year} {2011})}\BibitemShut {NoStop}%
\bibitem [{\citenamefont {Baumeier}\ \emph {et~al.}(2012)\citenamefont
  {Baumeier}, \citenamefont {Andrienko},\ and\ \citenamefont
  {Rohlfing}}]{Bau12}%
  \BibitemOpen
  \bibfield  {author} {\bibinfo {author} {\bibfnamefont {B.}~\bibnamefont
  {Baumeier}}, \bibinfo {author} {\bibfnamefont {D.}~\bibnamefont {Andrienko}},
  \ and\ \bibinfo {author} {\bibfnamefont {M.}~\bibnamefont {Rohlfing}},\
  }\href {\doibase 10.1021/ct300311x} {\bibfield  {journal} {\bibinfo
  {journal} {J. Chem. Theor. Comp.}\ }\textbf {\bibinfo {volume} {8}},\
  \bibinfo {pages} {2790} (\bibinfo {year} {2012})}\BibitemShut {NoStop}%
\bibitem [{\citenamefont {Bruneval}(2012)}]{Bru12}%
  \BibitemOpen
  \bibfield  {author} {\bibinfo {author} {\bibfnamefont {F.}~\bibnamefont
  {Bruneval}},\ }\href {\doibase 10.1063/1.4718428} {\bibfield  {journal}
  {\bibinfo  {journal} {J. Chem. Phys.}\ }\textbf {\bibinfo {volume} {136}},\
  \bibinfo {pages} {194107} (\bibinfo {year} {2012})}\BibitemShut {NoStop}%
\bibitem [{\citenamefont {van Setten}\ \emph {et~al.}(2013)\citenamefont {van
  Setten}, \citenamefont {Weigend},\ and\ \citenamefont {Evers}}]{Set13}%
  \BibitemOpen
  \bibfield  {author} {\bibinfo {author} {\bibfnamefont {M.~J.}\ \bibnamefont
  {van Setten}}, \bibinfo {author} {\bibfnamefont {F.}~\bibnamefont {Weigend}},
  \ and\ \bibinfo {author} {\bibfnamefont {F.}~\bibnamefont {Evers}},\ }\href
  {\doibase 10.1021/ct300648t} {\bibfield  {journal} {\bibinfo  {journal} {J.
  Chem. Theory Comput.}\ }\textbf {\bibinfo {volume} {9}},\ \bibinfo {pages}
  {232} (\bibinfo {year} {2013})}\BibitemShut {NoStop}%
\bibitem [{\citenamefont {Kaplan}\ \emph {et~al.}(2015)\citenamefont {Kaplan},
  \citenamefont {Weigend}, \citenamefont {Evers},\ and\ \citenamefont {van
  Setten}}]{Kap15}%
  \BibitemOpen
  \bibfield  {author} {\bibinfo {author} {\bibfnamefont {F.}~\bibnamefont
  {Kaplan}}, \bibinfo {author} {\bibfnamefont {F.}~\bibnamefont {Weigend}},
  \bibinfo {author} {\bibfnamefont {F.}~\bibnamefont {Evers}}, \ and\ \bibinfo
  {author} {\bibfnamefont {M.~J.}\ \bibnamefont {van Setten}},\ }\href
  {\doibase 10.1021/acs.jctc.5b00394} {\bibfield  {journal} {\bibinfo
  {journal} {J. Chem. Theory Comput.}\ }\textbf {\bibinfo {volume} {11}},\
  \bibinfo {pages} {5152} (\bibinfo {year} {2015})}\BibitemShut {NoStop}%
\bibitem [{\citenamefont {Tiago}\ \emph {et~al.}(2008)\citenamefont {Tiago},
  \citenamefont {Kent}, \citenamefont {Hood},\ and\ \citenamefont
  {Reboredo}}]{Tia08}%
  \BibitemOpen
  \bibfield  {author} {\bibinfo {author} {\bibfnamefont {M.~L.}\ \bibnamefont
  {Tiago}}, \bibinfo {author} {\bibfnamefont {P.~R.~C.}\ \bibnamefont {Kent}},
  \bibinfo {author} {\bibfnamefont {R.~Q.}\ \bibnamefont {Hood}}, \ and\
  \bibinfo {author} {\bibfnamefont {F.~A.}\ \bibnamefont {Reboredo}},\ }\href
  {\doibase 10.1063/1.2973627} {\bibfield  {journal} {\bibinfo  {journal} {J.
  Chem. Phys.}\ }\textbf {\bibinfo {volume} {129}},\ \bibinfo {pages} {084311}
  (\bibinfo {year} {2008})}\BibitemShut {NoStop}%
\bibitem [{\citenamefont {Palummo}\ \emph {et~al.}(2009)\citenamefont
  {Palummo}, \citenamefont {Hogan}, \citenamefont {Sottile}, \citenamefont
  {Bagalá},\ and\ \citenamefont {Rubio}}]{Pal09}%
  \BibitemOpen
  \bibfield  {author} {\bibinfo {author} {\bibfnamefont {M.}~\bibnamefont
  {Palummo}}, \bibinfo {author} {\bibfnamefont {C.}~\bibnamefont {Hogan}},
  \bibinfo {author} {\bibfnamefont {F.}~\bibnamefont {Sottile}}, \bibinfo
  {author} {\bibfnamefont {P.}~\bibnamefont {Bagalá}}, \ and\ \bibinfo
  {author} {\bibfnamefont {A.}~\bibnamefont {Rubio}},\ }\href {\doibase
  10.1063/1.3204938} {\bibfield  {journal} {\bibinfo  {journal} {J. Chem.
  Phys.}\ }\textbf {\bibinfo {volume} {131}},\ \bibinfo {pages} {084102}
  (\bibinfo {year} {2009})}\BibitemShut {NoStop}%
\bibitem [{\citenamefont {Ma}\ \emph {et~al.}(2009)\citenamefont {Ma},
  \citenamefont {Rohlfing},\ and\ \citenamefont {Molteni}}]{Ma09}%
  \BibitemOpen
  \bibfield  {author} {\bibinfo {author} {\bibfnamefont {Y.}~\bibnamefont
  {Ma}}, \bibinfo {author} {\bibfnamefont {M.}~\bibnamefont {Rohlfing}}, \ and\
  \bibinfo {author} {\bibfnamefont {C.}~\bibnamefont {Molteni}},\ }\href
  {\doibase 10.1103/PhysRevB.80.241405} {\bibfield  {journal} {\bibinfo
  {journal} {Phys. Rev. B}\ }\textbf {\bibinfo {volume} {80}},\ \bibinfo
  {pages} {241405} (\bibinfo {year} {2009})}\BibitemShut {NoStop}%
\bibitem [{\citenamefont {Foerster}\ \emph {et~al.}(2011)\citenamefont
  {Foerster}, \citenamefont {Koval},\ and\ \citenamefont
  {S\'{a}nchez-Portal}}]{Foe11}%
  \BibitemOpen
  \bibfield  {author} {\bibinfo {author} {\bibfnamefont {D.}~\bibnamefont
  {Foerster}}, \bibinfo {author} {\bibfnamefont {P.}~\bibnamefont {Koval}}, \
  and\ \bibinfo {author} {\bibfnamefont {D.}~\bibnamefont
  {S\'{a}nchez-Portal}},\ }\href {\doibase 10.1063/1.3624731} {\bibfield
  {journal} {\bibinfo  {journal} {J. Chem. Phys.}\ }\textbf {\bibinfo {volume}
  {135}},\ \bibinfo {pages} {074105} (\bibinfo {year} {2011})}\BibitemShut
  {NoStop}%
\bibitem [{\citenamefont {Marom}\ \emph {et~al.}(2012)\citenamefont {Marom},
  \citenamefont {Caruso}, \citenamefont {Ren}, \citenamefont {Hofmann},
  \citenamefont {K\"orzd\"orfer}, \citenamefont {Chelikowsky}, \citenamefont
  {Rubio}, \citenamefont {Scheffler},\ and\ \citenamefont {Rinke}}]{Mar12}%
  \BibitemOpen
  \bibfield  {author} {\bibinfo {author} {\bibfnamefont {N.}~\bibnamefont
  {Marom}}, \bibinfo {author} {\bibfnamefont {F.}~\bibnamefont {Caruso}},
  \bibinfo {author} {\bibfnamefont {X.}~\bibnamefont {Ren}}, \bibinfo {author}
  {\bibfnamefont {O.~T.}\ \bibnamefont {Hofmann}}, \bibinfo {author}
  {\bibfnamefont {T.}~\bibnamefont {K\"orzd\"orfer}}, \bibinfo {author}
  {\bibfnamefont {J.~R.}\ \bibnamefont {Chelikowsky}}, \bibinfo {author}
  {\bibfnamefont {A.}~\bibnamefont {Rubio}}, \bibinfo {author} {\bibfnamefont
  {M.}~\bibnamefont {Scheffler}}, \ and\ \bibinfo {author} {\bibfnamefont
  {P.}~\bibnamefont {Rinke}},\ }\href {\doibase 10.1103/PhysRevB.86.245127}
  {\bibfield  {journal} {\bibinfo  {journal} {Phys. Rev. B}\ }\textbf {\bibinfo
  {volume} {86}},\ \bibinfo {pages} {245127} (\bibinfo {year}
  {2012})}\BibitemShut {NoStop}%
\bibitem [{\citenamefont {Laflamme~Janssen}\ \emph {et~al.}(2015)\citenamefont
  {Laflamme~Janssen}, \citenamefont {Rousseau},\ and\ \citenamefont
  {C\^ot\'e}}]{Laf15}%
  \BibitemOpen
  \bibfield  {author} {\bibinfo {author} {\bibfnamefont {J.}~\bibnamefont
  {Laflamme~Janssen}}, \bibinfo {author} {\bibfnamefont {B.}~\bibnamefont
  {Rousseau}}, \ and\ \bibinfo {author} {\bibfnamefont {M.}~\bibnamefont
  {C\^ot\'e}},\ }\href {\doibase 10.1103/PhysRevB.91.125120} {\bibfield
  {journal} {\bibinfo  {journal} {Phys. Rev. B}\ }\textbf {\bibinfo {volume}
  {91}},\ \bibinfo {pages} {125120} (\bibinfo {year} {2015})}\BibitemShut
  {NoStop}%
\bibitem [{\citenamefont {Pham}\ \emph {et~al.}(2013)\citenamefont {Pham},
  \citenamefont {Nguyen}, \citenamefont {Rocca},\ and\ \citenamefont
  {Galli}}]{Anh13}%
  \BibitemOpen
  \bibfield  {author} {\bibinfo {author} {\bibfnamefont {T.~A.}\ \bibnamefont
  {Pham}}, \bibinfo {author} {\bibfnamefont {H.-V.}\ \bibnamefont {Nguyen}},
  \bibinfo {author} {\bibfnamefont {D.}~\bibnamefont {Rocca}}, \ and\ \bibinfo
  {author} {\bibfnamefont {G.}~\bibnamefont {Galli}},\ }\href {\doibase
  10.1103/PhysRevB.87.155148} {\bibfield  {journal} {\bibinfo  {journal} {Phys.
  Rev. B}\ }\textbf {\bibinfo {volume} {87}},\ \bibinfo {pages} {155148}
  (\bibinfo {year} {2013})}\BibitemShut {NoStop}%
\bibitem [{\citenamefont {van Setten}\ \emph {et~al.}(2015)\citenamefont {van
  Setten}, \citenamefont {Caruso}, \citenamefont {Sharifzadeh}, \citenamefont
  {Ren}, \citenamefont {Scheffler}, \citenamefont {Liu}, \citenamefont
  {Lischner}, \citenamefont {Lin}, \citenamefont {Deslippe}, \citenamefont
  {Louie}, \citenamefont {Yang}, \citenamefont {Weigend}, \citenamefont
  {Neaton}, \citenamefont {Evers},\ and\ \citenamefont {Rinke}}]{Set15}%
  \BibitemOpen
  \bibfield  {author} {\bibinfo {author} {\bibfnamefont {M.~J.}\ \bibnamefont
  {van Setten}}, \bibinfo {author} {\bibfnamefont {F.}~\bibnamefont {Caruso}},
  \bibinfo {author} {\bibfnamefont {S.}~\bibnamefont {Sharifzadeh}}, \bibinfo
  {author} {\bibfnamefont {X.}~\bibnamefont {Ren}}, \bibinfo {author}
  {\bibfnamefont {M.}~\bibnamefont {Scheffler}}, \bibinfo {author}
  {\bibfnamefont {F.}~\bibnamefont {Liu}}, \bibinfo {author} {\bibfnamefont
  {J.}~\bibnamefont {Lischner}}, \bibinfo {author} {\bibfnamefont
  {L.}~\bibnamefont {Lin}}, \bibinfo {author} {\bibfnamefont {J.~R.}\
  \bibnamefont {Deslippe}}, \bibinfo {author} {\bibfnamefont {S.~G.}\
  \bibnamefont {Louie}}, \bibinfo {author} {\bibfnamefont {C.}~\bibnamefont
  {Yang}}, \bibinfo {author} {\bibfnamefont {F.}~\bibnamefont {Weigend}},
  \bibinfo {author} {\bibfnamefont {J.~B.}\ \bibnamefont {Neaton}}, \bibinfo
  {author} {\bibfnamefont {F.}~\bibnamefont {Evers}}, \ and\ \bibinfo {author}
  {\bibfnamefont {P.}~\bibnamefont {Rinke}},\ }\href {\doibase
  10.1021/acs.jctc.5b00453} {\bibfield  {journal} {\bibinfo  {journal} {J.
  Chem. Theory Comput.}\ }\textbf {\bibinfo {volume} {11}},\ \bibinfo {pages}
  {5665} (\bibinfo {year} {2015})}\BibitemShut {NoStop}%
\bibitem [{\citenamefont {Rangel}\ \emph {et~al.}(2017)\citenamefont {Rangel},
  \citenamefont {Hamed}, \citenamefont {Bruneval},\ and\ \citenamefont
  {Neaton}}]{Ran17}%
  \BibitemOpen
  \bibfield  {author} {\bibinfo {author} {\bibfnamefont {T.}~\bibnamefont
  {Rangel}}, \bibinfo {author} {\bibfnamefont {S.~M.}\ \bibnamefont {Hamed}},
  \bibinfo {author} {\bibfnamefont {F.}~\bibnamefont {Bruneval}}, \ and\
  \bibinfo {author} {\bibfnamefont {J.~B.}\ \bibnamefont {Neaton}},\ }\href
  {\doibase 10.1021/acs.jctc.6b00163} {\bibfield  {journal} {\bibinfo
  {journal} {J. Chem. Theory Comput.}\ }\textbf {\bibinfo {volume} {12}},\
  \bibinfo {pages} {2834} (\bibinfo {year} {2017})}\BibitemShut {NoStop}%
\bibitem [{\citenamefont {Faber}\ \emph
  {et~al.}(2011{\natexlab{a}})\citenamefont {Faber}, \citenamefont
  {Attaccalite}, \citenamefont {Olevano}, \citenamefont {Runge},\ and\
  \citenamefont {Blase}}]{Fab11a}%
  \BibitemOpen
  \bibfield  {author} {\bibinfo {author} {\bibfnamefont {C.}~\bibnamefont
  {Faber}}, \bibinfo {author} {\bibfnamefont {C.}~\bibnamefont {Attaccalite}},
  \bibinfo {author} {\bibfnamefont {V.}~\bibnamefont {Olevano}}, \bibinfo
  {author} {\bibfnamefont {E.}~\bibnamefont {Runge}}, \ and\ \bibinfo {author}
  {\bibfnamefont {X.}~\bibnamefont {Blase}},\ }\href {\doibase
  10.1103/PhysRevB.83.115123} {\bibfield  {journal} {\bibinfo  {journal} {Phys.
  Rev. B}\ }\textbf {\bibinfo {volume} {83}},\ \bibinfo {pages} {115123}
  (\bibinfo {year} {2011}{\natexlab{a}})}\BibitemShut {NoStop}%
\bibitem [{\citenamefont {Krause}\ \emph {et~al.}(2015)\citenamefont {Krause},
  \citenamefont {Harding},\ and\ \citenamefont {Klopper}}]{Klo15}%
  \BibitemOpen
  \bibfield  {author} {\bibinfo {author} {\bibfnamefont {K.}~\bibnamefont
  {Krause}}, \bibinfo {author} {\bibfnamefont {M.~E.}\ \bibnamefont {Harding}},
  \ and\ \bibinfo {author} {\bibfnamefont {W.}~\bibnamefont {Klopper}},\ }\href
  {\doibase 10.1080/00268976.2015.1025113} {\bibfield  {journal} {\bibinfo
  {journal} {Mol. Phys.}\ }\textbf {\bibinfo {volume} {113}},\ \bibinfo {pages}
  {1952} (\bibinfo {year} {2015})}\BibitemShut {NoStop}%
\bibitem [{\citenamefont {Kaplan}\ \emph {et~al.}(2016)\citenamefont {Kaplan},
  \citenamefont {Harding}, \citenamefont {Seiler}, \citenamefont {Weigend},
  \citenamefont {Evers},\ and\ \citenamefont {van Setten}}]{Kap16}%
  \BibitemOpen
  \bibfield  {author} {\bibinfo {author} {\bibfnamefont {F.}~\bibnamefont
  {Kaplan}}, \bibinfo {author} {\bibfnamefont {M.~E.}\ \bibnamefont {Harding}},
  \bibinfo {author} {\bibfnamefont {C.}~\bibnamefont {Seiler}}, \bibinfo
  {author} {\bibfnamefont {F.}~\bibnamefont {Weigend}}, \bibinfo {author}
  {\bibfnamefont {F.}~\bibnamefont {Evers}}, \ and\ \bibinfo {author}
  {\bibfnamefont {M.~J.}\ \bibnamefont {van Setten}},\ }\href {\doibase
  10.1021/acs.jctc.5b01238} {\bibfield  {journal} {\bibinfo  {journal} {J.
  Chem. Theory Comput.}\ }\textbf {\bibinfo {volume} {12}},\ \bibinfo {pages}
  {2528} (\bibinfo {year} {2016})}\BibitemShut {NoStop}%
\bibitem [{\citenamefont {Knight}\ \emph {et~al.}(2016)\citenamefont {Knight},
  \citenamefont {Wang}, \citenamefont {Gallandi}, \citenamefont
  {Dolgounitcheva}, \citenamefont {Ren}, \citenamefont {Ortiz}, \citenamefont
  {Rinke}, \citenamefont {K\"orzd\"orfer},\ and\ \citenamefont
  {Marom}}]{Kni16}%
  \BibitemOpen
  \bibfield  {author} {\bibinfo {author} {\bibfnamefont {J.~W.}\ \bibnamefont
  {Knight}}, \bibinfo {author} {\bibfnamefont {X.}~\bibnamefont {Wang}},
  \bibinfo {author} {\bibfnamefont {L.}~\bibnamefont {Gallandi}}, \bibinfo
  {author} {\bibfnamefont {O.}~\bibnamefont {Dolgounitcheva}}, \bibinfo
  {author} {\bibfnamefont {X.}~\bibnamefont {Ren}}, \bibinfo {author}
  {\bibfnamefont {J.~V.}\ \bibnamefont {Ortiz}}, \bibinfo {author}
  {\bibfnamefont {P.}~\bibnamefont {Rinke}}, \bibinfo {author} {\bibfnamefont
  {T.}~\bibnamefont {K\"orzd\"orfer}}, \ and\ \bibinfo {author} {\bibfnamefont
  {N.}~\bibnamefont {Marom}},\ }\href {\doibase 10.1021/acs.jctc.5b00871}
  {\bibfield  {journal} {\bibinfo  {journal} {J. Chem. Theory Comput.}\
  }\textbf {\bibinfo {volume} {12}},\ \bibinfo {pages} {615} (\bibinfo {year}
  {2016})}\BibitemShut {NoStop}%
\bibitem [{\citenamefont {Duchemin}\ \emph {et~al.}(2012)\citenamefont
  {Duchemin}, \citenamefont {Deutsch},\ and\ \citenamefont {Blase}}]{Duc12}%
  \BibitemOpen
  \bibfield  {author} {\bibinfo {author} {\bibfnamefont {I.}~\bibnamefont
  {Duchemin}}, \bibinfo {author} {\bibfnamefont {T.}~\bibnamefont {Deutsch}}, \
  and\ \bibinfo {author} {\bibfnamefont {X.}~\bibnamefont {Blase}},\ }\href
  {\doibase 10.1103/PhysRevLett.109.167801} {\bibfield  {journal} {\bibinfo
  {journal} {Phys. Rev. Lett.}\ }\textbf {\bibinfo {volume} {109}},\ \bibinfo
  {pages} {167801} (\bibinfo {year} {2012})}\BibitemShut {NoStop}%
\bibitem [{\citenamefont {Niedzialek}\ \emph {et~al.}(2015)\citenamefont
  {Niedzialek}, \citenamefont {Duchemin}, \citenamefont {de~Queiroz},
  \citenamefont {Osella}, \citenamefont {Rao}, \citenamefont {Friend},
  \citenamefont {Blase}, \citenamefont {K\"ummel},\ and\ \citenamefont
  {Beljonne}}]{Nie15}%
  \BibitemOpen
  \bibfield  {author} {\bibinfo {author} {\bibfnamefont {D.}~\bibnamefont
  {Niedzialek}}, \bibinfo {author} {\bibfnamefont {I.}~\bibnamefont
  {Duchemin}}, \bibinfo {author} {\bibfnamefont {T.~B.}\ \bibnamefont
  {de~Queiroz}}, \bibinfo {author} {\bibfnamefont {S.}~\bibnamefont {Osella}},
  \bibinfo {author} {\bibfnamefont {A.}~\bibnamefont {Rao}}, \bibinfo {author}
  {\bibfnamefont {R.}~\bibnamefont {Friend}}, \bibinfo {author} {\bibfnamefont
  {X.}~\bibnamefont {Blase}}, \bibinfo {author} {\bibfnamefont
  {S.}~\bibnamefont {K\"ummel}}, \ and\ \bibinfo {author} {\bibfnamefont
  {D.}~\bibnamefont {Beljonne}},\ }\href {\doibase 10.1002/adfm.201402682}
  {\bibfield  {journal} {\bibinfo  {journal} {Adv. Funct. Mater.}\ }\textbf
  {\bibinfo {volume} {25}},\ \bibinfo {pages} {1972} (\bibinfo {year}
  {2015})}\BibitemShut {NoStop}%
\bibitem [{\citenamefont {Li}\ \emph {et~al.}(2017)\citenamefont {Li},
  \citenamefont {D'Avino}, \citenamefont {Pershin}, \citenamefont {Jacquemin},
  \citenamefont {Duchemin}, \citenamefont {Beljonne},\ and\ \citenamefont
  {Blase}}]{Li17}%
  \BibitemOpen
  \bibfield  {author} {\bibinfo {author} {\bibfnamefont {J.}~\bibnamefont
  {Li}}, \bibinfo {author} {\bibfnamefont {G.}~\bibnamefont {D'Avino}},
  \bibinfo {author} {\bibfnamefont {A.}~\bibnamefont {Pershin}}, \bibinfo
  {author} {\bibfnamefont {D.}~\bibnamefont {Jacquemin}}, \bibinfo {author}
  {\bibfnamefont {I.}~\bibnamefont {Duchemin}}, \bibinfo {author}
  {\bibfnamefont {D.}~\bibnamefont {Beljonne}}, \ and\ \bibinfo {author}
  {\bibfnamefont {X.}~\bibnamefont {Blase}},\ }\href {\doibase
  10.1103/PhysRevMaterials.1.025602} {\bibfield  {journal} {\bibinfo  {journal}
  {Phys. Rev. Materials}\ }\textbf {\bibinfo {volume} {1}},\ \bibinfo {pages}
  {025602} (\bibinfo {year} {2017})}\BibitemShut {NoStop}%
\bibitem [{\citenamefont {Neuhauser}\ \emph {et~al.}(2014)\citenamefont
  {Neuhauser}, \citenamefont {Gao}, \citenamefont {Arntsen}, \citenamefont
  {Karshenas}, \citenamefont {Rabani},\ and\ \citenamefont {Baer}}]{Neu14}%
  \BibitemOpen
  \bibfield  {author} {\bibinfo {author} {\bibfnamefont {D.}~\bibnamefont
  {Neuhauser}}, \bibinfo {author} {\bibfnamefont {Y.}~\bibnamefont {Gao}},
  \bibinfo {author} {\bibfnamefont {C.}~\bibnamefont {Arntsen}}, \bibinfo
  {author} {\bibfnamefont {C.}~\bibnamefont {Karshenas}}, \bibinfo {author}
  {\bibfnamefont {E.}~\bibnamefont {Rabani}}, \ and\ \bibinfo {author}
  {\bibfnamefont {R.}~\bibnamefont {Baer}},\ }\href {\doibase
  10.1103/PhysRevLett.113.076402} {\bibfield  {journal} {\bibinfo  {journal}
  {Phys. Rev. Lett.}\ }\textbf {\bibinfo {volume} {113}},\ \bibinfo {pages}
  {076402} (\bibinfo {year} {2014})}\BibitemShut {NoStop}%
\bibitem [{\citenamefont {Govoni}\ and\ \citenamefont {Galli}(2015)}]{Gov15}%
  \BibitemOpen
  \bibfield  {author} {\bibinfo {author} {\bibfnamefont {M.}~\bibnamefont
  {Govoni}}\ and\ \bibinfo {author} {\bibfnamefont {G.}~\bibnamefont {Galli}},\
  }\href {\doibase 10.1021/ct500958p} {\bibfield  {journal} {\bibinfo
  {journal} {J. Chem. Theory Comput.}\ }\textbf {\bibinfo {volume} {11}},\
  \bibinfo {pages} {2680} (\bibinfo {year} {2015})}\BibitemShut {NoStop}%
\bibitem [{\citenamefont {Baumeier}\ \emph {et~al.}(2014)\citenamefont
  {Baumeier}, \citenamefont {Rohlfing},\ and\ \citenamefont
  {Andrienko}}]{Bau14}%
  \BibitemOpen
  \bibfield  {author} {\bibinfo {author} {\bibfnamefont {B.}~\bibnamefont
  {Baumeier}}, \bibinfo {author} {\bibfnamefont {M.}~\bibnamefont {Rohlfing}},
  \ and\ \bibinfo {author} {\bibfnamefont {D.}~\bibnamefont {Andrienko}},\
  }\href {\doibase 10.1021/ct500479f} {\bibfield  {journal} {\bibinfo
  {journal} {J. Chem. Theory Comput.}\ }\textbf {\bibinfo {volume} {10}},\
  \bibinfo {pages} {3104} (\bibinfo {year} {2014})}\BibitemShut {NoStop}%
\bibitem [{\citenamefont {Duchemin}\ \emph {et~al.}(2016)\citenamefont
  {Duchemin}, \citenamefont {Jacquemin},\ and\ \citenamefont {Blase}}]{Duc16}%
  \BibitemOpen
  \bibfield  {author} {\bibinfo {author} {\bibfnamefont {I.}~\bibnamefont
  {Duchemin}}, \bibinfo {author} {\bibfnamefont {D.}~\bibnamefont {Jacquemin}},
  \ and\ \bibinfo {author} {\bibfnamefont {X.}~\bibnamefont {Blase}},\ }\href
  {\doibase 10.1063/1.4946778} {\bibfield  {journal} {\bibinfo  {journal} {J.
  Chem. Phys.}\ }\textbf {\bibinfo {volume} {144}},\ \bibinfo {pages} {164106}
  (\bibinfo {year} {2016})}\BibitemShut {NoStop}%
\bibitem [{\citenamefont {Tsiper}\ and\ \citenamefont {Soos}(2001)}]{Tsi01}%
  \BibitemOpen
  \bibfield  {author} {\bibinfo {author} {\bibfnamefont {E.~V.}\ \bibnamefont
  {Tsiper}}\ and\ \bibinfo {author} {\bibfnamefont {Z.~G.}\ \bibnamefont
  {Soos}},\ }\href {\doibase 10.1103/PhysRevB.64.195124} {\bibfield  {journal}
  {\bibinfo  {journal} {Phys. Rev. B}\ }\textbf {\bibinfo {volume} {64}},\
  \bibinfo {pages} {195124} (\bibinfo {year} {2001})}\BibitemShut {NoStop}%
\bibitem [{\citenamefont {Soos}\ \emph {et~al.}(2001)\citenamefont {Soos},
  \citenamefont {Tsiper},\ and\ \citenamefont {Pascal}}]{Tsi01b}%
  \BibitemOpen
  \bibfield  {author} {\bibinfo {author} {\bibfnamefont {Z.}~\bibnamefont
  {Soos}}, \bibinfo {author} {\bibfnamefont {E.~V.}\ \bibnamefont {Tsiper}}, \
  and\ \bibinfo {author} {\bibfnamefont {R.}~\bibnamefont {Pascal}},\ }\href
  {\doibase 10.1016/S0009-2614(01)00661-3} {\bibfield  {journal} {\bibinfo
  {journal} {Chem. Phys. Lett.}\ }\textbf {\bibinfo {volume} {342}},\ \bibinfo
  {pages} {652 } (\bibinfo {year} {2001})}\BibitemShut {NoStop}%
\bibitem [{\citenamefont {D'Avino}\ \emph {et~al.}(2014)\citenamefont
  {D'Avino}, \citenamefont {Muccioli}, \citenamefont {Zannoni}, \citenamefont
  {Beljonne},\ and\ \citenamefont {Soos}}]{Dav14}%
  \BibitemOpen
  \bibfield  {author} {\bibinfo {author} {\bibfnamefont {G.}~\bibnamefont
  {D'Avino}}, \bibinfo {author} {\bibfnamefont {L.}~\bibnamefont {Muccioli}},
  \bibinfo {author} {\bibfnamefont {C.}~\bibnamefont {Zannoni}}, \bibinfo
  {author} {\bibfnamefont {D.}~\bibnamefont {Beljonne}}, \ and\ \bibinfo
  {author} {\bibfnamefont {Z.~G.}\ \bibnamefont {Soos}},\ }\href {\doibase
  10.1021/ct500618w} {\bibfield  {journal} {\bibinfo  {journal} {J. Chem.
  Theory Comput.}\ }\textbf {\bibinfo {volume} {10}},\ \bibinfo {pages} {4959}
  (\bibinfo {year} {2014})}\BibitemShut {NoStop}%
\bibitem [{\citenamefont {D'Avino}\ \emph
  {et~al.}(2016{\natexlab{b}})\citenamefont {D'Avino}, \citenamefont {Vanzo},\
  and\ \citenamefont {Soos}}]{Dav16c}%
  \BibitemOpen
  \bibfield  {author} {\bibinfo {author} {\bibfnamefont {G.}~\bibnamefont
  {D'Avino}}, \bibinfo {author} {\bibfnamefont {D.}~\bibnamefont {Vanzo}}, \
  and\ \bibinfo {author} {\bibfnamefont {Z.~G.}\ \bibnamefont {Soos}},\ }\href
  {\doibase http://dx.doi.org/10.1063/1.4939840} {\bibfield  {journal}
  {\bibinfo  {journal} {J. Chem. Phys.}\ }\textbf {\bibinfo {volume} {144}},\
  \bibinfo {pages} {034702} (\bibinfo {year} {2016}{\natexlab{b}})}\BibitemShut
  {NoStop}%
\bibitem [{\citenamefont {Bounds}\ and\ \citenamefont {Munn}(1979)}]{Bou79}%
  \BibitemOpen
  \bibfield  {author} {\bibinfo {author} {\bibfnamefont {P.~J.}\ \bibnamefont
  {Bounds}}\ and\ \bibinfo {author} {\bibfnamefont {R.~W.}\ \bibnamefont
  {Munn}},\ }\href {\doibase http://dx.doi.org/10.1016/0301-0104(79)80067-1}
  {\bibfield  {journal} {\bibinfo  {journal} {Chem. Phys.}\ }\textbf {\bibinfo
  {volume} {44}},\ \bibinfo {pages} {103 } (\bibinfo {year}
  {1979})}\BibitemShut {NoStop}%
\bibitem [{\citenamefont {Bounds}\ and\ \citenamefont {Munn}(1981)}]{Bou81}%
  \BibitemOpen
  \bibfield  {author} {\bibinfo {author} {\bibfnamefont {P.~J.}\ \bibnamefont
  {Bounds}}\ and\ \bibinfo {author} {\bibfnamefont {R.~W.}\ \bibnamefont
  {Munn}},\ }\href {\doibase http://dx.doi.org/10.1016/0301-0104(81)80083-3}
  {\bibfield  {journal} {\bibinfo  {journal} {Chem. Phys.}\ }\textbf {\bibinfo
  {volume} {59}},\ \bibinfo {pages} {41 } (\bibinfo {year} {1981})}\BibitemShut
  {NoStop}%
\bibitem [{\citenamefont {Tsiper}\ \emph {et~al.}(2002)\citenamefont {Tsiper},
  \citenamefont {Soos}, \citenamefont {Gao},\ and\ \citenamefont
  {Kahn}}]{Tsi02}%
  \BibitemOpen
  \bibfield  {author} {\bibinfo {author} {\bibfnamefont {E.}~\bibnamefont
  {Tsiper}}, \bibinfo {author} {\bibfnamefont {Z.}~\bibnamefont {Soos}},
  \bibinfo {author} {\bibfnamefont {W.}~\bibnamefont {Gao}}, \ and\ \bibinfo
  {author} {\bibfnamefont {A.}~\bibnamefont {Kahn}},\ }\href {\doibase
  http://dx.doi.org/10.1016/S0009-2614(02)00774-1} {\bibfield  {journal}
  {\bibinfo  {journal} {Chem. Phys. Lett.}\ }\textbf {\bibinfo {volume}
  {360}},\ \bibinfo {pages} {47 } (\bibinfo {year} {2002})}\BibitemShut
  {NoStop}%
\bibitem [{\citenamefont {Topham}\ and\ \citenamefont {Soos}(2011)}]{Top11}%
  \BibitemOpen
  \bibfield  {author} {\bibinfo {author} {\bibfnamefont {B.~J.}\ \bibnamefont
  {Topham}}\ and\ \bibinfo {author} {\bibfnamefont {Z.~G.}\ \bibnamefont
  {Soos}},\ }\href {\doibase 10.1103/PhysRevB.84.165405} {\bibfield  {journal}
  {\bibinfo  {journal} {Phys. Rev. B}\ }\textbf {\bibinfo {volume} {84}},\
  \bibinfo {pages} {165405} (\bibinfo {year} {2011})}\BibitemShut {NoStop}%
\bibitem [{\citenamefont {D'Avino}\ \emph {et~al.}(2013)\citenamefont
  {D'Avino}, \citenamefont {Mothy}, \citenamefont {Muccioli}, \citenamefont
  {Zannoni}, \citenamefont {Wang}, \citenamefont {Cornil}, \citenamefont
  {Beljonne},\ and\ \citenamefont {Castet}}]{Dav13}%
  \BibitemOpen
  \bibfield  {author} {\bibinfo {author} {\bibfnamefont {G.}~\bibnamefont
  {D'Avino}}, \bibinfo {author} {\bibfnamefont {S.}~\bibnamefont {Mothy}},
  \bibinfo {author} {\bibfnamefont {L.}~\bibnamefont {Muccioli}}, \bibinfo
  {author} {\bibfnamefont {C.}~\bibnamefont {Zannoni}}, \bibinfo {author}
  {\bibfnamefont {L.}~\bibnamefont {Wang}}, \bibinfo {author} {\bibfnamefont
  {J.}~\bibnamefont {Cornil}}, \bibinfo {author} {\bibfnamefont
  {D.}~\bibnamefont {Beljonne}}, \ and\ \bibinfo {author} {\bibfnamefont
  {F.}~\bibnamefont {Castet}},\ }\href {\doibase 10.1021/jp402957g} {\bibfield
  {journal} {\bibinfo  {journal} {J. Phys. Chem. C}\ }\textbf {\bibinfo
  {volume} {117}},\ \bibinfo {pages} {12981} (\bibinfo {year}
  {2013})}\BibitemShut {NoStop}%
\bibitem [{\citenamefont {Poelking}\ \emph {et~al.}(2015)\citenamefont
  {Poelking}, \citenamefont {Tietze}, \citenamefont {Elschner}, \citenamefont
  {Olthof}, \citenamefont {Hertel}, \citenamefont {Baumeier}, \citenamefont
  {W{\"u}rthner}, \citenamefont {Meerholz}, \citenamefont {Leo},\ and\
  \citenamefont {Andrienko}}]{Poe15_nmat}%
  \BibitemOpen
  \bibfield  {author} {\bibinfo {author} {\bibfnamefont {C.}~\bibnamefont
  {Poelking}}, \bibinfo {author} {\bibfnamefont {M.}~\bibnamefont {Tietze}},
  \bibinfo {author} {\bibfnamefont {C.}~\bibnamefont {Elschner}}, \bibinfo
  {author} {\bibfnamefont {S.}~\bibnamefont {Olthof}}, \bibinfo {author}
  {\bibfnamefont {D.}~\bibnamefont {Hertel}}, \bibinfo {author} {\bibfnamefont
  {B.}~\bibnamefont {Baumeier}}, \bibinfo {author} {\bibfnamefont
  {F.}~\bibnamefont {W{\"u}rthner}}, \bibinfo {author} {\bibfnamefont
  {K.}~\bibnamefont {Meerholz}}, \bibinfo {author} {\bibfnamefont
  {K.}~\bibnamefont {Leo}}, \ and\ \bibinfo {author} {\bibfnamefont
  {D.}~\bibnamefont {Andrienko}},\ }\href {http://dx.doi.org/10.1038/nmat4167}
  {\bibfield  {journal} {\bibinfo  {journal} {Nat. Mater.}\ }\textbf {\bibinfo
  {volume} {14}},\ \bibinfo {pages} {434} (\bibinfo {year} {2015})}\BibitemShut
  {NoStop}%
\bibitem [{\citenamefont {Schwarze}\ \emph {et~al.}(2016)\citenamefont
  {Schwarze}, \citenamefont {Tress}, \citenamefont {Beyer}, \citenamefont
  {Gao}, \citenamefont {Scholz}, \citenamefont {Poelking}, \citenamefont
  {Ortstein}, \citenamefont {Gunther}, \citenamefont {Kasemann}, \citenamefont
  {Andrienko},\ and\ \citenamefont {Leo}}]{Sch16}%
  \BibitemOpen
  \bibfield  {author} {\bibinfo {author} {\bibfnamefont {M.}~\bibnamefont
  {Schwarze}}, \bibinfo {author} {\bibfnamefont {W.}~\bibnamefont {Tress}},
  \bibinfo {author} {\bibfnamefont {B.}~\bibnamefont {Beyer}}, \bibinfo
  {author} {\bibfnamefont {F.}~\bibnamefont {Gao}}, \bibinfo {author}
  {\bibfnamefont {R.}~\bibnamefont {Scholz}}, \bibinfo {author} {\bibfnamefont
  {C.}~\bibnamefont {Poelking}}, \bibinfo {author} {\bibfnamefont
  {K.}~\bibnamefont {Ortstein}}, \bibinfo {author} {\bibfnamefont {A.~A.}\
  \bibnamefont {Gunther}}, \bibinfo {author} {\bibfnamefont {D.}~\bibnamefont
  {Kasemann}}, \bibinfo {author} {\bibfnamefont {D.}~\bibnamefont {Andrienko}},
  \ and\ \bibinfo {author} {\bibfnamefont {K.}~\bibnamefont {Leo}},\ }\href
  {\doibase 10.1126/science.aaf0590} {\bibfield  {journal} {\bibinfo  {journal}
  {Science}\ }\textbf {\bibinfo {volume} {352}},\ \bibinfo {pages} {1446}
  (\bibinfo {year} {2016})}\BibitemShut {NoStop}%
\bibitem [{\citenamefont {Li}\ \emph {et~al.}(2016)\citenamefont {Li},
  \citenamefont {D'Avino}, \citenamefont {Duchemin}, \citenamefont {Beljonne},\
  and\ \citenamefont {Blase}}]{Li16}%
  \BibitemOpen
  \bibfield  {author} {\bibinfo {author} {\bibfnamefont {J.}~\bibnamefont
  {Li}}, \bibinfo {author} {\bibfnamefont {G.}~\bibnamefont {D'Avino}},
  \bibinfo {author} {\bibfnamefont {I.}~\bibnamefont {Duchemin}}, \bibinfo
  {author} {\bibfnamefont {D.}~\bibnamefont {Beljonne}}, \ and\ \bibinfo
  {author} {\bibfnamefont {X.}~\bibnamefont {Blase}},\ }\href {\doibase
  10.1021/acs.jpclett.6b01302} {\bibfield  {journal} {\bibinfo  {journal} {J.
  Phys. Chem. Lett.}\ }\textbf {\bibinfo {volume} {7}},\ \bibinfo {pages}
  {2814} (\bibinfo {year} {2016})}\BibitemShut {NoStop}%
\bibitem [{\citenamefont {Yoshida}\ \emph {et~al.}(2015)\citenamefont
  {Yoshida}, \citenamefont {Yamada}, \citenamefont {Tsutsumi},\ and\
  \citenamefont {Sato}}]{Yos15}%
  \BibitemOpen
  \bibfield  {author} {\bibinfo {author} {\bibfnamefont {H.}~\bibnamefont
  {Yoshida}}, \bibinfo {author} {\bibfnamefont {K.}~\bibnamefont {Yamada}},
  \bibinfo {author} {\bibfnamefont {J.}~\bibnamefont {Tsutsumi}}, \ and\
  \bibinfo {author} {\bibfnamefont {N.}~\bibnamefont {Sato}},\ }\href {\doibase
  10.1103/PhysRevB.92.075145} {\bibfield  {journal} {\bibinfo  {journal} {Phys.
  Rev. B}\ }\textbf {\bibinfo {volume} {92}},\ \bibinfo {pages} {075145}
  (\bibinfo {year} {2015})}\BibitemShut {NoStop}%
\bibitem [{\citenamefont {{Mosca Conte}}\ \emph {et~al.}(2009)\citenamefont
  {{Mosca Conte}}, \citenamefont {Ippoliti}, \citenamefont {Del~Sole},
  \citenamefont {Carloni},\ and\ \citenamefont {Pulci}}]{Mos09}%
  \BibitemOpen
  \bibfield  {author} {\bibinfo {author} {\bibfnamefont {M.}~\bibnamefont
  {{Mosca Conte}}}, \bibinfo {author} {\bibfnamefont {E.}~\bibnamefont
  {Ippoliti}}, \bibinfo {author} {\bibfnamefont {R.}~\bibnamefont {Del~Sole}},
  \bibinfo {author} {\bibfnamefont {P.}~\bibnamefont {Carloni}}, \ and\
  \bibinfo {author} {\bibfnamefont {O.}~\bibnamefont {Pulci}},\ }\href
  {\doibase 10.1021/ct800528e} {\bibfield  {journal} {\bibinfo  {journal} {J.
  Chem. Theory Comput.}\ }\textbf {\bibinfo {volume} {5}},\ \bibinfo {pages}
  {1822} (\bibinfo {year} {2009})}\BibitemShut {NoStop}%
\bibitem [{\citenamefont {Bagheri}\ \emph {et~al.}(2016)\citenamefont
  {Bagheri}, \citenamefont {Baumeier},\ and\ \citenamefont
  {Karttunen}}]{Bag16}%
  \BibitemOpen
  \bibfield  {author} {\bibinfo {author} {\bibfnamefont {B.}~\bibnamefont
  {Bagheri}}, \bibinfo {author} {\bibfnamefont {B.}~\bibnamefont {Baumeier}}, \
  and\ \bibinfo {author} {\bibfnamefont {M.}~\bibnamefont {Karttunen}},\ }\href
  {\doibase 10.1039/C6CP02944B} {\bibfield  {journal} {\bibinfo  {journal}
  {Phys. Chem. Chem. Phys.}\ }\textbf {\bibinfo {volume} {18}},\ \bibinfo
  {pages} {30297} (\bibinfo {year} {2016})}\BibitemShut {NoStop}%
\bibitem [{\citenamefont {Varsano}\ \emph {et~al.}(2017)\citenamefont
  {Varsano}, \citenamefont {Caprasecca},\ and\ \citenamefont {Coccia}}]{Var17}%
  \BibitemOpen
  \bibfield  {author} {\bibinfo {author} {\bibfnamefont {D.}~\bibnamefont
  {Varsano}}, \bibinfo {author} {\bibfnamefont {S.}~\bibnamefont {Caprasecca}},
  \ and\ \bibinfo {author} {\bibfnamefont {E.}~\bibnamefont {Coccia}},\ }\href
  {\doibase 10.1088/0953-8984/29/1/013002} {\bibfield  {journal} {\bibinfo
  {journal} {Journal of Physics: Condensed Matter}\ }\textbf {\bibinfo {volume}
  {29}},\ \bibinfo {pages} {013002} (\bibinfo {year} {2017})}\BibitemShut
  {NoStop}%
\bibitem [{\citenamefont {Morita}\ and\ \citenamefont {Kato}(1997)}]{Mor97}%
  \BibitemOpen
  \bibfield  {author} {\bibinfo {author} {\bibfnamefont {A.}~\bibnamefont
  {Morita}}\ and\ \bibinfo {author} {\bibfnamefont {S.}~\bibnamefont {Kato}},\
  }\href {\doibase 10.1021/ja9635342} {\bibfield  {journal} {\bibinfo
  {journal} {J. Am. Chem. Soc.}\ }\textbf {\bibinfo {volume} {119}},\ \bibinfo
  {pages} {4021} (\bibinfo {year} {1997})}\BibitemShut {NoStop}%
\bibitem [{\citenamefont {Tsutsumi}\ \emph {et~al.}(2009)\citenamefont
  {Tsutsumi}, \citenamefont {Yoshida}, \citenamefont {Murdey}, \citenamefont
  {Kato},\ and\ \citenamefont {Sato}}]{Tsu09}%
  \BibitemOpen
  \bibfield  {author} {\bibinfo {author} {\bibfnamefont {J.}~\bibnamefont
  {Tsutsumi}}, \bibinfo {author} {\bibfnamefont {H.}~\bibnamefont {Yoshida}},
  \bibinfo {author} {\bibfnamefont {R.}~\bibnamefont {Murdey}}, \bibinfo
  {author} {\bibfnamefont {S.}~\bibnamefont {Kato}}, \ and\ \bibinfo {author}
  {\bibfnamefont {N.}~\bibnamefont {Sato}},\ }\href {\doibase
  10.1021/jp903420w} {\bibfield  {journal} {\bibinfo  {journal} {J. Phys. Chem.
  A}\ }\textbf {\bibinfo {volume} {113}},\ \bibinfo {pages} {9207} (\bibinfo
  {year} {2009})}\BibitemShut {NoStop}%
\bibitem [{\citenamefont {Besler}\ \emph {et~al.}(1990)\citenamefont {Besler},
  \citenamefont {Merz},\ and\ \citenamefont {Kollman}}]{esp}%
  \BibitemOpen
  \bibfield  {author} {\bibinfo {author} {\bibfnamefont {B.~H.}\ \bibnamefont
  {Besler}}, \bibinfo {author} {\bibfnamefont {K.~M.}\ \bibnamefont {Merz}}, \
  and\ \bibinfo {author} {\bibfnamefont {P.~A.}\ \bibnamefont {Kollman}},\
  }\href {\doibase 10.1002/jcc.540110404} {\bibfield  {journal} {\bibinfo
  {journal} {J. Comput. Chem.}\ }\textbf {\bibinfo {volume} {11}},\ \bibinfo
  {pages} {431} (\bibinfo {year} {1990})}\BibitemShut {NoStop}%
\bibitem [{\citenamefont {Poelking}\ and\ \citenamefont
  {Andrienko}(2016)}]{Poe16}%
  \BibitemOpen
  \bibfield  {author} {\bibinfo {author} {\bibfnamefont {C.}~\bibnamefont
  {Poelking}}\ and\ \bibinfo {author} {\bibfnamefont {D.}~\bibnamefont
  {Andrienko}},\ }\href {\doibase 10.1021/acs.jctc.6b00599} {\bibfield
  {journal} {\bibinfo  {journal} {J. Chem. Theory Comput.}\ }\textbf {\bibinfo
  {volume} {12}},\ \bibinfo {pages} {4516} (\bibinfo {year}
  {2016})}\BibitemShut {NoStop}%
\bibitem [{\citenamefont {Martin}\ and\ \citenamefont
  {Schwinger}(1959)}]{Mar59}%
  \BibitemOpen
  \bibfield  {author} {\bibinfo {author} {\bibfnamefont {P.~C.}\ \bibnamefont
  {Martin}}\ and\ \bibinfo {author} {\bibfnamefont {J.}~\bibnamefont
  {Schwinger}},\ }\href {\doibase 10.1103/PhysRev.115.1342} {\bibfield
  {journal} {\bibinfo  {journal} {Phys. Rev.}\ }\textbf {\bibinfo {volume}
  {115}},\ \bibinfo {pages} {1342} (\bibinfo {year} {1959})}\BibitemShut
  {NoStop}%
\bibitem [{\citenamefont {Strinati}\ \emph {et~al.}(1982)\citenamefont
  {Strinati}, \citenamefont {Mattausch},\ and\ \citenamefont {Hanke}}]{Str82}%
  \BibitemOpen
  \bibfield  {author} {\bibinfo {author} {\bibfnamefont {G.}~\bibnamefont
  {Strinati}}, \bibinfo {author} {\bibfnamefont {H.}~\bibnamefont {Mattausch}},
  \ and\ \bibinfo {author} {\bibfnamefont {W.}~\bibnamefont {Hanke}},\
  }\href@noop {} {\bibfield  {journal} {\bibinfo  {journal} {Phys. Rev. B}\
  }\textbf {\bibinfo {volume} {25}},\ \bibinfo {pages} {2867} (\bibinfo {year}
  {1982})}\BibitemShut {NoStop}%
\bibitem [{\citenamefont {Aryasetiawan}\ and\ \citenamefont
  {Gunnarsson}(1998)}]{Ary98}%
  \BibitemOpen
  \bibfield  {author} {\bibinfo {author} {\bibfnamefont {F.}~\bibnamefont
  {Aryasetiawan}}\ and\ \bibinfo {author} {\bibfnamefont {O.}~\bibnamefont
  {Gunnarsson}},\ }\href@noop {} {\bibfield  {journal} {\bibinfo  {journal}
  {Rep. Prog. Phys.}\ }\textbf {\bibinfo {volume} {61}},\ \bibinfo {pages}
  {237} (\bibinfo {year} {1998})}\BibitemShut {NoStop}%
\bibitem [{\citenamefont {Farid}(1999)}]{Far99}%
  \BibitemOpen
  \bibfield  {author} {\bibinfo {author} {\bibfnamefont {B.}~\bibnamefont
  {Farid}},\ }\href@noop {} {\emph {\bibinfo {title} {Ground and low-lying
  excited states of interacting electron systems; a survey and some critical
  analyses}}}\ (\bibinfo  {publisher} {N. H. March, editor, Electron
  Correlation in the Solid State, Imperial College Press},\ \bibinfo {year}
  {1999})\BibitemShut {NoStop}%
\bibitem [{\citenamefont {K\"orzd\"orfer}\ and\ \citenamefont
  {Marom}(2012)}]{Kor12}%
  \BibitemOpen
  \bibfield  {author} {\bibinfo {author} {\bibfnamefont {T.}~\bibnamefont
  {K\"orzd\"orfer}}\ and\ \bibinfo {author} {\bibfnamefont {N.}~\bibnamefont
  {Marom}},\ }\href {\doibase 10.1103/PhysRevB.86.041110} {\bibfield  {journal}
  {\bibinfo  {journal} {Phys. Rev. B}\ }\textbf {\bibinfo {volume} {86}},\
  \bibinfo {pages} {041110} (\bibinfo {year} {2012})}\BibitemShut {NoStop}%
\bibitem [{\citenamefont {Bruneval}\ and\ \citenamefont
  {Marques}(2013)}]{Bru13}%
  \BibitemOpen
  \bibfield  {author} {\bibinfo {author} {\bibfnamefont {F.}~\bibnamefont
  {Bruneval}}\ and\ \bibinfo {author} {\bibfnamefont {M.~A.~L.}\ \bibnamefont
  {Marques}},\ }\href {\doibase 10.1021/ct300835h} {\bibfield  {journal}
  {\bibinfo  {journal} {J. Chem. Theory Comput.}\ }\textbf {\bibinfo {volume}
  {9}},\ \bibinfo {pages} {324} (\bibinfo {year} {2013})}\BibitemShut {NoStop}%
\bibitem [{\citenamefont {Dressel}\ \emph {et~al.}(2008)\citenamefont
  {Dressel}, \citenamefont {Gompf}, \citenamefont {Faltermeier}, \citenamefont
  {Tripathi}, \citenamefont {Pflaum},\ and\ \citenamefont {Schubert}}]{Dre08}%
  \BibitemOpen
  \bibfield  {author} {\bibinfo {author} {\bibfnamefont {M.}~\bibnamefont
  {Dressel}}, \bibinfo {author} {\bibfnamefont {B.}~\bibnamefont {Gompf}},
  \bibinfo {author} {\bibfnamefont {D.}~\bibnamefont {Faltermeier}}, \bibinfo
  {author} {\bibfnamefont {A.}~\bibnamefont {Tripathi}}, \bibinfo {author}
  {\bibfnamefont {J.}~\bibnamefont {Pflaum}}, \ and\ \bibinfo {author}
  {\bibfnamefont {M.}~\bibnamefont {Schubert}},\ }\href {\doibase
  10.1364/OE.16.019770} {\bibfield  {journal} {\bibinfo  {journal} {Opt.
  Express}\ }\textbf {\bibinfo {volume} {16}},\ \bibinfo {pages} {19770}
  (\bibinfo {year} {2008})}\BibitemShut {NoStop}%
\bibitem [{\citenamefont {Bruneval}\ \emph {et~al.}(2006)\citenamefont
  {Bruneval}, \citenamefont {Vast},\ and\ \citenamefont {Reining}}]{Bru06}%
  \BibitemOpen
  \bibfield  {author} {\bibinfo {author} {\bibfnamefont {F.}~\bibnamefont
  {Bruneval}}, \bibinfo {author} {\bibfnamefont {N.}~\bibnamefont {Vast}}, \
  and\ \bibinfo {author} {\bibfnamefont {L.}~\bibnamefont {Reining}},\ }\href
  {\doibase 10.1103/PhysRevB.74.045102} {\bibfield  {journal} {\bibinfo
  {journal} {Phys. Rev. B}\ }\textbf {\bibinfo {volume} {74}},\ \bibinfo
  {pages} {045102} (\bibinfo {year} {2006})}\BibitemShut {NoStop}%
\bibitem [{Note1()}]{Note1}%
  \BibitemOpen
  \bibinfo {note} {\protect \leavevmode {\protect \color {red} We recall that
  auxiliary functions $\protect \{\beta \protect \}$ describe charge densities
  corresponding to an electrical monopoles ($l=0$), dipoles ($l=1$),
  quadrupoles ($l=2$), etc. The $R^{-(1+l+l')}$ dependence of reaction field
  matrix elements follows from classical electrostatics under the hypothesis
  that $R$ largely exceeds the size of the QM region}}\BibitemShut {NoStop}%
\bibitem [{\citenamefont {Faber}\ \emph
  {et~al.}(2011{\natexlab{b}})\citenamefont {Faber}, \citenamefont {Janssen},
  \citenamefont {C\^ot\'e}, \citenamefont {Runge},\ and\ \citenamefont
  {Blase}}]{Fab11b}%
  \BibitemOpen
  \bibfield  {author} {\bibinfo {author} {\bibfnamefont {C.}~\bibnamefont
  {Faber}}, \bibinfo {author} {\bibfnamefont {J.~L.}\ \bibnamefont {Janssen}},
  \bibinfo {author} {\bibfnamefont {M.}~\bibnamefont {C\^ot\'e}}, \bibinfo
  {author} {\bibfnamefont {E.}~\bibnamefont {Runge}}, \ and\ \bibinfo {author}
  {\bibfnamefont {X.}~\bibnamefont {Blase}},\ }\href {\doibase
  10.1103/PhysRevB.84.155104} {\bibfield  {journal} {\bibinfo  {journal} {Phys.
  Rev. B}\ }\textbf {\bibinfo {volume} {84}},\ \bibinfo {pages} {155104}
  (\bibinfo {year} {2011}{\natexlab{b}})}\BibitemShut {NoStop}%
\bibitem [{\citenamefont {Blase}\ and\ \citenamefont
  {Attaccalite}(2011)}]{Bla11b}%
  \BibitemOpen
  \bibfield  {author} {\bibinfo {author} {\bibfnamefont {X.}~\bibnamefont
  {Blase}}\ and\ \bibinfo {author} {\bibfnamefont {C.}~\bibnamefont
  {Attaccalite}},\ }\href {\doibase 10.1063/1.3655352} {\bibfield  {journal}
  {\bibinfo  {journal} {Appl. Phys. Lett.}\ }\textbf {\bibinfo {volume} {99}},\
  \bibinfo {pages} {171909} (\bibinfo {year} {2011})}\BibitemShut {NoStop}%
\bibitem [{\citenamefont {Jacquemin}\ \emph
  {et~al.}(2015{\natexlab{a}})\citenamefont {Jacquemin}, \citenamefont
  {Duchemin},\ and\ \citenamefont {Blase}}]{Jac15a}%
  \BibitemOpen
  \bibfield  {author} {\bibinfo {author} {\bibfnamefont {D.}~\bibnamefont
  {Jacquemin}}, \bibinfo {author} {\bibfnamefont {I.}~\bibnamefont {Duchemin}},
  \ and\ \bibinfo {author} {\bibfnamefont {X.}~\bibnamefont {Blase}},\ }\href
  {\doibase 10.1021/acs.jctc.5b00304} {\bibfield  {journal} {\bibinfo
  {journal} {J. Chem. Theory Comput.}\ }\textbf {\bibinfo {volume} {11}},\
  \bibinfo {pages} {3290} (\bibinfo {year} {2015}{\natexlab{a}})}\BibitemShut
  {NoStop}%
\bibitem [{\citenamefont {Jacquemin}\ \emph
  {et~al.}(2015{\natexlab{b}})\citenamefont {Jacquemin}, \citenamefont
  {Duchemin},\ and\ \citenamefont {Blase}}]{Jac15b}%
  \BibitemOpen
  \bibfield  {author} {\bibinfo {author} {\bibfnamefont {D.}~\bibnamefont
  {Jacquemin}}, \bibinfo {author} {\bibfnamefont {I.}~\bibnamefont {Duchemin}},
  \ and\ \bibinfo {author} {\bibfnamefont {X.}~\bibnamefont {Blase}},\ }\href
  {\doibase 10.1021/acs.jctc.5b00619} {\bibfield  {journal} {\bibinfo
  {journal} {J. Chem. Theory Comput.}\ }\textbf {\bibinfo {volume} {11}},\
  \bibinfo {pages} {5340} (\bibinfo {year} {2015}{\natexlab{b}})}\BibitemShut
  {NoStop}%
\bibitem [{\citenamefont {Dunning}(1989)}]{Dun89}%
  \BibitemOpen
  \bibfield  {author} {\bibinfo {author} {\bibfnamefont {T.~H.}\ \bibnamefont
  {Dunning}},\ }\href {\doibase http://dx.doi.org/10.1063/1.456153} {\bibfield
  {journal} {\bibinfo  {journal} {J. Chem. Phys.}\ }\textbf {\bibinfo {volume}
  {90}},\ \bibinfo {pages} {1007} (\bibinfo {year} {1989})}\BibitemShut
  {NoStop}%
\bibitem [{\citenamefont {Weigend}(2006)}]{Wei06}%
  \BibitemOpen
  \bibfield  {author} {\bibinfo {author} {\bibfnamefont {F.}~\bibnamefont
  {Weigend}},\ }\href {\doibase 10.1039/B515623H} {\bibfield  {journal}
  {\bibinfo  {journal} {Phys. Chem. Chem. Phys.}\ }\textbf {\bibinfo {volume}
  {8}},\ \bibinfo {pages} {1057} (\bibinfo {year} {2006})}\BibitemShut
  {NoStop}%
\bibitem [{\citenamefont {Zerner}\ \emph {et~al.}(1980)\citenamefont {Zerner},
  \citenamefont {Loew}, \citenamefont {Kirchner},\ and\ \citenamefont
  {Mueller-Westerhoff}}]{zindo}%
  \BibitemOpen
  \bibfield  {author} {\bibinfo {author} {\bibfnamefont {M.~C.}\ \bibnamefont
  {Zerner}}, \bibinfo {author} {\bibfnamefont {G.~H.}\ \bibnamefont {Loew}},
  \bibinfo {author} {\bibfnamefont {R.~F.}\ \bibnamefont {Kirchner}}, \ and\
  \bibinfo {author} {\bibfnamefont {U.~T.}\ \bibnamefont
  {Mueller-Westerhoff}},\ }\href {\doibase 10.1021/ja00522a025} {\bibfield
  {journal} {\bibinfo  {journal} {J. Am. Chem. Soc.}\ }\textbf {\bibinfo
  {volume} {102}},\ \bibinfo {pages} {589} (\bibinfo {year}
  {1980})}\BibitemShut {NoStop}%
\bibitem [{\citenamefont {Frisch}\ \emph {et~al.}()\citenamefont {Frisch},
  \citenamefont {Trucks}, \citenamefont {Schlegel}, \citenamefont {Scuseria},
  \citenamefont {Robb}, \citenamefont {Cheeseman}, \citenamefont {Scalmani},
  \citenamefont {Barone}, \citenamefont {Mennucci}, \citenamefont {Petersson},
  \citenamefont {Nakatsuji}, \citenamefont {Caricato}, \citenamefont {Li},
  \citenamefont {Hratchian}, \citenamefont {Izmaylov}, \citenamefont {Bloino},
  \citenamefont {Zheng}, \citenamefont {Sonnenberg}, \citenamefont {Hada},
  \citenamefont {Ehara}, \citenamefont {Toyota}, \citenamefont {Fukuda},
  \citenamefont {Hasegawa}, \citenamefont {Ishida}, \citenamefont {Nakajima},
  \citenamefont {Honda}, \citenamefont {Kitao}, \citenamefont {Nakai},
  \citenamefont {Vreven}, \citenamefont {Montgomery}, \citenamefont {Peralta},
  \citenamefont {Ogliaro}, \citenamefont {Bearpark}, \citenamefont {Heyd},
  \citenamefont {Brothers}, \citenamefont {Kudin}, \citenamefont {Staroverov},
  \citenamefont {Kobayashi}, \citenamefont {Normand}, \citenamefont
  {Raghavachari}, \citenamefont {Rendell}, \citenamefont {Burant},
  \citenamefont {Iyengar}, \citenamefont {Tomasi}, \citenamefont {Cossi},
  \citenamefont {Rega}, \citenamefont {Millam}, \citenamefont {Klene},
  \citenamefont {Knox}, \citenamefont {Cross}, \citenamefont {Bakken},
  \citenamefont {Adamo}, \citenamefont {Jaramillo}, \citenamefont {Gomperts},
  \citenamefont {Stratmann}, \citenamefont {Yazyev}, \citenamefont {Austin},
  \citenamefont {Cammi}, \citenamefont {Pomelli}, \citenamefont {Ochterski},
  \citenamefont {Martin}, \citenamefont {Morokuma}, \citenamefont {Zakrzewski},
  \citenamefont {Voth}, \citenamefont {Salvador}, \citenamefont {Dannenberg},
  \citenamefont {Dapprich}, \citenamefont {Daniels}, \citenamefont {Farkas},
  \citenamefont {Foresman}, \citenamefont {Ortiz}, \citenamefont {Cioslowski},\
  and\ \citenamefont {Fox}}]{g09d1}%
  \BibitemOpen
  \bibfield  {author} {\bibinfo {author} {\bibfnamefont {M.~J.}\ \bibnamefont
  {Frisch}}, \bibinfo {author} {\bibfnamefont {G.~W.}\ \bibnamefont {Trucks}},
  \bibinfo {author} {\bibfnamefont {H.~B.}\ \bibnamefont {Schlegel}}, \bibinfo
  {author} {\bibfnamefont {G.~E.}\ \bibnamefont {Scuseria}}, \bibinfo {author}
  {\bibfnamefont {M.~A.}\ \bibnamefont {Robb}}, \bibinfo {author}
  {\bibfnamefont {J.~R.}\ \bibnamefont {Cheeseman}}, \bibinfo {author}
  {\bibfnamefont {G.}~\bibnamefont {Scalmani}}, \bibinfo {author}
  {\bibfnamefont {V.}~\bibnamefont {Barone}}, \bibinfo {author} {\bibfnamefont
  {B.}~\bibnamefont {Mennucci}}, \bibinfo {author} {\bibfnamefont {G.~A.}\
  \bibnamefont {Petersson}}, \bibinfo {author} {\bibfnamefont {H.}~\bibnamefont
  {Nakatsuji}}, \bibinfo {author} {\bibfnamefont {M.}~\bibnamefont {Caricato}},
  \bibinfo {author} {\bibfnamefont {X.}~\bibnamefont {Li}}, \bibinfo {author}
  {\bibfnamefont {H.~P.}\ \bibnamefont {Hratchian}}, \bibinfo {author}
  {\bibfnamefont {A.~F.}\ \bibnamefont {Izmaylov}}, \bibinfo {author}
  {\bibfnamefont {J.}~\bibnamefont {Bloino}}, \bibinfo {author} {\bibfnamefont
  {G.}~\bibnamefont {Zheng}}, \bibinfo {author} {\bibfnamefont {J.~L.}\
  \bibnamefont {Sonnenberg}}, \bibinfo {author} {\bibfnamefont
  {M.}~\bibnamefont {Hada}}, \bibinfo {author} {\bibfnamefont {M.}~\bibnamefont
  {Ehara}}, \bibinfo {author} {\bibfnamefont {K.}~\bibnamefont {Toyota}},
  \bibinfo {author} {\bibfnamefont {R.}~\bibnamefont {Fukuda}}, \bibinfo
  {author} {\bibfnamefont {J.}~\bibnamefont {Hasegawa}}, \bibinfo {author}
  {\bibfnamefont {M.}~\bibnamefont {Ishida}}, \bibinfo {author} {\bibfnamefont
  {T.}~\bibnamefont {Nakajima}}, \bibinfo {author} {\bibfnamefont
  {Y.}~\bibnamefont {Honda}}, \bibinfo {author} {\bibfnamefont
  {O.}~\bibnamefont {Kitao}}, \bibinfo {author} {\bibfnamefont
  {H.}~\bibnamefont {Nakai}}, \bibinfo {author} {\bibfnamefont
  {T.}~\bibnamefont {Vreven}}, \bibinfo {author} {\bibfnamefont {J.~A.}\
  \bibnamefont {Montgomery}, \bibfnamefont {{Jr.}}}, \bibinfo {author}
  {\bibfnamefont {J.~E.}\ \bibnamefont {Peralta}}, \bibinfo {author}
  {\bibfnamefont {F.}~\bibnamefont {Ogliaro}}, \bibinfo {author} {\bibfnamefont
  {M.}~\bibnamefont {Bearpark}}, \bibinfo {author} {\bibfnamefont {J.~J.}\
  \bibnamefont {Heyd}}, \bibinfo {author} {\bibfnamefont {E.}~\bibnamefont
  {Brothers}}, \bibinfo {author} {\bibfnamefont {K.~N.}\ \bibnamefont {Kudin}},
  \bibinfo {author} {\bibfnamefont {V.~N.}\ \bibnamefont {Staroverov}},
  \bibinfo {author} {\bibfnamefont {R.}~\bibnamefont {Kobayashi}}, \bibinfo
  {author} {\bibfnamefont {J.}~\bibnamefont {Normand}}, \bibinfo {author}
  {\bibfnamefont {K.}~\bibnamefont {Raghavachari}}, \bibinfo {author}
  {\bibfnamefont {A.}~\bibnamefont {Rendell}}, \bibinfo {author} {\bibfnamefont
  {J.~C.}\ \bibnamefont {Burant}}, \bibinfo {author} {\bibfnamefont {S.~S.}\
  \bibnamefont {Iyengar}}, \bibinfo {author} {\bibfnamefont {J.}~\bibnamefont
  {Tomasi}}, \bibinfo {author} {\bibfnamefont {M.}~\bibnamefont {Cossi}},
  \bibinfo {author} {\bibfnamefont {N.}~\bibnamefont {Rega}}, \bibinfo {author}
  {\bibfnamefont {J.~M.}\ \bibnamefont {Millam}}, \bibinfo {author}
  {\bibfnamefont {M.}~\bibnamefont {Klene}}, \bibinfo {author} {\bibfnamefont
  {J.~E.}\ \bibnamefont {Knox}}, \bibinfo {author} {\bibfnamefont {J.~B.}\
  \bibnamefont {Cross}}, \bibinfo {author} {\bibfnamefont {V.}~\bibnamefont
  {Bakken}}, \bibinfo {author} {\bibfnamefont {C.}~\bibnamefont {Adamo}},
  \bibinfo {author} {\bibfnamefont {J.}~\bibnamefont {Jaramillo}}, \bibinfo
  {author} {\bibfnamefont {R.}~\bibnamefont {Gomperts}}, \bibinfo {author}
  {\bibfnamefont {R.~E.}\ \bibnamefont {Stratmann}}, \bibinfo {author}
  {\bibfnamefont {O.}~\bibnamefont {Yazyev}}, \bibinfo {author} {\bibfnamefont
  {A.~J.}\ \bibnamefont {Austin}}, \bibinfo {author} {\bibfnamefont
  {R.}~\bibnamefont {Cammi}}, \bibinfo {author} {\bibfnamefont
  {C.}~\bibnamefont {Pomelli}}, \bibinfo {author} {\bibfnamefont {J.~W.}\
  \bibnamefont {Ochterski}}, \bibinfo {author} {\bibfnamefont {R.~L.}\
  \bibnamefont {Martin}}, \bibinfo {author} {\bibfnamefont {K.}~\bibnamefont
  {Morokuma}}, \bibinfo {author} {\bibfnamefont {V.~G.}\ \bibnamefont
  {Zakrzewski}}, \bibinfo {author} {\bibfnamefont {G.~A.}\ \bibnamefont
  {Voth}}, \bibinfo {author} {\bibfnamefont {P.}~\bibnamefont {Salvador}},
  \bibinfo {author} {\bibfnamefont {J.~J.}\ \bibnamefont {Dannenberg}},
  \bibinfo {author} {\bibfnamefont {S.}~\bibnamefont {Dapprich}}, \bibinfo
  {author} {\bibfnamefont {A.~D.}\ \bibnamefont {Daniels}}, \bibinfo {author}
  {\bibfnamefont {O.}~\bibnamefont {Farkas}}, \bibinfo {author} {\bibfnamefont
  {J.~B.}\ \bibnamefont {Foresman}}, \bibinfo {author} {\bibfnamefont {J.~V.}\
  \bibnamefont {Ortiz}}, \bibinfo {author} {\bibfnamefont {J.}~\bibnamefont
  {Cioslowski}}, \ and\ \bibinfo {author} {\bibfnamefont {D.~J.}\ \bibnamefont
  {Fox}},\ }\href@noop {} {\enquote {\bibinfo {title} {Gaussian09 revision
  d.01},}\ }\bibinfo {note} {Gaussian Inc. Wallingford CT 2009}\BibitemShut
  {NoStop}%
\bibitem [{\citenamefont {Siegrist}\ \emph {et~al.}(2001)\citenamefont
  {Siegrist}, \citenamefont {Kloc}, \citenamefont {Sch\"{o}n}, \citenamefont
  {Batlogg}, \citenamefont {Haddon}, \citenamefont {Berg},\ and\ \citenamefont
  {Thomas}}]{Sie01}%
  \BibitemOpen
  \bibfield  {author} {\bibinfo {author} {\bibfnamefont {T.}~\bibnamefont
  {Siegrist}}, \bibinfo {author} {\bibfnamefont {C.}~\bibnamefont {Kloc}},
  \bibinfo {author} {\bibfnamefont {J.~H.}\ \bibnamefont {Sch\"{o}n}}, \bibinfo
  {author} {\bibfnamefont {B.}~\bibnamefont {Batlogg}}, \bibinfo {author}
  {\bibfnamefont {R.~C.}\ \bibnamefont {Haddon}}, \bibinfo {author}
  {\bibfnamefont {S.}~\bibnamefont {Berg}}, \ and\ \bibinfo {author}
  {\bibfnamefont {G.~A.}\ \bibnamefont {Thomas}},\ }\href {\doibase
  10.1002/1521-3773(20010504)40:9<1732::AID-ANIE17320>3.0.CO;2-7} {\bibfield
  {journal} {\bibinfo  {journal} {Angew. Chem. Int. Ed.}\ }\textbf {\bibinfo
  {volume} {40}},\ \bibinfo {pages} {1732} (\bibinfo {year}
  {2001})}\BibitemShut {NoStop}%
\bibitem [{\citenamefont {Sakamoto}\ \emph {et~al.}(2004)\citenamefont
  {Sakamoto}, \citenamefont {Suzuki}, \citenamefont {Kobayashi}, \citenamefont
  {Gao}, \citenamefont {Fukai}, \citenamefont {Inoue}, \citenamefont {Sato},\
  and\ \citenamefont {Tokito}}]{Sak04}%
  \BibitemOpen
  \bibfield  {author} {\bibinfo {author} {\bibfnamefont {Y.}~\bibnamefont
  {Sakamoto}}, \bibinfo {author} {\bibfnamefont {T.}~\bibnamefont {Suzuki}},
  \bibinfo {author} {\bibfnamefont {M.}~\bibnamefont {Kobayashi}}, \bibinfo
  {author} {\bibfnamefont {Y.}~\bibnamefont {Gao}}, \bibinfo {author}
  {\bibfnamefont {Y.}~\bibnamefont {Fukai}}, \bibinfo {author} {\bibfnamefont
  {Y.}~\bibnamefont {Inoue}}, \bibinfo {author} {\bibfnamefont
  {F.}~\bibnamefont {Sato}}, \ and\ \bibinfo {author} {\bibfnamefont
  {S.}~\bibnamefont {Tokito}},\ }\href {\doibase 10.1021/ja0476258} {\bibfield
  {journal} {\bibinfo  {journal} {J. Am. Chem. Soc.}\ }\textbf {\bibinfo
  {volume} {126}},\ \bibinfo {pages} {8138} (\bibinfo {year}
  {2004})}\BibitemShut {NoStop}%
\bibitem [{\citenamefont {Kera}\ \emph {et~al.}(2013)\citenamefont {Kera},
  \citenamefont {Hosoumi}, \citenamefont {Sato}, \citenamefont {Fukagawa},
  \citenamefont {Nagamatsu}, \citenamefont {Sakamoto}, \citenamefont {Suzuki},
  \citenamefont {Huang}, \citenamefont {Chen}, \citenamefont {Wee},
  \citenamefont {Coropceanu},\ and\ \citenamefont {Ueno}}]{Ker13}%
  \BibitemOpen
  \bibfield  {author} {\bibinfo {author} {\bibfnamefont {S.}~\bibnamefont
  {Kera}}, \bibinfo {author} {\bibfnamefont {S.}~\bibnamefont {Hosoumi}},
  \bibinfo {author} {\bibfnamefont {K.}~\bibnamefont {Sato}}, \bibinfo {author}
  {\bibfnamefont {H.}~\bibnamefont {Fukagawa}}, \bibinfo {author}
  {\bibfnamefont {S.-i.}\ \bibnamefont {Nagamatsu}}, \bibinfo {author}
  {\bibfnamefont {Y.}~\bibnamefont {Sakamoto}}, \bibinfo {author}
  {\bibfnamefont {T.}~\bibnamefont {Suzuki}}, \bibinfo {author} {\bibfnamefont
  {H.}~\bibnamefont {Huang}}, \bibinfo {author} {\bibfnamefont
  {W.}~\bibnamefont {Chen}}, \bibinfo {author} {\bibfnamefont {A.~T.~S.}\
  \bibnamefont {Wee}}, \bibinfo {author} {\bibfnamefont {V.}~\bibnamefont
  {Coropceanu}}, \ and\ \bibinfo {author} {\bibfnamefont {N.}~\bibnamefont
  {Ueno}},\ }\href {\doibase 10.1021/jp4032089} {\bibfield  {journal} {\bibinfo
   {journal} {J. Phys. Chem. C}\ }\textbf {\bibinfo {volume} {117}},\ \bibinfo
  {pages} {22428} (\bibinfo {year} {2013})}\BibitemShut {NoStop}%
\bibitem [{\citenamefont {Chen}\ and\ \citenamefont {Chao}(2005)}]{Che05}%
  \BibitemOpen
  \bibfield  {author} {\bibinfo {author} {\bibfnamefont {H.-Y.}\ \bibnamefont
  {Chen}}\ and\ \bibinfo {author} {\bibfnamefont {I.}~\bibnamefont {Chao}},\
  }\href {\doibase http://dx.doi.org/10.1016/j.cplett.2004.11.125} {\bibfield
  {journal} {\bibinfo  {journal} {Chem. Phys. Lett.}\ }\textbf {\bibinfo
  {volume} {401}},\ \bibinfo {pages} {539 } (\bibinfo {year}
  {2005})}\BibitemShut {NoStop}%
\bibitem [{\citenamefont {Ruiz~Delgado}\ \emph {et~al.}(2009)\citenamefont
  {Ruiz~Delgado}, \citenamefont {Pigg}, \citenamefont {da~Silva~Filho},
  \citenamefont {Gruhn}, \citenamefont {Sakamoto}, \citenamefont {Suzuki},
  \citenamefont {Osuna}, \citenamefont {Casado}, \citenamefont {Hernández},
  \citenamefont {Navarrete}, \citenamefont {Martinelli}, \citenamefont
  {Cornil}, \citenamefont {Sánchez-Carrera}, \citenamefont {Coropceanu},\ and\
  \citenamefont {Brédas}}]{Del09}%
  \BibitemOpen
  \bibfield  {author} {\bibinfo {author} {\bibfnamefont {M.~C.}\ \bibnamefont
  {Ruiz~Delgado}}, \bibinfo {author} {\bibfnamefont {K.~R.}\ \bibnamefont
  {Pigg}}, \bibinfo {author} {\bibfnamefont {D.~A.}\ \bibnamefont
  {da~Silva~Filho}}, \bibinfo {author} {\bibfnamefont {N.~E.}\ \bibnamefont
  {Gruhn}}, \bibinfo {author} {\bibfnamefont {Y.}~\bibnamefont {Sakamoto}},
  \bibinfo {author} {\bibfnamefont {T.}~\bibnamefont {Suzuki}}, \bibinfo
  {author} {\bibfnamefont {R.~M.}\ \bibnamefont {Osuna}}, \bibinfo {author}
  {\bibfnamefont {J.}~\bibnamefont {Casado}}, \bibinfo {author} {\bibfnamefont
  {V.}~\bibnamefont {Hernández}}, \bibinfo {author} {\bibfnamefont {J.~T.~L.}\
  \bibnamefont {Navarrete}}, \bibinfo {author} {\bibfnamefont {N.~G.}\
  \bibnamefont {Martinelli}}, \bibinfo {author} {\bibfnamefont
  {J.}~\bibnamefont {Cornil}}, \bibinfo {author} {\bibfnamefont {R.~S.}\
  \bibnamefont {Sánchez-Carrera}}, \bibinfo {author} {\bibfnamefont
  {V.}~\bibnamefont {Coropceanu}}, \ and\ \bibinfo {author} {\bibfnamefont
  {J.-L.}\ \bibnamefont {Brédas}},\ }\href {\doibase 10.1021/ja807528w}
  {\bibfield  {journal} {\bibinfo  {journal} {J. Am. Chem. Soc.}\ }\textbf
  {\bibinfo {volume} {131}},\ \bibinfo {pages} {1502} (\bibinfo {year}
  {2009})}\BibitemShut {NoStop}%
\bibitem [{\citenamefont {Ryno}\ \emph {et~al.}(2013)\citenamefont {Ryno},
  \citenamefont {Lee}, \citenamefont {Sears}, \citenamefont {Risko},\ and\
  \citenamefont {Brédas}}]{Ryn13}%
  \BibitemOpen
  \bibfield  {author} {\bibinfo {author} {\bibfnamefont {S.~M.}\ \bibnamefont
  {Ryno}}, \bibinfo {author} {\bibfnamefont {S.~R.}\ \bibnamefont {Lee}},
  \bibinfo {author} {\bibfnamefont {J.~S.}\ \bibnamefont {Sears}}, \bibinfo
  {author} {\bibfnamefont {C.}~\bibnamefont {Risko}}, \ and\ \bibinfo {author}
  {\bibfnamefont {J.-L.}\ \bibnamefont {Brédas}},\ }\href {\doibase
  10.1021/jp402991z} {\bibfield  {journal} {\bibinfo  {journal} {J. Phys. Chem.
  C}\ }\textbf {\bibinfo {volume} {117}},\ \bibinfo {pages} {13853} (\bibinfo
  {year} {2013})}\BibitemShut {NoStop}%
\bibitem [{\citenamefont {Faber}\ \emph {et~al.}(2014)\citenamefont {Faber},
  \citenamefont {Boulanger}, \citenamefont {Attaccalite}, \citenamefont
  {Duchemin},\ and\ \citenamefont {Blase}}]{Fab14}%
  \BibitemOpen
  \bibfield  {author} {\bibinfo {author} {\bibfnamefont {C.}~\bibnamefont
  {Faber}}, \bibinfo {author} {\bibfnamefont {P.}~\bibnamefont {Boulanger}},
  \bibinfo {author} {\bibfnamefont {C.}~\bibnamefont {Attaccalite}}, \bibinfo
  {author} {\bibfnamefont {I.}~\bibnamefont {Duchemin}}, \ and\ \bibinfo
  {author} {\bibfnamefont {X.}~\bibnamefont {Blase}},\ }\href {\doibase
  10.1098/rsta.2013.0271} {\bibfield  {journal} {\bibinfo  {journal} {Phil.
  Trans. R. Soc. A}\ }\textbf {\bibinfo {volume} {372}},\ \bibinfo {pages}
  {20130271} (\bibinfo {year} {2014})}\BibitemShut {NoStop}%
\bibitem [{Note2()}]{Note2}%
  \BibitemOpen
  \bibinfo {note} {The difference in the $GW_\protect \mathrm
  {e}|$DFT$_\protect \mathrm {g}$ gap with respect to the value we recently
  reported (3.05 eV in Ref.~\protect \rev@citealpnum {Li16}) is due to the
  different atomic coordinates used for the QM molecule. In this work atomic
  positions are used as provided in the cif file,\cite {Sie01} while in
  Ref.~\protect \rev@citealpnum {Li16} hydrogen atoms were optimized in a
  preliminary gas-phase DFT calculation}\BibitemShut {NoStop}%
\bibitem [{\citenamefont {Tsiper}\ and\ \citenamefont {Soos}(2003)}]{Tsi03}%
  \BibitemOpen
  \bibfield  {author} {\bibinfo {author} {\bibfnamefont {E.~V.}\ \bibnamefont
  {Tsiper}}\ and\ \bibinfo {author} {\bibfnamefont {Z.~G.}\ \bibnamefont
  {Soos}},\ }\href {\doibase 10.1103/PhysRevB.68.085301} {\bibfield  {journal}
  {\bibinfo  {journal} {Phys. Rev. B}\ }\textbf {\bibinfo {volume} {68}},\
  \bibinfo {pages} {085301} (\bibinfo {year} {2003})}\BibitemShut {NoStop}%
\bibitem [{Note3()}]{Note3}%
  \BibitemOpen
  \bibinfo {note} {\protect \leavevmode {\protect \color {red} A polarization
  energy of 0.96 (1.31) eV has been obtained for PEN HOMO (LUMO) with our $GW$
  approach combined with the standard polarizable continuum model,\cite {Duc16}
  using an isotropic dielectric constant of 3.5, representative of bulk
  PEN.\cite {Dav14} This compares well with the present embedded-$GW$ results
  employing the CR model\cite {Tsi01} for the MM subsystem: $\Delta ^{\protect
  \mathrm {COHSEX}}=$1.00 and 1.13 eV for PEN HOMO and LUMO, respectively. We
  emphasize, however, that the polarizable continuum model cannot account for
  the electrostatic crystal field effects that can shift the energy levels by
  as much as one eV with respect to the vacuum level }}\BibitemShut {NoStop}%
\bibitem [{\citenamefont {Refaely-Abramson}\ \emph {et~al.}(2013)\citenamefont
  {Refaely-Abramson}, \citenamefont {Sharifzadeh}, \citenamefont {Jain},
  \citenamefont {Baer}, \citenamefont {Neaton},\ and\ \citenamefont
  {Kronik}}]{Abr13}%
  \BibitemOpen
  \bibfield  {author} {\bibinfo {author} {\bibfnamefont {S.}~\bibnamefont
  {Refaely-Abramson}}, \bibinfo {author} {\bibfnamefont {S.}~\bibnamefont
  {Sharifzadeh}}, \bibinfo {author} {\bibfnamefont {M.}~\bibnamefont {Jain}},
  \bibinfo {author} {\bibfnamefont {R.}~\bibnamefont {Baer}}, \bibinfo {author}
  {\bibfnamefont {J.~B.}\ \bibnamefont {Neaton}}, \ and\ \bibinfo {author}
  {\bibfnamefont {L.}~\bibnamefont {Kronik}},\ }\href {\doibase
  10.1103/PhysRevB.88.081204} {\bibfield  {journal} {\bibinfo  {journal} {Phys.
  Rev. B}\ }\textbf {\bibinfo {volume} {88}},\ \bibinfo {pages} {081204}
  (\bibinfo {year} {2013})}\BibitemShut {NoStop}%
\bibitem [{\citenamefont {Campbell}\ \emph {et~al.}(1962)\citenamefont
  {Campbell}, \citenamefont {Robertson},\ and\ \citenamefont
  {Trotter}}]{Cam62}%
  \BibitemOpen
  \bibfield  {author} {\bibinfo {author} {\bibfnamefont {R.~B.}\ \bibnamefont
  {Campbell}}, \bibinfo {author} {\bibfnamefont {J.~M.}\ \bibnamefont
  {Robertson}}, \ and\ \bibinfo {author} {\bibfnamefont {J.}~\bibnamefont
  {Trotter}},\ }\href {\doibase 10.1107/S0365110X62000699} {\bibfield
  {journal} {\bibinfo  {journal} {Acta Crystallographica}\ }\textbf {\bibinfo
  {volume} {15}},\ \bibinfo {pages} {289} (\bibinfo {year} {1962})}\BibitemShut
  {NoStop}%
\bibitem [{\citenamefont {Schiefer}\ \emph {et~al.}(2007)\citenamefont
  {Schiefer}, \citenamefont {Huth}, \citenamefont {Dobrinevski},\ and\
  \citenamefont {Nickel}}]{Sch07}%
  \BibitemOpen
  \bibfield  {author} {\bibinfo {author} {\bibfnamefont {S.}~\bibnamefont
  {Schiefer}}, \bibinfo {author} {\bibfnamefont {M.}~\bibnamefont {Huth}},
  \bibinfo {author} {\bibfnamefont {A.}~\bibnamefont {Dobrinevski}}, \ and\
  \bibinfo {author} {\bibfnamefont {B.}~\bibnamefont {Nickel}},\ }\href
  {\doibase 10.1021/ja0730516} {\bibfield  {journal} {\bibinfo  {journal} {J.
  Am. Chem. Soc.}\ }\textbf {\bibinfo {volume} {129}},\ \bibinfo {pages}
  {10316} (\bibinfo {year} {2007})}\BibitemShut {NoStop}%
\bibitem [{\citenamefont {Coropceanu}\ \emph {et~al.}(2002)\citenamefont
  {Coropceanu}, \citenamefont {Malagoli}, \citenamefont {da~Silva~Filho},
  \citenamefont {Gruhn}, \citenamefont {Bill},\ and\ \citenamefont
  {Brédas}}]{Cor02}%
  \BibitemOpen
  \bibfield  {author} {\bibinfo {author} {\bibfnamefont {V.}~\bibnamefont
  {Coropceanu}}, \bibinfo {author} {\bibfnamefont {M.}~\bibnamefont
  {Malagoli}}, \bibinfo {author} {\bibfnamefont {D.~A.}\ \bibnamefont
  {da~Silva~Filho}}, \bibinfo {author} {\bibfnamefont {N.~E.}\ \bibnamefont
  {Gruhn}}, \bibinfo {author} {\bibfnamefont {T.~G.}\ \bibnamefont {Bill}}, \
  and\ \bibinfo {author} {\bibfnamefont {J.~L.}\ \bibnamefont {Brédas}},\
  }\href {\doibase 10.1103/PhysRevLett.89.275503} {\bibfield  {journal}
  {\bibinfo  {journal} {Phys. Rev. Lett.}\ }\textbf {\bibinfo {volume} {89}},\
  \bibinfo {pages} {275503} (\bibinfo {year} {2002})}\BibitemShut {NoStop}%
\bibitem [{\citenamefont {Crocker}\ \emph {et~al.}(1993)\citenamefont
  {Crocker}, \citenamefont {Wang},\ and\ \citenamefont {Kebarle}}]{Cro93}%
  \BibitemOpen
  \bibfield  {author} {\bibinfo {author} {\bibfnamefont {L.}~\bibnamefont
  {Crocker}}, \bibinfo {author} {\bibfnamefont {T.}~\bibnamefont {Wang}}, \
  and\ \bibinfo {author} {\bibfnamefont {P.}~\bibnamefont {Kebarle}},\ }\href
  {\doibase 10.1021/ja00070a030} {\bibfield  {journal} {\bibinfo  {journal} {J.
  Am. Chem. Soc.}\ }\textbf {\bibinfo {volume} {115}},\ \bibinfo {pages} {7818}
  (\bibinfo {year} {1993})}\BibitemShut {NoStop}%
\bibitem [{\citenamefont {Kakuta}\ \emph {et~al.}(2007)\citenamefont {Kakuta},
  \citenamefont {Hirahara}, \citenamefont {Matsuda}, \citenamefont {Nagao},
  \citenamefont {Hasegawa}, \citenamefont {Ueno},\ and\ \citenamefont
  {Sakamoto}}]{Kak07}%
  \BibitemOpen
  \bibfield  {author} {\bibinfo {author} {\bibfnamefont {H.}~\bibnamefont
  {Kakuta}}, \bibinfo {author} {\bibfnamefont {T.}~\bibnamefont {Hirahara}},
  \bibinfo {author} {\bibfnamefont {I.}~\bibnamefont {Matsuda}}, \bibinfo
  {author} {\bibfnamefont {T.}~\bibnamefont {Nagao}}, \bibinfo {author}
  {\bibfnamefont {S.}~\bibnamefont {Hasegawa}}, \bibinfo {author}
  {\bibfnamefont {N.}~\bibnamefont {Ueno}}, \ and\ \bibinfo {author}
  {\bibfnamefont {K.}~\bibnamefont {Sakamoto}},\ }\href {\doibase
  10.1103/PhysRevLett.98.247601} {\bibfield  {journal} {\bibinfo  {journal}
  {Phys. Rev. Lett.}\ }\textbf {\bibinfo {volume} {98}},\ \bibinfo {pages}
  {247601} (\bibinfo {year} {2007})}\BibitemShut {NoStop}%
\bibitem [{\citenamefont {Ciuchi}\ \emph {et~al.}(2012)\citenamefont {Ciuchi},
  \citenamefont {Hatch}, \citenamefont {H\"ochst}, \citenamefont {Faber},
  \citenamefont {Blase},\ and\ \citenamefont {Fratini}}]{Ciu12}%
  \BibitemOpen
  \bibfield  {author} {\bibinfo {author} {\bibfnamefont {S.}~\bibnamefont
  {Ciuchi}}, \bibinfo {author} {\bibfnamefont {R.~C.}\ \bibnamefont {Hatch}},
  \bibinfo {author} {\bibfnamefont {H.}~\bibnamefont {H\"ochst}}, \bibinfo
  {author} {\bibfnamefont {C.}~\bibnamefont {Faber}}, \bibinfo {author}
  {\bibfnamefont {X.}~\bibnamefont {Blase}}, \ and\ \bibinfo {author}
  {\bibfnamefont {S.}~\bibnamefont {Fratini}},\ }\href {\doibase
  10.1103/PhysRevLett.108.256401} {\bibfield  {journal} {\bibinfo  {journal}
  {Phys. Rev. Lett.}\ }\textbf {\bibinfo {volume} {108}},\ \bibinfo {pages}
  {256401} (\bibinfo {year} {2012})}\BibitemShut {NoStop}%
\bibitem [{\citenamefont {Valeev}\ \emph {et~al.}(2006)\citenamefont {Valeev},
  \citenamefont {Coropceanu}, \citenamefont {da~Silva~Filho}, \citenamefont
  {Salman},\ and\ \citenamefont {Brédas}}]{Val06}%
  \BibitemOpen
  \bibfield  {author} {\bibinfo {author} {\bibfnamefont {E.~F.}\ \bibnamefont
  {Valeev}}, \bibinfo {author} {\bibfnamefont {V.}~\bibnamefont {Coropceanu}},
  \bibinfo {author} {\bibfnamefont {D.~A.}\ \bibnamefont {da~Silva~Filho}},
  \bibinfo {author} {\bibfnamefont {S.}~\bibnamefont {Salman}}, \ and\ \bibinfo
  {author} {\bibfnamefont {J.-L.}\ \bibnamefont {Brédas}},\ }\href {\doibase
  10.1021/ja061827h} {\bibfield  {journal} {\bibinfo  {journal} {J. Am. Chem.
  Soc.}\ }\textbf {\bibinfo {volume} {128}},\ \bibinfo {pages} {9882} (\bibinfo
  {year} {2006})}\BibitemShut {NoStop}%
\bibitem [{\citenamefont {Li}\ \emph {et~al.}(2012)\citenamefont {Li},
  \citenamefont {Coropceanu},\ and\ \citenamefont {Brédas}}]{Li12}%
  \BibitemOpen
  \bibfield  {author} {\bibinfo {author} {\bibfnamefont {Y.}~\bibnamefont
  {Li}}, \bibinfo {author} {\bibfnamefont {V.}~\bibnamefont {Coropceanu}}, \
  and\ \bibinfo {author} {\bibfnamefont {J.-L.}\ \bibnamefont {Brédas}},\
  }\href {\doibase 10.1021/jz301575u} {\bibfield  {journal} {\bibinfo
  {journal} {J. Phys. Chem. Lett.}\ }\textbf {\bibinfo {volume} {3}},\ \bibinfo
  {pages} {3325} (\bibinfo {year} {2012})}\BibitemShut {NoStop}%
\end{thebibliography}%


\end{document}